\documentclass[journal]{IEEEtran}
\IEEEoverridecommandlockouts
\usepackage{setspace}
\usepackage{cite}
\usepackage{amsmath,amssymb,amsfonts,mathtools,bm}
\usepackage{amsthm}
\usepackage{algorithm}
\usepackage{algpseudocode}
\usepackage{graphicx}
\usepackage{textcomp}
\usepackage{xcolor,float}
\usepackage{tabularx}
\usepackage{subcaption}
\usepackage{textcomp}
\usepackage{diagbox}
\usepackage{svg}
\usepackage{booktabs}
\usepackage{hyperref}

\newtheorem{problem}{Problem}
\usepackage{graphicx}
\usepackage{makecell}
\usepackage{amsthm}
\usepackage{caption}
\usepackage{sidecap}
\usepackage{microtype}
\usepackage{booktabs} 
\usepackage{multirow}
\usepackage{adjustbox} 
\usepackage{amsmath}
\usepackage{amssymb}
\usepackage{colortbl}
\usepackage{makecell}
\usepackage{url}
\usepackage{balance}
\usepackage{bbding}
\usepackage{hyperref}
\usepackage{graphicx}
\usepackage{sidecap}
\usepackage{microtype}
\usepackage{graphicx}
\usepackage{booktabs} 
\usepackage{multirow}
\usepackage{array}
\usepackage{graphicx}
\usepackage{makecell}
\usepackage{amsmath}
\usepackage{amssymb}

\usepackage{tikz}

\usepackage{changes}
\usepackage{ifthen}
\newboolean{showchanges}
\setboolean{showchanges}{false}  

\newcommand{\add}[1]{%
    \ifthenelse{\boolean{showchanges}}%
        {\textcolor{blue}{#1}}
        {#1\relax}
}

\definecolor{lime}{HTML}{A6CE39}
\DeclareRobustCommand{\orcidicon}{%
    \begin{tikzpicture}
    \draw[lime, fill=lime] (0,0) 
    circle [radius=0.16] 
    node[white] {{\fontfamily{qag}\selectfont \tiny ID}};    \draw[white, fill=white] (-0.0625,0.095) 
    circle [radius=0.007];    \end{tikzpicture}
    \hspace{-2mm}}
\foreach \x in {A, ..., Z}{%
    \expandafter\xdef\csname orcid\x\endcsname{\noexpand\href{https://orcid.org/\csname orcidauthor\x\endcsname}{\noexpand\orcidicon}}
    }

\allowdisplaybreaks
\renewcommand{\baselinestretch}{1}

\begin{document}

\title{RadioDiff-FS: Physics-Informed Manifold Alignment in Few-Shot Diffusion Models for High-Fidelity Radio Map Construction}

\author{
Xiucheng Wang\orcidA{},~\IEEEmembership{Graduate Student Member,~IEEE,}
Zixuan Guo\orcidB{},
Nan Cheng\orcidC{},~\IEEEmembership{Senior Member,~IEEE,} 
Zhisheng Yin\orcidH{},~\IEEEmembership{Member,~IEEE,}
Ruijin Sun\orcidD{},~\IEEEmembership{Member,~IEEE,}
Xuemin (Sherman) Shen\orcidG{},~\IEEEmembership{Fellow,~IEEE}

\thanks{
\par This work was supported by the National Key Research and Development Program of China (2024YFB907500).
\par Xiucheng Wang, Nan Cheng, Zhisheng Yin, and Ruijin Sun are with the State Key Laboratory of ISN and School of Telecommunications Engineering, Xidian University, Xi’an 710071, China (e-mail: xcwang\_1@stu.xidian.edu.cn; dr.nan.cheng@ieee.org, \{zsyin, sunruijin\}@xidian.edu.cn);\textit{(Corresponding author: Nan Cheng.)}.
\par Zixuan Guo is with the School of Physics, Xidian University, Xi’an 710071, China (e-mail: 24179100067@stu.xidian.edu.cn)
\par Xuemin (Sherman) Shen is with the Department of Electrical and Computer Engineering, University of Waterloo, Waterloo, N2L 3G1, Canada (e-mail: sshen@uwaterloo.ca).
}

} 

    \maketitle

\IEEEdisplaynontitleabstractindextext

\IEEEpeerreviewmaketitle

\begin{abstract}
Radio maps (RMs) provide spatially continuous propagation characterizations essential for 6G network planning, but high-fidelity RM construction remains challenging. Rigorous electromagnetic solvers incur prohibitive computational latency, while data-driven models demand massive labeled datasets and generalize poorly from simplified simulations to complex multipath environments. This paper proposes RadioDiff-FS, a few-shot diffusion framework that adapts a pretrained main-path generator to multipath-rich target domains with only a small number of high-fidelity samples. The adaptation is grounded in a theoretical decomposition of the multipath RM into a dominant main-path component and a directionally sparse residual. This decomposition shows that the cross-domain shift corresponds to a bounded and geometrically structured feature translation rather than an arbitrary distribution change. A direction-consistency loss (DCL) is then introduced to constrain diffusion score updates along physically plausible propagation directions, thereby suppressing phase-inconsistent artifacts that arise in the low-data regime. Experiments show that RadioDiff-FS reduces NMSE by 59.5\% on static RMs and by 74.0\% on dynamic RMs relative to the vanilla diffusion baseline, achieving an SSIM of 0.9752 and a PSNR of 36.37 dB under severely limited supervision. Even in a one-shot setting with a single target-domain sample per scene, RadioDiff-FS outperforms all fully supervised baselines, confirming that the directional constraint provides an effective inductive bias under extreme data scarcity. Code is available at \url{https://github.com/UNIC-Lab/RadioDiff-FS}.
\end{abstract}

\section{Introduction}
Radio maps (RMs) provide a spatially continuous abstraction of large-scale wireless propagation and are increasingly recognized as a core spatial infrastructure for 6G intelligence \cite{wang2026tutorial,shen2023toward}. By decoupling channel state information (CSI) acquisition from real-time pilot overhead, RMs serve as a unified \cite{wang2024tutorial}, queryable substrate for latency-critical functionalities including user-centric coverage optimization \cite{Yang2023a}, proactive beamforming \cite{11314850}, and electromagnetic digital twin construction \cite{11152929}. Unlike transient pointwise channel estimates, RMs explicitly encode global spatial layouts and capture mesoscopic propagation features that govern network behavior, from occlusion boundaries defined by building geometries to diffraction at sharp edges and high-transmission apertures \cite{11278649,deschamps1972ray}. This structural awareness is particularly valuable at high frequencies. In dense millimeter-wave (mmWave) deployments \cite{8373698}, a pre-computed RM enables anticipation of non-line-of-sight (NLoS) blockage \cite{liu2025foundation,11036155}, facilitating proactive UAV-assisted relay planning and robust connectivity under dynamic environmental conditions \cite{wang2022joint}.

Despite this potential, high-fidelity RM construction is constrained by two fundamental bottlenecks regarding computational efficiency \cite{deschamps1972ray,jones2013theory} and distributional generalization \cite{wang2024radiodiff,11278649,vapnik1998statistical}. Rigorous electromagnetic solvers, including full-wave methods and high-fidelity ray-tracing engines, provide authoritative approximations of real-world propagation physics \cite{deschamps1972ray,jones2013theory}. However, when modeling carrier-frequency dependence, fine spatial resolution, and higher-order multipath interactions in material-heterogeneous environments, these solvers frequently incur minute-level latency even for modest scene scales \cite{deschamps1972ray}. Such overhead precludes rapid design-space exploration and the real-time scene updates required by dynamic network digital twins \cite{11152929}. Data-driven generative models, including U-Net \cite{levie2021radiounet}, Transformer architectures \cite{9753644}, and diffusion-based backbones, can synthesize kilometer-scale RMs within sub-second runtimes \cite{11278649,wang2024radiodiff,11282987}. Yet their generalization is limited by the training distribution. Because scalable training often relies on simplified propagation surrogates, these models frequently fail to capture nonstationary phenomena that dominate real-world environments, including high-order reflections and edge diffraction \cite{goodfellow2016deep,11278649}. Diffusion models further exacerbate this data demand, as their iterative reverse process requires comprehensive coverage of the data manifold for stable trajectory learning, intensifying the dependence on massive multipath datasets that are prohibitively expensive to acquire.
\begin{figure*}[ht]
    \centering
    \includegraphics[width=1.0\linewidth]{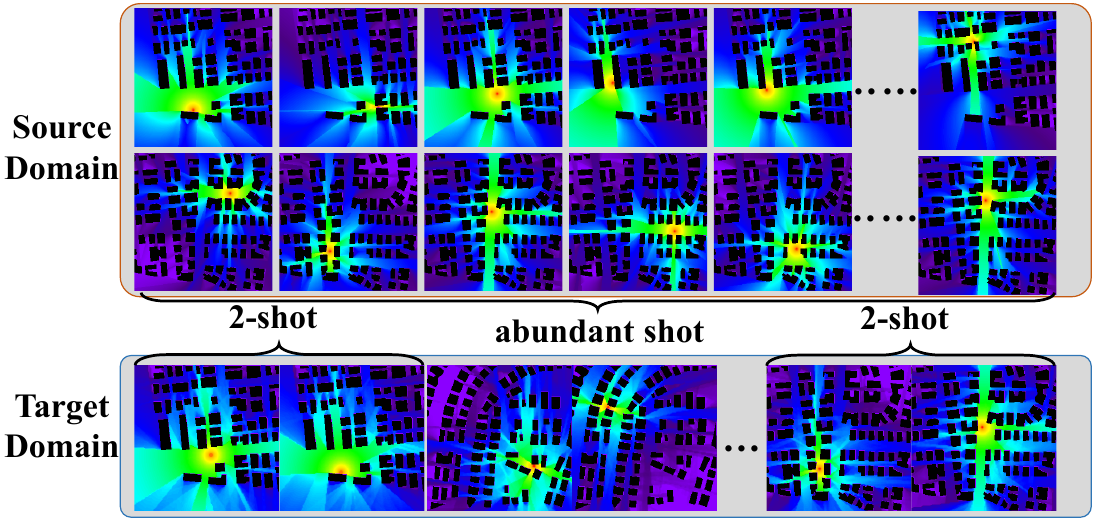}
    \caption{Illustration of the few-shot learning for RM construction.}
    \vspace{-12pt}
    \label{fig-demo}
\end{figure*}

The underlying obstacle is the unfavorable economics of high-fidelity data acquisition, as illustrated in Fig.~\ref{fig-demo}. Field measurements incur prohibitive costs in skilled labor, specialized equipment, and logistical coordination, rendering large-scale collection across diverse cities, material classes, and architectural morphologies impractical. High-fidelity ray-tracing simulation faces comparable scalability barriers. Generating fifty thousand samples via rigorous multipath tracing would typically require over eight hundred compute-hours \cite{wolfle1997intelligent}, and this burden increases substantially with higher-order reflections, realistic material catalogs, and vertical scene complexity. Exhaustively expanding datasets to cover all relevant multipath interactions is neither scalable nor economically viable \cite{zhang2024metadiff}. These constraints motivate a paradigm shift from indiscriminate data accumulation to principled cross-domain adaptation. The central hypothesis is that a tractable solution should leverage abundant, computationally inexpensive main-path data to learn global spatial layouts, while relying on only a small set of high-fidelity samples to correct structural biases and inject domain-specific multipath details. This few-shot strategy reframes the objective as transfer learning that couples generative modeling with physical consistency.

To address these challenges, we propose RadioDiff-FS, a framework that reformulates high-fidelity RM construction as a physics-informed few-shot diffusion adaptation problem. Rather than treating RM generation as a black-box image-to-image translation task, our approach builds on a rigorous decomposition of the electromagnetic field. The multipath RM is formalized as a dominant main-path component augmented by a structured, directionally sparse residual. This decomposition provides a strong physical inductive bias, indicating that adaptation from a main-path prior to a multipath target involves learning a bounded geometric correction rather than synthesizing a new distribution from scratch. RadioDiff-FS therefore transfers global spatial layout knowledge from the pre-trained backbone and concentrates generative capacity on localized diffraction and reflection effects. A Direction-Consistency Loss (DCL) is introduced to regularize the fine-tuning process. The DCL constrains score function updates to evolve along physically plausible propagation directions by penalizing gradient components orthogonal to the valid propagation manifold. This suppresses phase-inconsistent artifacts and mode hallucinations in low-data regimes while maintaining geometric fidelity at diffraction edges and shadow boundaries. The main contributions are summarized as follows.

\begin{enumerate}
\item We propose RadioDiff-FS, which is, to the best of our knowledge, the first framework to formulate high-fidelity RM construction as a few-shot diffusion adaptation problem. By transferring knowledge from a pre-trained backbone learned on computationally inexpensive main-path priors to multipath-rich target domains, it bridges the distributional gap caused by complex propagation environments and circumvents prohibitive data acquisition costs.

\item We establish a rigorous theoretical decomposition of the multipath RM into a dominant main-path component and a directionally sparse residual. This formulation provides a physical inductive bias and demonstrate that the cross-domain transition is characterized by learnable geometric shifts rather than chaotic distribution changes, and lays the theoretical foundation for efficient few-shot adaptation.

\item We design a Direction-Consistency Loss that functions as a lightweight manifold projector within the diffusion process. By constraining score function updates to physically plausible propagation directions, DCL penalizes gradient components associated with mode hallucinations, thereby ensuring that synthesized maps maintain global spatial coherence while accurately recovering fine-grained diffraction and reflection details.

\item Extensive experiments on the RadioMapSeer benchmark demonstrate that RadioDiff-FS achieves state-of-the-art performance in both static and dynamic scenarios, with superior edge preservation and significantly lower reconstruction error compared to existing baselines under extremely limited supervision. Ablation studies validate the critical role of DCL in stabilizing the adaptation trajectory.
\end{enumerate}

\section{Related Work and Preliminary}
\subsection{RM Construction Methods}
RMs offer a spatially continuous abstraction of wireless propagation environments and serve as a fundamental enabler for environment-aware 6G applications, such as coverage optimization, localization, and digital twin construction \cite{wang2026tutorial}. In recent years, learning-based approaches have shown great promise in improving the accuracy and efficiency of RM construction, particularly in complex urban and indoor environments where traditional statistical models based on radial symmetry fail to account for diffraction, shadowing, and occlusion effects. However, despite their effectiveness, these neural models often rely on large-scale training data that is expensive to obtain either through simulation or measurement.

Initial studies, such as RadioUNet \cite{levie2021radiounet}, demonstrated that convolutional encoder–decoder architectures can produce accurate two-dimensional (2D) pathloss maps from structured urban layouts, significantly reducing inference latency compared with ray tracing. Subsequent works like PMNet improved generalization and sample efficiency through advanced vision architectures. Although these discriminative models remain computationally efficient and interpretable, they typically require dense supervision and are difficult to adapt to out-of-distribution environments or scenarios with sparse sensor coverage. Moreover, most of these methods focus on planar pathloss estimation, ignoring vertical structure and multi-metric coupling \cite{11282987,wang2024radiodiff}. To overcome the limitations of pointwise regression under sparse measurements, conditional generative models have been explored. RME-GAN \cite{zhang2023rme} and DeepREM \cite{deeprem} propose using conditional generative adversarial networks and completion autoencoders to reconstruct dense RMs from limited measurements. These models integrate global propagation patterns and local texture variations to enhance estimation performance. While they significantly outperform interpolation baselines, they still rely heavily on simulation data and do not incorporate electromagnetic consistency into the learning process. Graph neural networks (GNNs) offer another route by encoding structural priors derived from building layouts and sensor topology \cite{chen2023graph}. Recent methods formulate sparse measurements as nodes and use edge structures informed by connectivity or spatial correlation to facilitate map completion. RadioGAT \cite{radiogat} extends this by capturing spatial–spectral dependencies across multiple frequency bands. These models offer improved robustness to sampling irregularities and can generalize under semi-supervised settings. However, without explicit electromagnetic constraints, their behavior around complex propagation features such as diffraction edges and shadow zones remains difficult to guarantee.

Diffusion models have recently emerged as a powerful alternative due to their generative structure and sample flexibility. RadioDiff \cite{wang2024radiodiff} models RM generation as a conditional diffusion process and introduces attention mechanisms and fast Fourier modules to enhance feature extraction in dynamic environments. RadioDiff-3D builds upon this by incorporating three-dimensional (3D) convolutional networks and utilizing the UrbanRadio3D dataset, which contains multi-height, multi-parameter RMs including pathloss, direction-of-arrival (DoA), and time-of-arrival (ToA) \cite{11083758}. These models achieve state-of-the-art performance in high-dimensional map estimation. However, without embedding physical priors, they may fail to reconstruct physically meaningful structures in unseen domains, especially in low-data regimes. To address this, physics-informed models have been introduced to embed electromagnetic theory into neural estimators. PEFNet integrates the volume integral equation into the learning objective, constraining the relationship between incident and total electric fields. More advanced methods, such as PhyRMDM \cite{jia2025rmdm} and RadioDiff-$k^2$ \cite{11278649}, incorporate Helmholtz residuals, source terms, and boundary conditions into diffusion model training. RadioDiff-$k^2$ in particular establishes a theoretical correspondence between wave propagation singularities and negative wave-number domains, enabling the network to identify and reconstruct diffraction-sensitive structures via a two-stage generative process. These approaches significantly improve physical plausibility and generalization, but they often assume access to accurate simulations or well-calibrated measurements, which remain costly and difficult to obtain at scale. Task-specific strategies such as RM inpainting based on propagation priority and radio depth maps have also been proposed to address areas where sensing is physically restricted \cite{wang2025radiodiff}. In parallel, the development of large-scale datasets such as UrbanRadio3D has greatly expanded the coverage of height, geometry, and multi-metric labeling, providing an essential benchmark for high-fidelity 3D×3D RM construction. Nevertheless, the cost and time required for data generation via ray tracing or controlled field measurements remain prohibitive in most deployment scenarios, and the resulting models often exhibit limited cross-region transferability.

In summary, although learning-based methods for RM construction have achieved impressive results across a variety of modeling paradigms, they remain heavily reliant on large volumes of high-quality data, either from simulation or measurement. However, due to the inherently high cost of generating such data, especially for diverse urban environments and 3D geometries, reliance on large-scale supervision represents a critical bottleneck. This reality motivates a paradigm shift toward few-shot adaptation, wherein a pre-trained model is efficiently fine-tuned to a new deployment scenario using only a handful of labeled samples. 
\begin{figure*}
    \centering
    \includegraphics[width=1.0\linewidth]{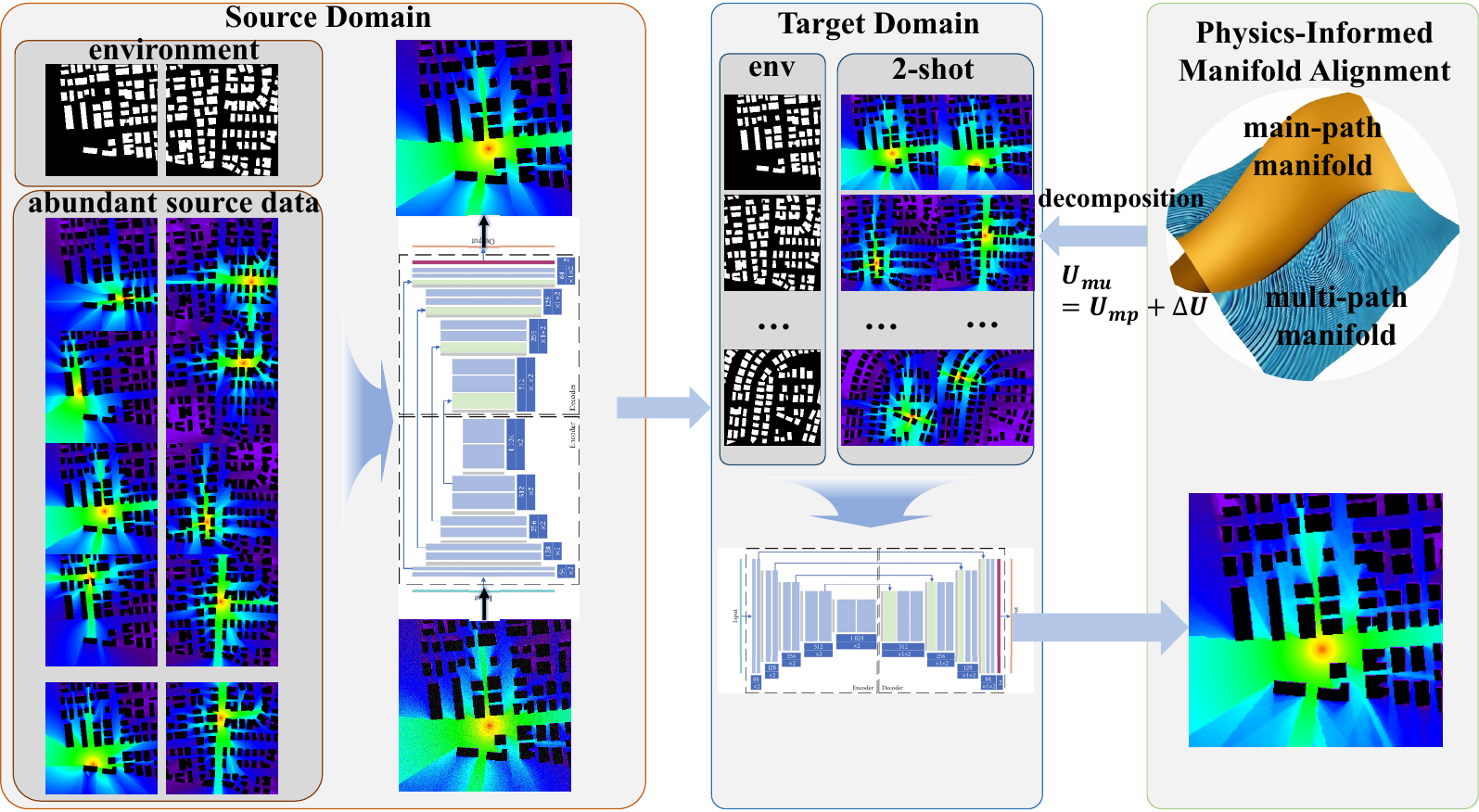}
    \caption{The RadioDiff-FS framework. The model pre-trains on abundant main-path data (left) and adapts to the multipath target domain using few-shot supervision (middle), whose transfer is enabled by physics-informed manifold alignment.}
    \label{fig-method}
\end{figure*}
\subsection{EM Computing Methods for RM Construction}
Recent advances in deep learning-based RM construction have underscored the importance of accurate and scalable ground truth generation. In practice, the dominant path model (DPM) \cite{dpm} has emerged as the standard solver for large-scale RM datasets due to its computational efficiency and fast rendering capabilities. By modeling only the strongest propagation path between transmitter and receiver while omitting multipath effects, DPM enables rapid data generation across dense spatial grids and diverse environmental layouts. As a result, mainstream benchmarks such as RadioMapSeer, UrbanRadio3D, and UrbanMIMOMap have all adopted DPM as their primary simulation engine. These datasets typically comprise tens of thousands of labeled samples, with UrbanRadio3D notably scaling up to over ten million images, making it the largest publicly available RM dataset to date. The widespread use of DPM reflects a practical trade-off: while it sacrifices full physical fidelity, its low computational cost makes it ideal for supporting supervised training of large neural architectures. Shooting-and-bouncing ray (SBR) \cite{bhalla1996three} methods, which adopt a ray-launching approach that allows for multiple specular reflections, provide a richer geometric representation than DPM. However, their sampling-based mechanism does not guarantee that rays will reach all regions of interest, particularly in dense, occluded, or non-line-of-sight (NLOS) environments. This coverage uncertainty introduces spatial sparsity that is undesirable in dense RM estimation tasks, particularly when complete spatial coverage is required for training or evaluation. As a result, although SBR is used in some site-specific simulations, it has seen limited adoption in large-scale RM dataset generation compared to DPM.

In contrast, intelligent ray tracing (IRT) \cite{irt} offers a more rigorous modeling paradigm that more faithfully adheres to electromagnetic propagation principles. By solving high-frequency approximations of Maxwell’s equations, IRT explicitly accounts for reflection, diffraction, and scattering, thereby generating multipath-aware RMs that more accurately represent complex propagation effects. This makes IRT particularly suitable for reconstructing signal behavior near sharp edges, shadow boundaries, and other critical regions. However, the substantial computational overhead associated with full multipath tracing has constrained the scale at which IRT can be applied. As a result, only a small number of publicly available datasets incorporate IRT-simulated labels. For instance, RadioMapSeer includes an IRT subset of just over 1000 samples, which remains insufficient for training contemporary deep neural networks without relying on transfer learning or extensive data augmentation.

\subsection{Denoising Diffusion Model}
DMs have emerged as highly effective generative frameworks capable of learning complex data distributions and synthesizing high-fidelity samples with remarkable precision \cite{ho2020denoising}. Among these, score-based DMs provide a principled approach grounded in stochastic differential equations (SDEs), wherein data is progressively perturbed by noise in the forward process and subsequently recovered by learning the score function, defined as the gradient of the log-probability density of the noisy data \cite{song2020score}. Unlike conventional DMs that rely on discrete-time Markov chains, score-based models employ continuous-time dynamics, making them particularly suitable for inverse problems such as RM reconstruction under the Bayesian sampling paradigm. Formally, the forward SDE is defined as
\begin{align}
d\bm{x} = f(\bm{x}, t) dt + g(t) d\bm{w},
\end{align}
where $t$ is the diffusion time index, and \( f(\bm{x}, t) \) denotes the drift term, \( g(t) \) is a time-dependent diffusion coefficient, and \( d\bm{w} \) represents a standard Wiener process. As \( t \) increases, the perturbed distribution \( p_t(\bm{x}) \) asymptotically approaches an isotropic Gaussian. The generative process involves solving the corresponding reverse-time SDE:
\begin{align}
d\bm{x} = \left[ f(\bm{x}, t) - g^2(t) \nabla_{\bm{x}} \log p_t(\bm{x}) \right] dt + g(t) d\bm{\bar{w}},
\end{align}
where the score function \( \nabla_{\bm{x}} \log p_t(\bm{x}) \) is typically approximated by a neural network \( s_\theta(\bm{x}, t) \), trained to satisfy
\begin{align}
s_\theta(\bm{x}, t) \approx \nabla_{\bm{x}} \log p_t(\bm{x}).
\end{align}

For deterministic sampling, the equivalent probability flow ordinary differential equation (ODE) is
\begin{align}
d\bm{x} = \left[ f(\bm{x}, t) - \frac{1}{2} g^2(t) \nabla_{\bm{x}} \log p_t(\bm{x}) \right] dt,
\end{align}
which removes stochasticity while preserving sample quality, thereby drawing close connections to denoising diffusion probabilistic models (DDPMs). In the DDPM framework, the forward process is discretized as
\begin{align}
q(\bm{x}_t | \bm{x}_{t-1}) = \mathcal{N}(\bm{x}_t; \alpha_t \bm{x}_{t-1}, \beta_t \bm{I}),
\end{align}
with the score function approximated via a learned denoiser \( \epsilon_\theta \) as
\begin{align}
s_\theta(\bm{x}, t) = -\frac{\epsilon_\theta(\bm{x}, t)}{\sqrt{1 - \bar{\alpha}_t}}.
\end{align}
Thus, DDPM can be viewed as a discretized instance of a score-based continuous model under a variance-preserving schedule.

In the context of RM generation, a decoupled formulation has recently been proposed to enhance generative stability, notably in the RadioDiff framework \cite{wang2024radiodiff}. Rather than injecting noise directly, the decoupled diffusion model (DDM) first attenuates the input signal to a baseline before introducing noise. The forward transition from clean input \( \bm{x}_0 \) to noisy state \( \bm{x}_t \) is governed by
\begin{align}
q\left(\bm{x}_t \mid \bm{x}_0\right) = \mathcal{N}\left(\gamma_t \bm{x}_0, \delta_t^2 \bm{I}\right),
\end{align}
where \( \gamma_t \) controls signal decay and \( \delta_t \) determines noise variance. This process can be equivalently expressed through the following SDE:
\begin{align}
&d \bm{x}_t = \bm{f}_t \bm{x}_t dt + g_t d\bm{\epsilon}_t,\\
&\bm{f}_t = \frac{d \log \gamma_t}{dt},\\
&\int_{0}^{t}\bm{f}_t dt = -x_0,\\
&g_t^2 = \frac{d \delta_t^2}{dt} - 2 \bm{f}_t \delta_t^2,
\end{align}
where \( \bm{f}_t \) defines the rate of attenuation and \( g_t^2 \) governs the rate of noise accumulation.

To generate samples, the reverse-time SDE is solved:
\begin{align}
d \bm{x}_t = \left[\bm{f}_t \bm{x}_t - g_t^2 \nabla_{\bm{x}} \log q\left(\bm{x}_t\right)\right] dt + g_t d\overline{\bm{\epsilon}}_t.
\end{align}
The structured separation between signal attenuation and noise addition enables more stable training and higher-quality synthesis. Notably, the deterministic component of DDM simplifies the forward mapping as
\begin{align}
q(\bm{x}_t|\bm{x}_0) = \mathcal{N}\left(\bm{x}_0 + \int_0^t \bm{f}_\tau d\tau, t\bm{I}\right),
\end{align}
and permits an efficient reverse-step formulation:
\begin{align}
q\left(\bm{z}_{t-\Delta t} \mid \bm{z}_t, \bm{z}_0\right)  &=\mathcal{N}\left(\bm{z}_{t} +\int_t^{t-\Delta t} \bm{f}_t \mathrm{~d} t\right. \notag\\
& \left.\qquad\qquad-\frac{\Delta t}{\sqrt{t}} \bm{\epsilon}, \frac{\Delta t(t-\Delta t)}{t} \bm{I}\right).\label{ddm-reverse}
\end{align}
By decoupling signal dynamics and stochasticity, the DDM approach significantly improves generative quality, robustness, and computational efficiency. These properties render it highly advantageous for high-resolution and structure-aware RM reconstruction tasks under sparse supervision, where stability and controllability are essential.

\section{System Model and Problem Formulation}

We consider a wireless propagation scenario over a discretized planar region represented as an $N \times N$ uniform grid. Each cell in this grid corresponds to a spatial resolution element of width $\Delta$. The cell size is assumed small enough that the pathloss within it can be treated as constant. A RM over this region is defined as a matrix $\bm{P} \in \mathbb{R}^{N \times N}$, where each entry $P(i,j)$ denotes the pathloss at the center of cell $(i,j)$. The pathloss is measured at a fixed carrier frequency $f_c$ with the corresponding wavelength $\lambda = c / f_c$, where $c$ denotes the speed of light in free space. A single transmitter, modeled as a point radiator, is positioned at $\bm{r} = \langle d_x, d_y, d_z \rangle$, where $(d_x, d_y)$ denotes the horizontal coordinates in the 2D region and $d_z > 0$ indicates the height above the ground. The transmitter emits electromagnetic energy isotropically or according to a specified antenna gain pattern, and serves as the sole radiation source. Although we focus on the single-transmitter setting, the formulation can be extended to multi-antenna or multi-site deployments by considering multiple independent transmitter locations.

The surrounding environment is composed entirely of static structures such as buildings. The geometric layout of these structures is captured by a binary occupancy matrix $\bm{H} \in \{0,1\}^{N \times N}$. An entry $H(i,j) = 1$ indicates that the cell at location $(i,j)$ is fully occupied by a non-penetrable obstacle, while $H(i,j) = 0$ represents free space. These obstacles are assumed to fully block direct wave propagation. Given a transmitter position $\bm{r}$ and a static environment $\bm{H}$, the goal is to estimate the pathloss field $\bm{P}_\star(\bm{r}) = \mathcal{F}(\bm{H}; \bm{r})$, where $\mathcal{F}$ denotes the environment-to-field operator. In this work, we consider two distinct types of RMs. The first type is the static RM (SRM), which models only the dominant propagation path and captures large-scale attenuation effects such as free-space spreading and shadowing due to occlusions. The second type is the multi-path RM (MRM), which provides a more detailed characterization of the propagation field by incorporating multipath effects. These effects include reflections and diffractions induced by the surrounding static geometry. The MRM offers a more physically complete description of the electromagnetic field, particularly in non-line-of-sight regions.

To support the construction of these RMs, we assume the availability of two labeled datasets. The first dataset is denoted as $\mathcal{D}_{\mathrm{SRM}} = \{ \bm{H}_k, \bm{r}_k, \bm{P}_k^{\mathrm{SRM}} \}_{k=1}^{K}$ and contains a large number of examples generated under the SRM setting. These samples are efficient to compute and cover a wide variety of layouts and transmitter placements. The second dataset is denoted as $\mathcal{D}_{\mathrm{MRM}} = \{ \bm{H}_\ell, \bm{r}_\ell, \bm{P}_\ell^{\mathrm{MRM}} \}_{\ell=1}^{L}$ and contains a small number of high-fidelity samples that capture multipath propagation effects. Due to the computational overhead of accurate multipath modeling, the size of this dataset is limited, i.e., $L \ll K$. The objective of the above problem can be formulated as follows.
\begin{problem}
    \begin{align}
        &\min_{\bm{\theta}} \sum_{\ell=1}^{L} \frac{1}{L} \mathcal{L}\left(\bm{P}_\ell^{\mathrm{MRM}},\hat{\bm{P}}_\ell^{\mathrm{MRM}}\right),\label{obj}\\
        &\text{s.t.} \quad \hat{\bm{P}}_\ell^{\mathrm{MRM}} = \bm{\mu}_{\bm{\theta}}\left(\bm{H}, \bm{r}\mid\mathcal{D}_\mathrm{SRM},\mathcal{D}_\mathrm{MRM}\right),\tag{\ref{obj}a}
    \end{align}
\end{problem}
where $\bm{\theta}$ denotes the trainable parameters of the neural network $\bm{\mu}(\cdot)$.

\section{Few-Shot DM for RM Construction}

\subsection{Main-Path Decomposition}

Consider a bounded planar domain $\Omega\subset\mathbb{R}^2$ and a fixed transmitter--environment configuration. At a receiver location $\bm{x}\in\Omega$, let $\mathcal{P}(\bm{x})$ denote a finite set of propagation paths induced by the environment. Under the power-incoherent superposition model commonly adopted in RM construction, the received power field is expressed as the nonnegative sum
\begin{align}
U(\bm{x})=\sum_{p\in\mathcal{P}(\bm{x})} P_p(\bm{x}),\qquad P_p(\bm{x})\ge 0,
\label{eq:rm_incoherent_sum}
\end{align}
where $P_p(\bm{x})$ aggregates geometric spreading and interaction-dependent losses along path $p$. Let $p^\star(\bm{x})\in\arg\max_{p\in\mathcal{P}(\bm{x})}P_p(\bm{x})$ denote a strongest path at $\bm{x}$. The main-path RM (MP-RM) is then defined as
\begin{align}
U_{\mathrm{MP}}(\bm{x})=P_{p^\star(\bm{x})}(\bm{x}).
\label{eq:mp_rm_def}
\end{align}
Let $\mathcal{P}_k(\bm{x})\subseteq\mathcal{P}(\bm{x})$ denote the subset that collects all paths whose interaction order does not exceed $k$, and assume $p^\star(\bm{x})\in\mathcal{P}_k(\bm{x})$. The multipath RM (MU-RM) is defined by aggregating all such paths:
\begin{align}
U_{\mathrm{MU}}(\bm{x})=\sum_{p\in\mathcal{P}_k(\bm{x})}P_p(\bm{x}).
\label{eq:mu_rm_def}
\end{align}
Subtracting \eqref{eq:mp_rm_def} from \eqref{eq:mu_rm_def} yields the fundamental decomposition
\begin{align}
&U_{\mathrm{MU}}(\bm{x})
=U_{\mathrm{MP}}(\bm{x})+\Delta U(\bm{x}), \label{eq:mp_plus_residual_main}\\
&\Delta U(\bm{x})\triangleq\sum_{p\in\mathcal{P}_k(\bm{x})\setminus\{p^\star(\bm{x})\}}P_p(\bm{x}).
\label{eq:mp_plus_residual}
\end{align}
Since every summand in \eqref{eq:mp_plus_residual} is nonnegative, the transition from MP-RM to MU-RM is pointwise monotone:
\begin{align}
&U_{\mathrm{MU}}(\bm{x})\ge U_{\mathrm{MP}}(\bm{x}),\\
&\Delta U(\bm{x})\ge 0,\quad \forall\, \bm{x}\in\Omega,
\label{eq:monotone}
\end{align}
and $\Delta U(\bm{x})=0$ holds if and only if no additional path contributes power at $\bm{x}$ within $\mathcal{P}_k(\bm{x})$.

To describe the spatial regularity of the residual, we introduce a visibility indicator $V_p(\bm{x})\in\{0,1\}$ for each path $p$ and factorize the per-path power as
\begin{align}
&P_p(\bm{x})=V_p(\bm{x})\,\widetilde P_p(\bm{x}),\\
&\widetilde P_p(\bm{x})\ge 0,
\label{eq:visibility_factorization}
\end{align}
where $\widetilde P_p(\bm{x})$ captures the interaction losses along path $p$. This factor varies smoothly with $\bm{x}$ whenever the interaction sequence defining $p$ remains valid. We define the visibility state as $S(\bm{x})=\{V_p(\bm{x})\}_{p\in\mathcal{P}_k(\bm{x})}$ and partition $\Omega$ into regions of constant visibility state
\begin{align}
\mathcal{R}_s=\{\bm{x}\in\Omega:\ S(\bm{x})=s\}.
\label{eq:visibility_partition}
\end{align}
On each region $\mathcal{R}_s$, every indicator $V_p(\bm{x})$ is constant. Therefore, both $U_{\mathrm{MP}}$ and $U_{\mathrm{MU}}$ reduce to finite sums of smooth functions $\widetilde P_p(\bm{x})$ and inherit the same smoothness within $\mathcal{R}_s$. Non-smooth behavior of spatial derivatives can only occur on the union of visibility boundaries $\partial\mathcal{R}_s$, where at least one indicator $V_p(\bm{x})$ toggles. Under standard geometric optics rules, these boundaries form a finite union of curves and occupy zero area measure in $\Omega$. The MU-RM therefore differs from the MP-RM through a nonnegative residual $\Delta U$ that is smooth almost everywhere and changes sharply only across visibility boundaries. The decomposition \eqref{eq:mp_plus_residual_main} implies an additive relation under any linear integral operator. Let $\mathcal{L}_\sigma$ be a low-pass operator with bandwidth parameter $\sigma$, represented by a nonnegative kernel $\kappa_\sigma$ of unit mass:
\begin{align}
&\mathcal{L}_\sigma[U](\bm{x})=\int_{\Omega}\kappa_\sigma(\bm{x}-\bm{y})\,U(\bm{y})\,\mathrm{d}\bm{y},\\
&\int_{\mathbb{R}^2}\kappa_\sigma(\bm{u})\,\mathrm{d}\bm{u}=1,\quad \kappa_\sigma(\bm{u})\ge 0.
\label{eq:lowpass_operator}
\end{align}
By linearity and \eqref{eq:mp_plus_residual_main},
\begin{align}
\mathcal{L}_\sigma[U_{\mathrm{MU}}]
=\mathcal{L}_\sigma[U_{\mathrm{MP}}]+\mathcal{L}_\sigma[\Delta U].
\label{eq:lowpass_decomp}
\end{align}
A quantitative bound on the low-frequency increment follows from standard convolution inequalities. We denote the $L^p$ norm of a function $f$ over $\Omega$ by $\|f\|_p$. Applying Young's convolution inequality yields
\begin{align}
\|\mathcal{L}_\sigma[\Delta U]\|_2
\le \|\kappa_\sigma\|_1\,\|\Delta U\|_2
=\|\Delta U\|_2,
\label{eq:young_l2}
\end{align}
and
\begin{align}
\|\mathcal{L}_\sigma[\Delta U]\|_\infty
\le \|\kappa_\sigma\|_\infty\,\|\Delta U\|_1.
\label{eq:young_linf}
\end{align}
Suppose $\Delta U$ is effectively supported on a subset $\mathcal{S}\subseteq\Omega$ with small area and is bounded by $\Delta U(\bm{x})\le U_{\max}$. Then the $L^1$ norm of the residual satisfies
\begin{align}
\|\Delta U\|_1=\int_{\Omega}\Delta U(\bm{x})\,\mathrm{d}\bm{x}
=\int_{\mathcal{S}}\Delta U(\bm{x})\,\mathrm{d}\bm{x}
\le U_{\max}\,|\mathcal{S}|,
\label{eq:l1_support_bound}
\end{align}
where $|\mathcal{S}|$ denotes the area of $\mathcal{S}$. Combining \eqref{eq:l1_support_bound} with \eqref{eq:young_linf} yields
\begin{align}
\|\mathcal{L}_\sigma[\Delta U]\|_\infty
\le \|\kappa_\sigma\|_\infty\,U_{\max}\,|\mathcal{S}|.
\label{eq:lowfreq_increment_bound}
\end{align}
For typical low-pass kernels, $\|\kappa_\sigma\|_\infty$ decreases as $\sigma$ increases. Thus the coarse-scale effect of a spatially localized residual is attenuated by smoothing. This result formalizes a central structural implication of \eqref{eq:mp_plus_residual_main}: the MP-RM dominates the large-scale skeleton of the RM, while the MU-RM introduces structured positive increments concentrated in geometric neighborhoods where additional paths become visible.

\subsection{Feature Shift Geometry}

The MP-RM and the MU-RM satisfy the residual decomposition
\begin{align}
U_{\mathrm{MU}}(\bm{x}) = U_{\mathrm{MP}}(\bm{x}) + \Delta U(\bm{x}),\quad
\Delta U(\bm{x})\ge 0,\ \forall\, \bm{x}\in\Omega.
\label{eq:iva2_residual}
\end{align}
To connect this physical structure with learning, we introduce a fixed feature encoder
\begin{align}
\Phi:\ L^2(\Omega)\rightarrow \mathbb{R}^d,\qquad \bm{z}=\Phi(U),
\label{eq:iva2_encoder}
\end{align}
which maps a RM field to a $d$-dimensional representation. The encoder $\Phi$ is kept fixed during adaptation and serves only to measure domain shift and structural consistency. We assume that the representation varies stably with the RM field, as stated below.

\textbf{Assumption 1.} There exists a constant $L_{\Phi}>0$ such that
\begin{align}
\|\Phi(U)-\Phi(\widetilde U)\|_2 \le L_{\Phi}\,\|U-\widetilde U\|_{2},\qquad \forall\, U,\widetilde U\in L^2(\Omega).
\label{eq:iva2_lipschitz}
\end{align}

Applying \eqref{eq:iva2_lipschitz} to \eqref{eq:iva2_residual} yields a deterministic bound on the feature increment induced by multipath contributions:
\begin{align}
\|\Phi(U_{\mathrm{MU}})-\Phi(U_{\mathrm{MP}})\|_2
&= \|\Phi(U_{\mathrm{MP}}+\Delta U)-\Phi(U_{\mathrm{MP}})\|_2\notag\\
&\le L_{\Phi}\,\|\Delta U\|_2.
\label{eq:iva2_increment_bound}
\end{align}
If $\Delta U$ is spatially localized or energy-limited as established in Section~IV-A, the induced feature shift remains controlled. This result supports stable transfer from MP-RM to MU-RM in the learned representation.

Let $U^{\mathrm{MP}}$ and $U^{\mathrm{MU}}$ be random fields representing the MP-RM and MU-RM domains under matched transmitter and environment semantics. Define the corresponding feature vectors as
\begin{align}
\bm{z}^{\mathrm{MP}}=\Phi(U^{\mathrm{MP}}),\qquad
\bm{z}^{\mathrm{MU}}=\Phi(U^{\mathrm{MU}}),
\label{eq:iva2_z_defs}
\end{align}
and denote the domain means by
\begin{align}
\bm{\mu}_{\mathrm{MP}}=\mathbb{E}[\bm{z}^{\mathrm{MP}}],\qquad
\bm{\mu}_{\mathrm{MU}}=\mathbb{E}[\bm{z}^{\mathrm{MU}}].
\label{eq:iva2_means}
\end{align}
The average feature shift induced by multipath is defined as
\begin{align}
\bm{w}\triangleq \bm{\mu}_{\mathrm{MU}}-\bm{\mu}_{\mathrm{MP}}.
\label{eq:iva2_w_def}
\end{align}
Using \eqref{eq:iva2_residual} and \eqref{eq:iva2_z_defs}, the vector $\bm{w}$ in \eqref{eq:iva2_w_def} admits the equivalent representation
\begin{align}
\bm{w}
=\mathbb{E}\!\left[\Phi(U^{\mathrm{MP}}+\Delta U)-\Phi(U^{\mathrm{MP}})\right],
\label{eq:iva2_w_equiv}
\end{align}
which shows that $\bm{w}$ is the mean feature displacement caused by the multipath residual. The structural implication of Section~IV-A is that the MP-RM skeleton is preserved while $\Delta U$ injects localized increments. In a representation that captures global geometry and coarse power layout, this observation motivates a translation-dominant approximation in the feature space.

Consider paired samples $\{(U_i^{\mathrm{MP}},U_i^{\mathrm{MU}})\}_{i=1}^{N}$ sharing the same underlying configuration, and define
\begin{align}
\bm{z}_i^{\mathrm{MP}}=\Phi(U_i^{\mathrm{MP}}),\qquad
\bm{z}_i^{\mathrm{MU}}=\Phi(U_i^{\mathrm{MU}}).
\label{eq:iva2_paired_features}
\end{align}
We model the cross-domain relation as
\begin{align}
&\bm{z}_i^{\mathrm{MU}} = \bm{z}_i^{\mathrm{MP}}+\bm{w}+\bm{\delta}_i, \label{eq:iva2_translation_model}\\
&\mathbb{E}[\bm{\delta}_i]=\bm{0},\quad
\|\bm{\delta}_i\|_2\le \eta,
\label{eq:iva2_translation_model_bound}
\end{align}
where $\bm{\delta}_i$ captures sample-specific multipath variability not explained by the common direction $\bm{w}$.

A first-order justification of \eqref{eq:iva2_translation_model} follows from local linearization. Assume that $\Phi$ is Fr\'echet differentiable in a neighborhood of $U_i^{\mathrm{MP}}$, and write
\begin{align}
\Phi(U_i^{\mathrm{MP}}+\Delta U_i)
=\Phi(U_i^{\mathrm{MP}}) + \mathcal{J}_{\Phi}(U_i^{\mathrm{MP}})[\Delta U_i] + \bm{r}_i,
\label{eq:iva2_taylor}
\end{align}
where $\mathcal{J}_{\Phi}(U)$ denotes the Fr\'echet derivative as a bounded linear operator and $\bm{r}_i$ is the higher-order remainder. Taking expectations and using \eqref{eq:iva2_w_equiv} yields
\begin{align}
\bm{w}
= \mathbb{E}\!\left[\mathcal{J}_{\Phi}(U^{\mathrm{MP}})[\Delta U]\right] + \mathbb{E}[\bm{r}].
\label{eq:iva2_w_taylor_mean}
\end{align}
The translation model \eqref{eq:iva2_translation_model} then follows by identifying $\bm{\delta}_i$ with the deviation of $\mathcal{J}_{\Phi}(U_i^{\mathrm{MP}})[\Delta U_i]+\bm{r}_i$ from its mean.

Under \eqref{eq:iva2_translation_model}, pairwise relations are stable up to the fluctuation level $\eta$. Then, for any $i$ and $j$, we have
\begin{align}
\bm{z}_i^{\mathrm{MU}}-\bm{z}_j^{\mathrm{MU}}
= \bm{z}_i^{\mathrm{MP}}-\bm{z}_j^{\mathrm{MP}} + \bm{\delta}_i-\bm{\delta}_j.
\label{eq:iva2_pair_diff}
\end{align}
By the reverse triangle inequality,
\begin{align}
&\Big|\,\|\bm{z}_i^{\mathrm{MU}}-\bm{z}_j^{\mathrm{MU}}\|_2
-\|\bm{z}_i^{\mathrm{MP}}-\bm{z}_j^{\mathrm{MP}}\|_2\,\Big|\notag\\
&\quad\le \|\bm{\delta}_i-\bm{\delta}_j\|_2
\le \|\bm{\delta}_i\|_2+\|\bm{\delta}_j\|_2
\le 2\eta.
\label{eq:iva2_distance_stability}
\end{align}
Equation~\eqref{eq:iva2_distance_stability} formalizes that the relational geometry of MP-RM features is largely preserved in the MU-RM domain. The dominant inter-domain difference is captured by the common shift direction $\bm{w}$.

Given the paired features in \eqref{eq:iva2_paired_features}, define empirical means as
\begin{align}
\widehat{\bm{\mu}}_{\mathrm{MP}}=\frac{1}{N}\sum_{i=1}^{N}\bm{z}_i^{\mathrm{MP}},\qquad
\widehat{\bm{\mu}}_{\mathrm{MU}}=\frac{1}{N}\sum_{i=1}^{N}\bm{z}_i^{\mathrm{MU}},
\label{eq:iva2_emp_means}
\end{align}
and define the empirical shift direction as
\begin{align}
\widehat{\bm{w}}=\widehat{\bm{\mu}}_{\mathrm{MU}}-\widehat{\bm{\mu}}_{\mathrm{MP}}.
\label{eq:iva2_w_hat}
\end{align}
From \eqref{eq:iva2_translation_model} and \eqref{eq:iva2_emp_means}, it follows that
\begin{align}
\widehat{\bm{w}}=\bm{w}+\frac{1}{N}\sum_{i=1}^{N}\bm{\delta}_i,
\label{eq:iva2_w_hat_error}
\end{align}
and therefore
\begin{align}
\|\widehat{\bm{w}}-\bm{w}\|_2
\le \frac{1}{N}\sum_{i=1}^{N}\|\bm{\delta}_i\|_2
\le \eta.
\label{eq:iva2_w_hat_bound}
\end{align}
Equation~\eqref{eq:iva2_w_hat_bound} shows that the shift direction can be reliably estimated from a small number of paired samples. This guarantee holds when the translation-dominant approximation is valid with bounded fluctuations.

\subsection{Directional Consistency Principle}

Let $\mathcal{G}_{\theta}$ denote a conditional generator that maps an MP-RM input $U^{\mathrm{MP}}$ and condition $c$ to a predicted MU-RM output
\begin{align}
\widehat U^{\mathrm{MU}}=\mathcal{G}_{\theta}\!\left(U^{\mathrm{MP}},c\right).
\label{eq:iva3_generator}
\end{align}
Let $\Phi$ be the fixed feature encoder defined in \eqref{eq:iva2_encoder}. The feature displacement induced by the generator is defined as
\begin{align}
\Delta \bm{z}
\triangleq \Phi\!\left(\widehat U^{\mathrm{MU}}\right)-\Phi\!\left(U^{\mathrm{MP}}\right).
\label{eq:iva3_dz_def}
\end{align}
Let $\bm{w}=\bm{\mu}_{\mathrm{MU}}-\bm{\mu}_{\mathrm{MP}}$ be the mean shift direction defined in \eqref{eq:iva2_w_def}. The central requirement is that the generator update should align with the dominant cross-domain direction. At the same time, the alignment should allow sample-dependent magnitudes that reflect spatially varying multipath increments.

\textbf{Definition 1.} The directionally consistent set associated with $\bm{w}$ is
\begin{align}
\mathcal{C}(\bm{w})
=\left\{\bm{v}\in\mathbb{R}^{d}:\ \bm{v}=\alpha \bm{w}+\bm{e},\ \alpha\ge 0,\ \bm{e}^\top \bm{w} = 0\right\},
\label{eq:iva3_cone_set}
\end{align}
where $\alpha$ determines the amount of multipath injection and $\bm{e}$ represents orthogonal distortion. Directional consistency requires suppressing the orthogonal component $\bm{e}$.

To enforce this principle, we decompose $\Delta\bm{z}$ into its projection onto $\bm{w}$ and its orthogonal residual. Define
\begin{align}
&\alpha(\Delta\bm{z})
=\frac{\Delta\bm{z}^\top \bm{w}}{\|\bm{w}\|_2^2},\\
&\Delta\bm{z}_{\parallel}
=\alpha(\Delta\bm{z})\,\bm{w},\\
&\Delta\bm{z}_{\perp}
=\Delta\bm{z}-\Delta\bm{z}_{\parallel}.
\label{eq:iva3_proj}
\end{align}
The directional consistency loss penalizes $\Delta\bm{z}_{\perp}$ and discourages negative alignment with $\bm{w}$. For a paired dataset $\{(U_i^{\mathrm{MP}},U_i^{\mathrm{MU}},c_i)\}_{i=1}^{N}$, define
\begin{align}
&\widehat U_i^{\mathrm{MU}}=\mathcal{G}_{\theta}\!\left(U_i^{\mathrm{MP}},c_i\right),\\
&\Delta\bm{z}_i=\Phi(\widehat U_i^{\mathrm{MU}})-\Phi(U_i^{\mathrm{MP}}).
\label{eq:iva3_batch_dz}
\end{align}
The proposed directional consistency loss is
\begin{align}
\mathcal{L}_{\mathrm{dir}}(\theta)
=\frac{1}{N}&\sum_{i=1}^{N}\left\|
\Delta\bm{z}_i-\frac{\Delta\bm{z}_i^\top \bm{w}}{\|\bm{w}\|_2^2}\,\bm{w}
\right\|_2^2\notag \\
&+\frac{\beta}{N}\sum_{i=1}^{N}\left[\max\!\left(0,-\Delta\bm{z}_i^\top \bm{w}\right)\right]^2,
\label{eq:iva3_ldir}
\end{align}
where $\beta\ge 0$ controls the penalty on negative alignment. The first term in \eqref{eq:iva3_ldir} enforces that the feature displacement lies in the one-dimensional subspace spanned by $\bm{w}$, thereby suppressing orthogonal distortions. The second term enforces nonnegative alignment, which is consistent with the nonnegative multipath increment property established in Section~IV-A.

The necessity of a directional constraint follows from the non-identifiability of distance-based alignment in Euclidean feature spaces. Consider any objective that matches only pairwise distances between generated features and MP-RM features. Such an objective admits solutions related by orthogonal transformations, since for any orthogonal matrix $\bm{O}$,
\begin{align}
\|\bm{O}\bm{a}-\bm{O}\bm{b}\|_2=\|\bm{a}-\bm{b}\|_2,\qquad \forall\, \bm{a},\bm{b}\in\mathbb{R}^{d}.
\label{eq:iva3_orth_invariance}
\end{align}
A generator can therefore satisfy distance constraints while rotating the feature configuration. Such rotation changes the semantic axes along which the MP-RM structure is represented and can distort the inherited skeleton. In contrast, minimizing \eqref{eq:iva3_ldir} suppresses the orthogonal component for every sample. This fixes the cross-domain displacement direction and eliminates the rotational ambiguity.

The directional constraint also yields a stability property consistent with Section~IV-B. Under the translation-dominant model, a directionally consistent generator satisfies
\begin{align}
&\Delta\bm{z}_i=\alpha_i\,\bm{w}+\bm{e}_i,\quad
\bm{e}_i^\top \bm{w}=0,\quad
\|\bm{e}_i\|_2 \le \varepsilon.
\label{eq:iva3_model_dir}
\end{align}
Then, for any $i$ and $j$, we have
\begin{align}
\bm{z}_i^{\mathrm{gen}}-\bm{z}_j^{\mathrm{gen}}
=\bm{z}_i^{\mathrm{MP}}-\bm{z}_j^{\mathrm{MP}}+(\alpha_i-\alpha_j)\,\bm{w}+(\bm{e}_i-\bm{e}_j),
\label{eq:iva3_pairwise_gen}
\end{align}
where $\bm{z}_i^{\mathrm{gen}}=\Phi(\widehat U_i^{\mathrm{MU}})$. When $\bm{e}_i$ is controlled by the first term in \eqref{eq:iva3_ldir}, the deformation of the relational geometry is confined to a one-dimensional modulation along $\bm{w}$. This preserves the dominant MP-RM skeleton encoded by directions orthogonal to $\bm{w}$ and allocates the cross-domain change to the physically meaningful axis of multipath injection.

The overall fine-tuning objective combines a task fidelity term with the directional consistency regularizer. Let $\mathcal{L}_{\mathrm{fit}}(\theta)$ denote a supervised or self-consistent reconstruction loss between $\widehat U_i^{\mathrm{MU}}$ and $U_i^{\mathrm{MU}}$ in the MU-RM domain. The final training objective is
\begin{align}
\min_{\theta}\ \mathcal{L}_{\mathrm{fit}}(\theta)+\lambda_{\mathrm{dir}}\,\mathcal{L}_{\mathrm{dir}}(\theta),
\label{eq:iva3_total}
\end{align}
where $\lambda_{\mathrm{dir}}>0$ controls the regularization strength. This regularizer steers adaptation along the mean shift direction from MP-RM to MU-RM while suppressing orthogonal distortions. It thereby aligns cross-domain statistics without sacrificing the geometric integrity of the RM skeleton.

\subsection{Few-Shot Fine-Tuning}

The objective is to adapt a pretrained generative model from the MP-RM domain to the MU-RM domain using only a small number of labeled MU-RM samples. The MP-RM domain provides abundant supervision that captures stable large-scale spatial priors and global propagation skeletons. In contrast, the MU-RM domain exhibits additional spatially localized power increments induced by multipath interactions. The adaptation should preserve the pretrained global structure and inject MU-RM-specific details in a controlled and physically consistent manner. Let $c$ denote the conditioning input that encodes scene geometry and transmitter attributes. The fine-tuning set is composed of paired samples
\begin{align}
\mathcal{D}_{\mathrm{ft}}=\left\{(x_{0,i}^{\mathrm{MP}},\,x_{0,i}^{\mathrm{MU}},\,c_i)\right\}_{i=1}^{N},
\label{eq:ft_dataset}
\end{align}
where $N$ is small and each pair shares the same configuration specified by $c_i$.

As shown in Fig.~\ref{fig-method}, we employ a variance-preserving diffusion model to parameterize the conditional generator. Let $x_0$ denote a clean MU-RM sample and let $q(x_t\mid x_0)$ be the forward noising process indexed by $t\in[0,1]$. The perturbed sample is written as
\begin{align}
x_t=\alpha(t)\,x_0+\sigma(t)\,\varepsilon,\qquad \varepsilon\sim\mathcal{N}(0,I),
\label{eq:ft_forward}
\end{align}
where $\alpha(t)$ is a monotonically decreasing signal factor and $\sigma(t)$ is a monotonically increasing noise factor. The denoiser predicts the injected noise $\varepsilon$ from the tuple $(x_t,t,c)$. To allow different propagation components to be modeled with separate heads, we write the network output as $\varepsilon_{\theta,r}(x_t,t,c)$ for $r\in\mathcal{R}$ and define $\varepsilon_r$ as the corresponding ground-truth noise in \eqref{eq:ft_forward}. The base objective is the denoising score matching loss
\begin{align}
\mathcal{L}_{\mathrm{base}}
=\mathbb{E}_{x_0,\,t,\,\{\varepsilon_r\}_{r\in\mathcal{R}}}
\left[
\sum_{r\in\mathcal{R}}
\left\|
\varepsilon_r-\varepsilon_{\theta,r}(x_t,t,c)
\right\|_2^2
\right].
\label{eq:ft_lbase}
\end{align}

To guide few-shot adaptation toward physically consistent updates, we introduce a directional constraint in a frozen semantic feature space. Let $\phi:\mathbb{R}^{H\times W}\rightarrow\mathbb{R}^d$ be a fixed encoder that maps a RM to a $d$-dimensional embedding. The mean embeddings of the MP-RM and MU-RM domains are defined as
\begin{align}
\bm{\mu}_{\mathrm{MP}}=\mathbb{E}\!\left[\phi(x_0^{\mathrm{MP}})\right],\qquad
\bm{\mu}_{\mathrm{MU}}=\mathbb{E}\!\left[\phi(x_0^{\mathrm{MU}})\right].
\label{eq:ft_means}
\end{align}
The mean shift vector and its unit direction are given by
\begin{align}
\bm{w}=\bm{\mu}_{\mathrm{MU}}-\bm{\mu}_{\mathrm{MP}},\qquad
\bm{v}=\frac{\bm{w}}{\|\bm{w}\|_2}.
\label{eq:ft_dir}
\end{align}
For each triplet $(x_{0}^{\mathrm{MP}},x_{0}^{\mathrm{MU}},c)\in\mathcal{D}_{\mathrm{ft}}$, the target-aligned shift magnitude in feature space is defined as
\begin{align}
\eta(c)=\bm{v}^{\top}\!\Big(\phi(x_{0}^{\mathrm{MU}})-\phi(x_{0}^{\mathrm{MP}})\Big),\qquad \eta(c)\ge 0.
\label{eq:ft_eta}
\end{align}
Let $\widehat x_{0,\theta}(x_t,t,c)$ denote the predicted clean sample produced by the denoiser at diffusion time $t$. The feature displacement relative to the MP-RM reference is
\begin{align}
\Delta\bm{h}=\phi\!\left(\widehat x_{0,\theta}(x_t,t,c)\right)-\phi\!\left(x_0^{\mathrm{MP}}\right).
\label{eq:ft_dh}
\end{align}
We decompose $\Delta\bm{h}$ into components parallel and orthogonal to $\bm{v}$:
\begin{align}
\Delta\bm{h}_{\parallel}=\left(\Delta\bm{h}^\top \bm{v}\right)\bm{v},\qquad
\Delta\bm{h}_{\perp}=\Delta\bm{h}-\Delta\bm{h}_{\parallel}.
\label{eq:ft_proj}
\end{align}
The directional consistency loss suppresses orthogonal distortions and aligns the parallel component to the target magnitude:
\begin{align}
\mathcal{L}_{\mathrm{dir}}
=\left\|\Delta\bm{h}_{\perp}\right\|_2^2
+\beta\left(\Delta\bm{h}^\top \bm{v}-\eta(c)\right)^2,
\label{eq:ft_ldir}
\end{align}
where $\beta\ge 0$ controls the strength of magnitude matching. The encoder $\phi$ remains fixed during fine-tuning.

Directional regularization is most effective when the denoised prediction $\widehat x_{0,\theta}$ is informative. We therefore weight $\mathcal{L}_{\mathrm{dir}}$ by a schedule that increases with the signal-to-noise ratio (SNR). The SNR at diffusion time $t$ is defined as
\begin{align}
\mathrm{SNR}(t)=\frac{\alpha(t)^2}{\sigma(t)^2},
\label{eq:ft_snr}
\end{align}
and the normalized increasing weight is given by
\begin{align}
\lambda_{\mathrm{dir}}(t)=\lambda_{\max}\,
\frac{\mathrm{SNR}(t)}{\mathrm{SNR}(t)+1},
\label{eq:ft_lam}
\end{align}
where $\lambda_{\max}>0$ is the maximum regularization strength. The final fine-tuning objective is
\begin{align}
\mathcal{L}_{\mathrm{total}}
=\mathbb{E}\!\left[\mathcal{L}_{\mathrm{base}}\right]
+\mathbb{E}\!\left[\lambda_{\mathrm{dir}}(t)\,\mathcal{L}_{\mathrm{dir}}\right].
\label{eq:ft_ltotal}
\end{align}
This objective preserves the pretrained MP-RM structural priors through $\mathcal{L}_{\mathrm{base}}$ while enforcing propagation-consistent feature shifts through $\mathcal{L}_{\mathrm{dir}}$. It enables data-efficient transfer from the MP-RM domain to the MU-RM domain with limited labeled samples.

\begin{table*}[!ht]
\captionsetup{font={small}, skip=16pt}
\centering
\caption{\textbf{Quantitative Comparison on IRT4.} Results in bold red and underlined blue highlight the best and second best, respectively. The last column denotes the percentage improvement of RadioDiff-FS over the best baseline without fine-tuning.}
\vspace{-9pt}
\resizebox{0.85\linewidth}{!}{
\begin{tabular}{@{}c|cccccc|c@{}}
\toprule
Methods & RadioDiff-FS & \begin{tabular}{@{}c@{}}RadioDiff-FS \\ (one-shot)\end{tabular} & PhyRMDM & RadioDiff & RME-GAN & RadioUNet & 
\begin{tabular}{@{}c@{}}Rate vs. \\ w/o Fine-Tuning (\%) \end{tabular}
 \\ \midrule
NMSE & {\color[HTML]{9A0000} \textbf{0.0049}} & 0.0204 & {\color[HTML]{00009B} \underline{0.0100}} & 0.0121 & 0.0155 & 0.0159 & {\color[HTML]{00adb5}$\downarrow$ 59.50\%} \\
RMSE & {\color[HTML]{9A0000} \textbf{0.0196}} & 0.0278 & {\color[HTML]{00009B} \underline{0.0278}} & 0.0309 & 0.0340 & 0.0344 & {\color[HTML]{00adb5}$\downarrow$ 36.57\%} \\
SSIM $\uparrow$ & {\color[HTML]{9A0000} \textbf{0.9752}} & {\color[HTML]{00009B} \underline{0.9285}} & 0.9180 & 0.9268 & 0.9123 & 0.9102 & {\color[HTML]{f08a5d}$\uparrow$ 5.22\%} \\
PSNR $\uparrow$ & {\color[HTML]{9A0000} \textbf{36.37}} & {\color[HTML]{00009B} \underline{0.9285}} & 31.40 & 30.44 & 29.74 & 29.64 & {\color[HTML]{f08a5d}$\uparrow$ 19.48\%} \\ \bottomrule
\end{tabular}
}
\vspace{-9pt}
\label{tab:comp_irt4}
\end{table*}
 
\begin{table*}[!ht]
\captionsetup{font={small}, skip=16pt}
\centering
\caption{\textbf{Quantitative Comparison on IRT4 With Car.} Results in bold red and underlined blue highlight the best and second best, respectively. The last column denotes the percentage improvement of RadioDiff-FS over the best baseline without fine-tuning.}
\vspace{-9pt}
\resizebox{0.85\linewidth}{!}{
\begin{tabular}{@{}c|cccccc|c@{}}
\toprule
Methods & RadioDiff-FS & \begin{tabular}{@{}c@{}}RadioDiff-FS \\ (one-shot)\end{tabular} & PhyRMDM & RadioDiff & RME-GAN & RadioUNet & 
\begin{tabular}{@{}c@{}}Rate vs. \\ w/o Fine-Tuning (\%) \end{tabular}
 \\ \midrule
NMSE & {\color[HTML]{9A0000} \textbf{0.0121}} & 0.0214 & 0.1113 & 0.0465 & {\color[HTML]{00009B} \underline{0.0183}} & 0.0381 & {\color[HTML]{00adb5}$\downarrow$ 73.98\%} \\
RMSE & {\color[HTML]{9A0000} \textbf{0.0214}} & {\color[HTML]{00009B} \underline{0.0296}} & 0.2178 & 0.1319 & 0.0358 & 0.0307 & {\color[HTML]{00adb5}$\downarrow$ 83.78\%} \\
SSIM $\uparrow$ & {\color[HTML]{9A0000} \textbf{0.9510}} & {\color[HTML]{00009B} \underline{0.8969}} & 0.7182 & 0.7968 & 0.8823 & 0.8487 & {\color[HTML]{f08a5d}$\uparrow$ 19.35\%} \\
PSNR $\uparrow$ & {\color[HTML]{9A0000} \textbf{32.68}} & 30.61 & 25.23 & 27.86 & {\color[HTML]{00009B} \underline{30.74}} & 29.57 & {\color[HTML]{f08a5d}$\uparrow$ 17.30\%} \\ \bottomrule
\end{tabular}
}
\vspace{-9pt}
\label{tab:comp_irt4_car}
\end{table*}

 
\section{Experimental Results}
 
\subsection{Datasets}
 
\begin{figure*}[!ht]
\centering
\captionsetup{font={small}, skip=16pt}
\renewcommand{\thesubfigure}{\arabic{subfigure}}
\renewcommand{\arraystretch}{0.8}
\setlength{\tabcolsep}{1.5pt}
\begin{tabular}{c c c c c c c}
  \subcaptionbox{}{\includegraphics[width=0.13\linewidth]{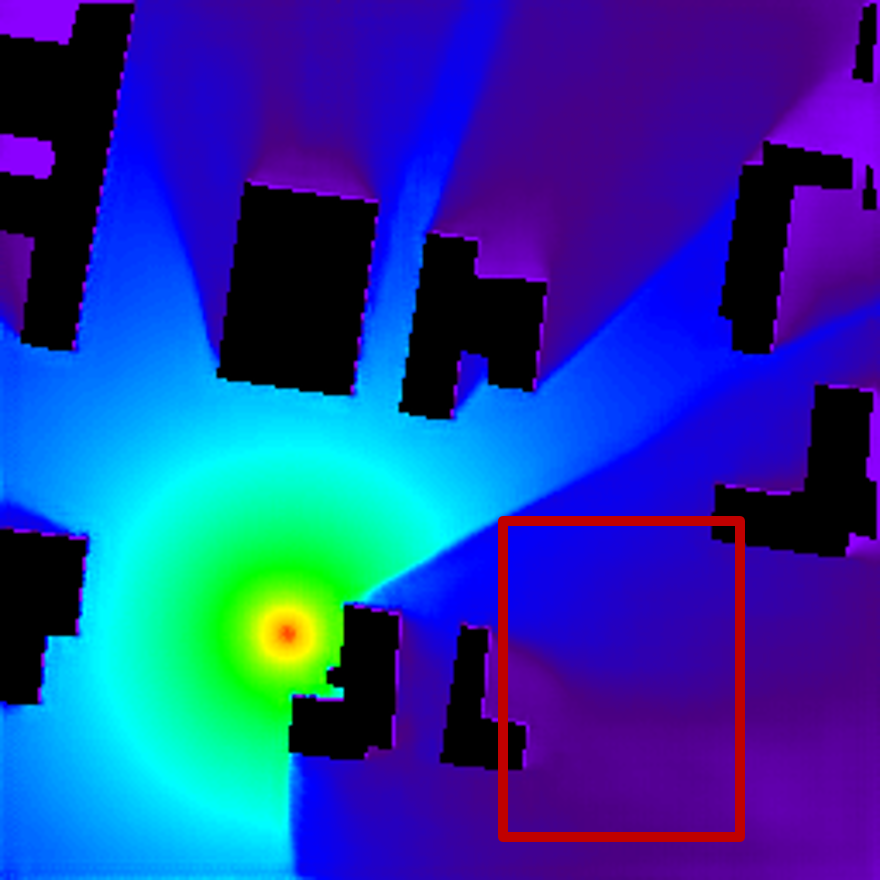}} &
  \subcaptionbox{}{\includegraphics[width=0.13\linewidth]{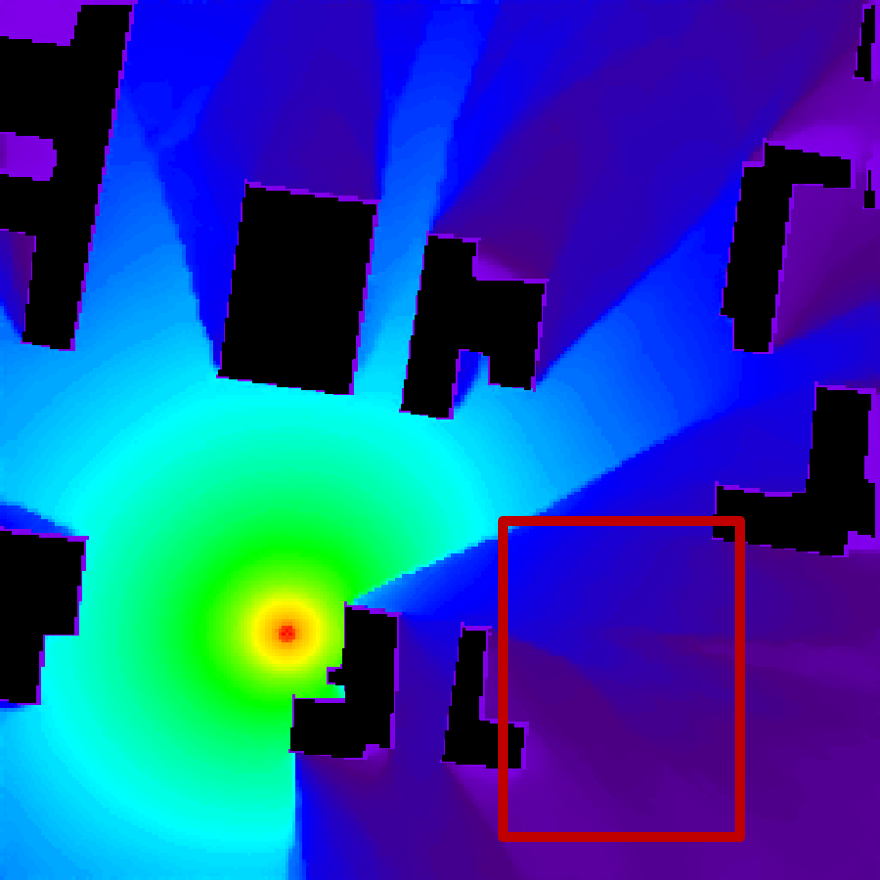}} &
  \subcaptionbox{}{\includegraphics[width=0.13\linewidth]{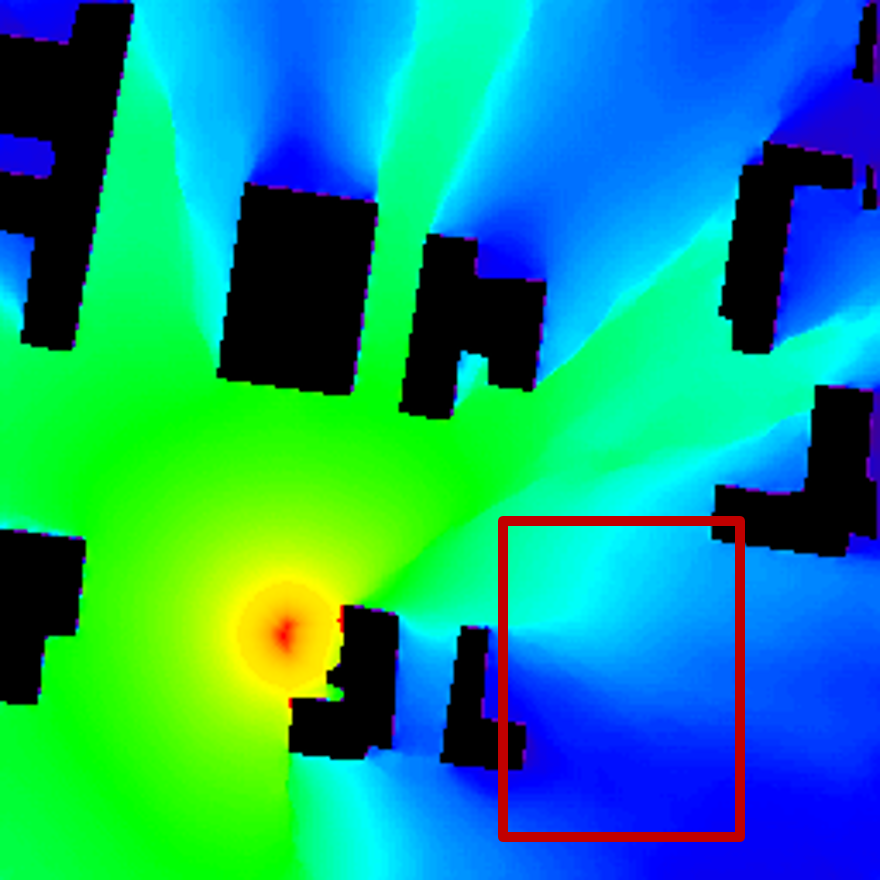}} &
  \subcaptionbox{}{\includegraphics[width=0.13\linewidth]{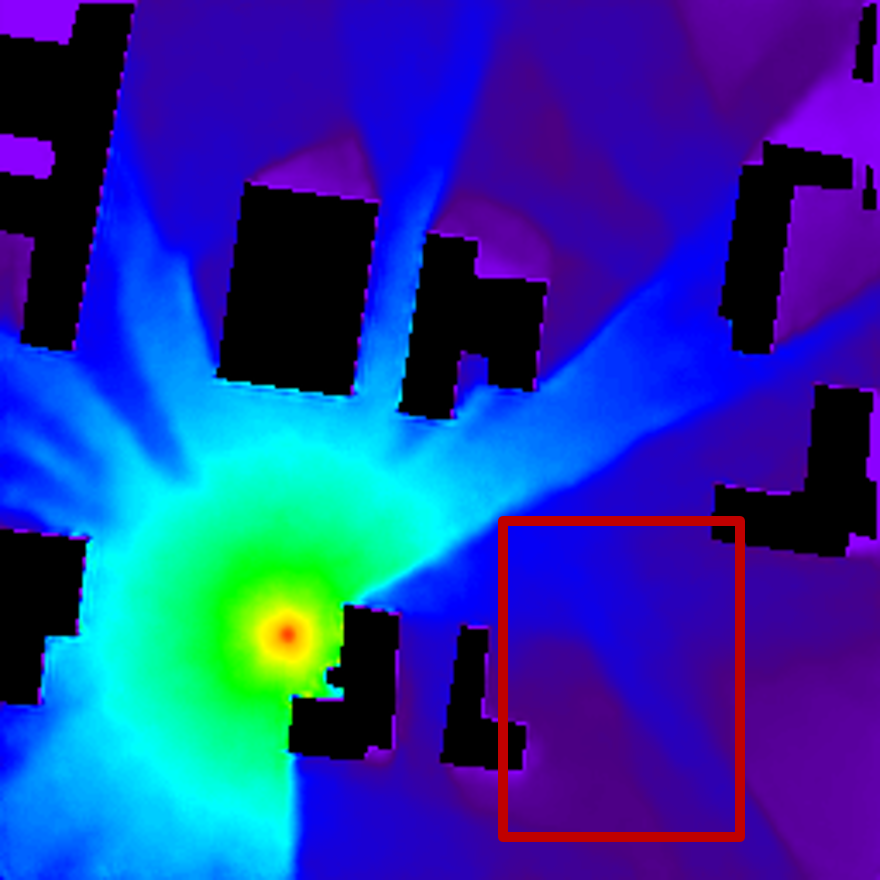}} &
  \subcaptionbox{}{\includegraphics[width=0.13\linewidth]{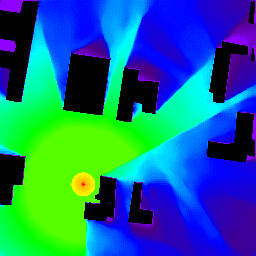}} &
  \subcaptionbox{}{\includegraphics[width=0.13\linewidth]{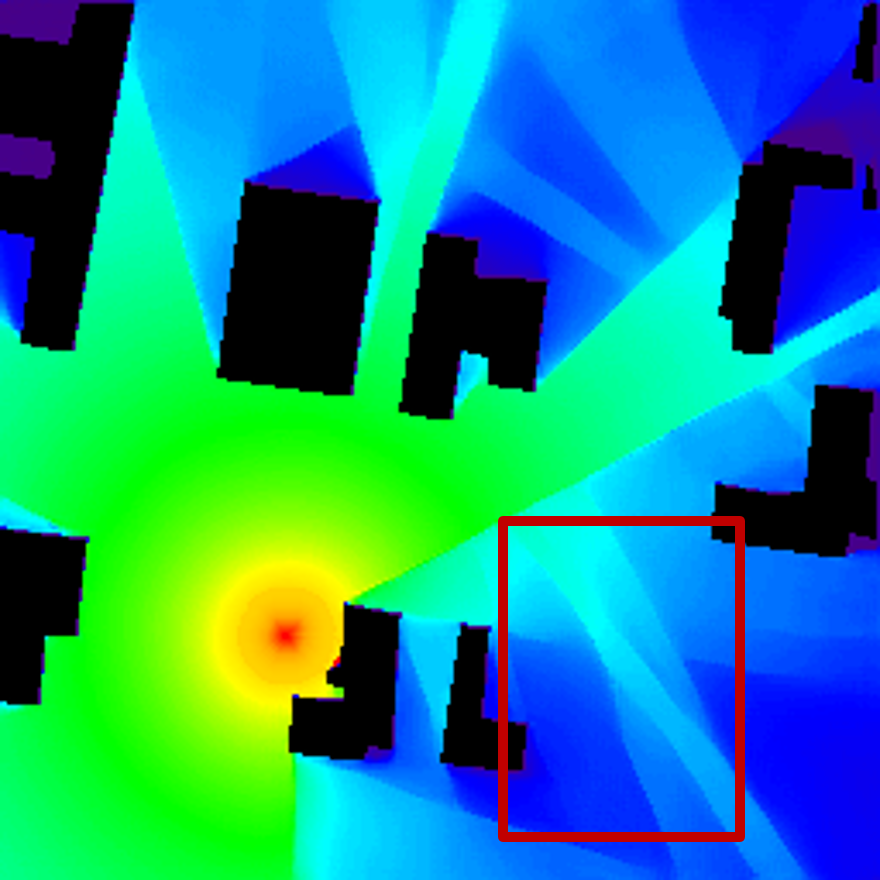}} &
  \subcaptionbox{}{\includegraphics[width=0.13\linewidth]{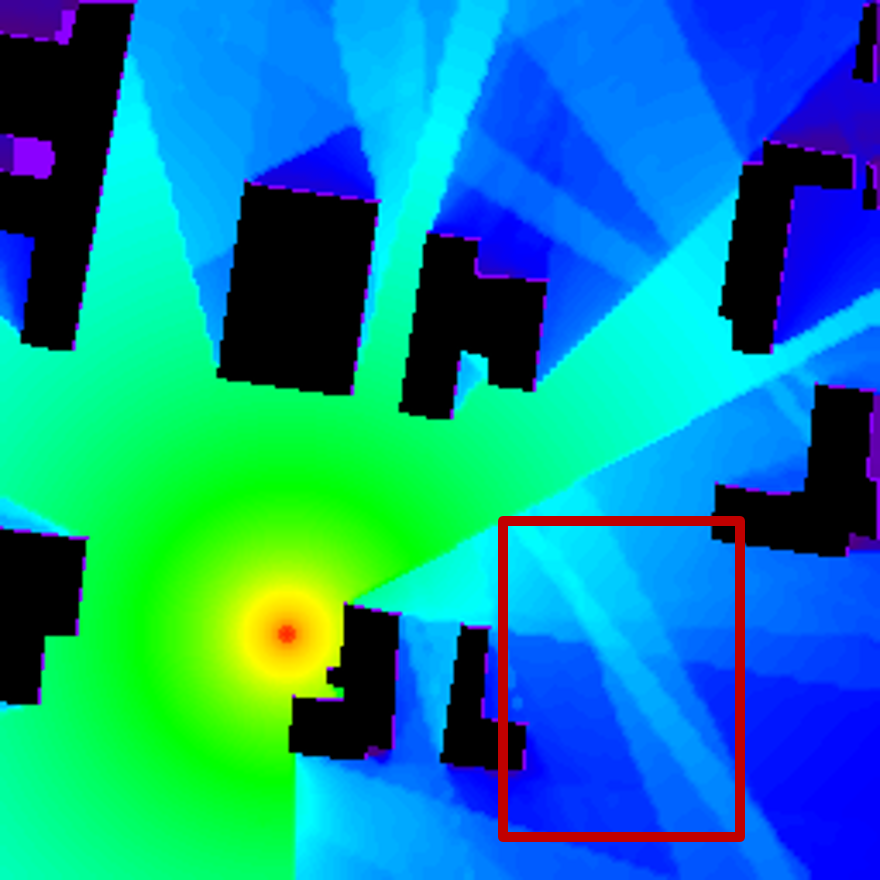}} \\
  \subcaptionbox{}{\includegraphics[width=0.13\linewidth]{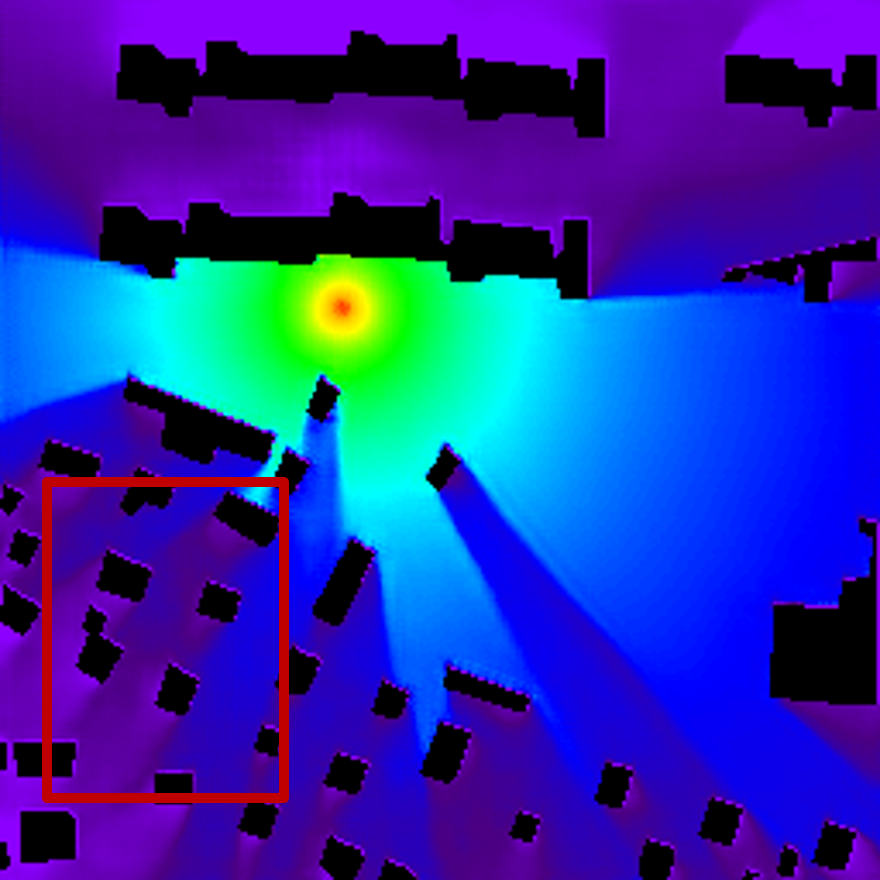}} &
  \subcaptionbox{}{\includegraphics[width=0.13\linewidth]{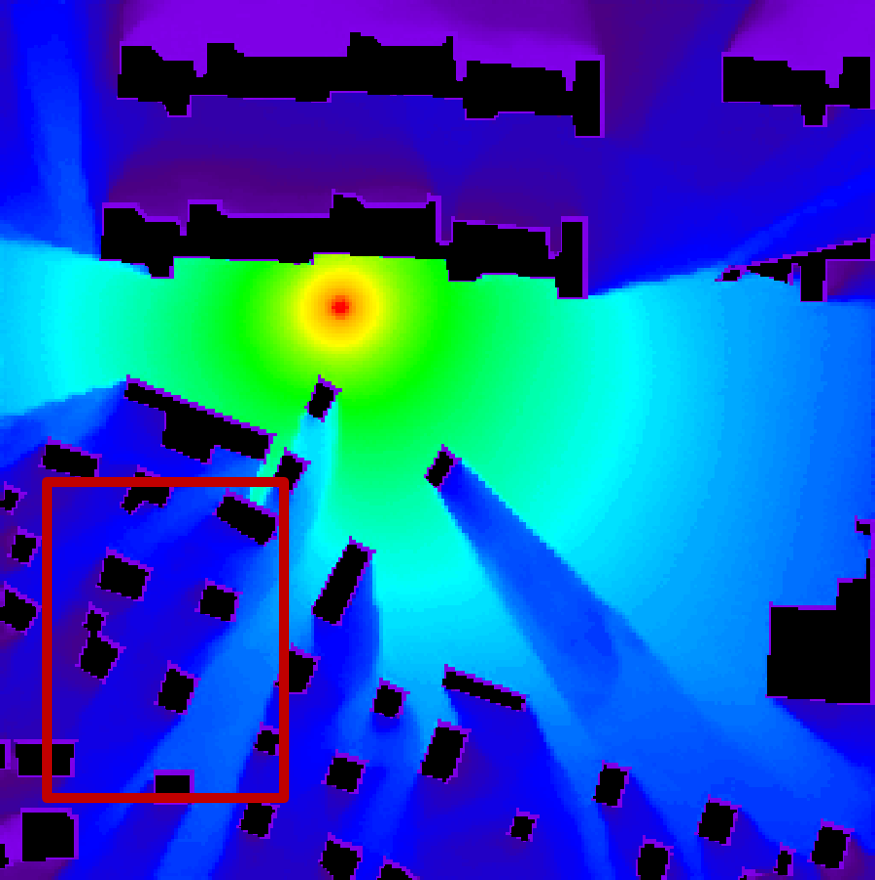}} &
  \subcaptionbox{}{\includegraphics[width=0.13\linewidth]{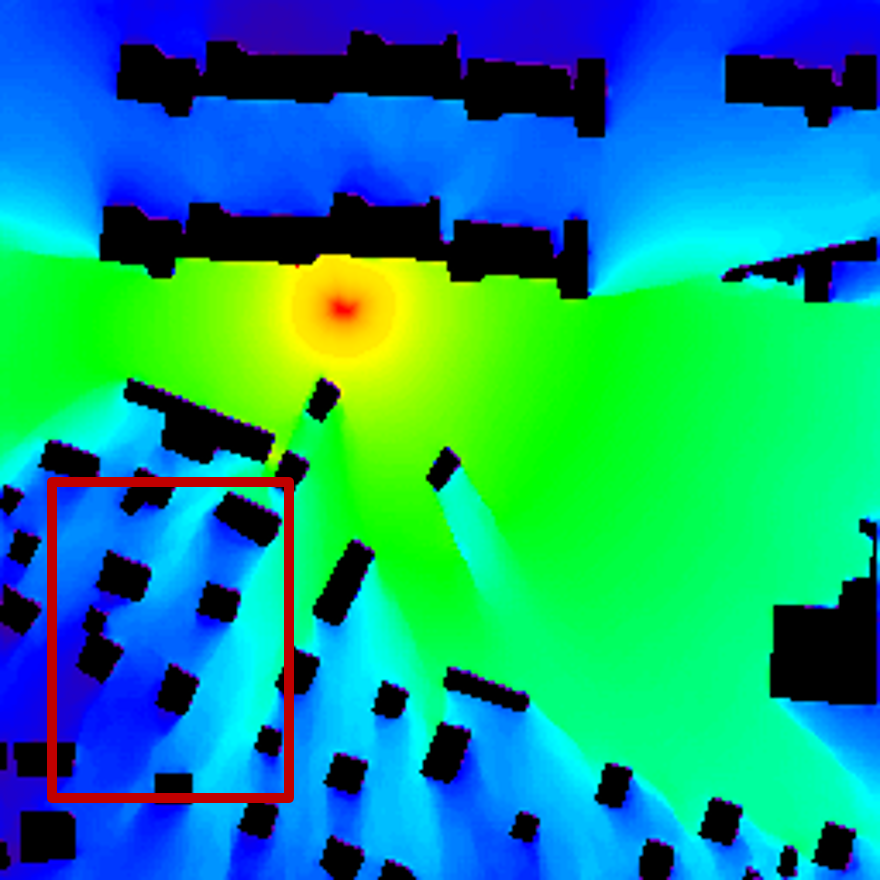}} &
  \subcaptionbox{}{\includegraphics[width=0.13\linewidth]{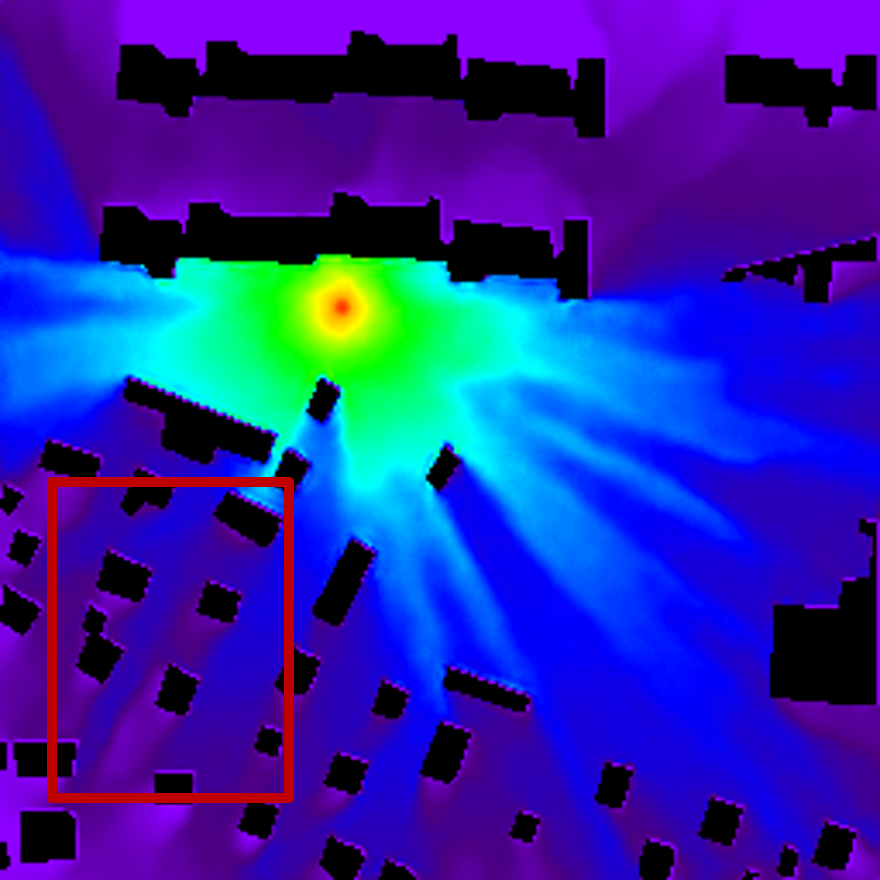}} &
  \subcaptionbox{}{\includegraphics[width=0.13\linewidth]{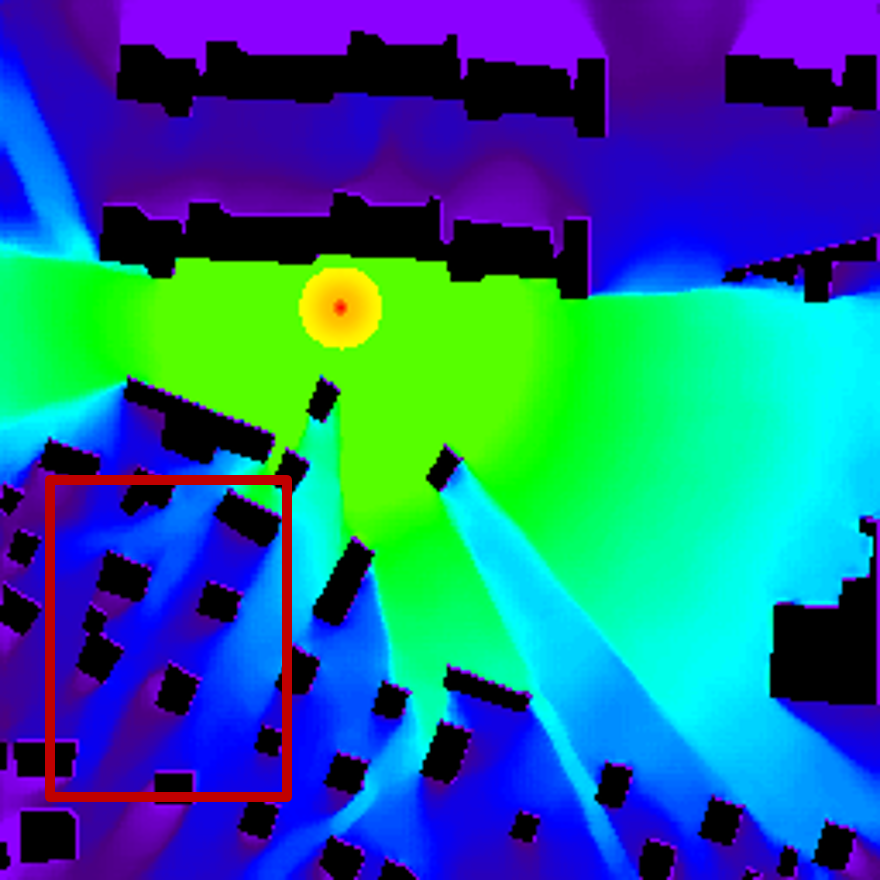}} &
  \subcaptionbox{}{\includegraphics[width=0.13\linewidth]{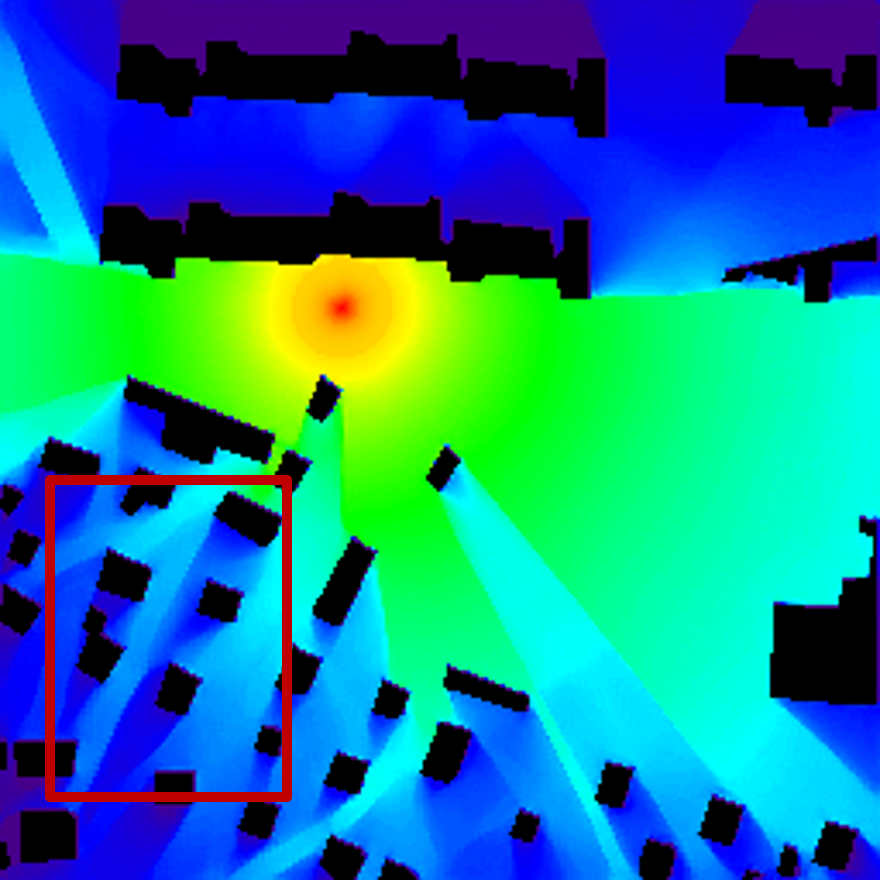}} &
  \subcaptionbox{}{\includegraphics[width=0.13\linewidth]{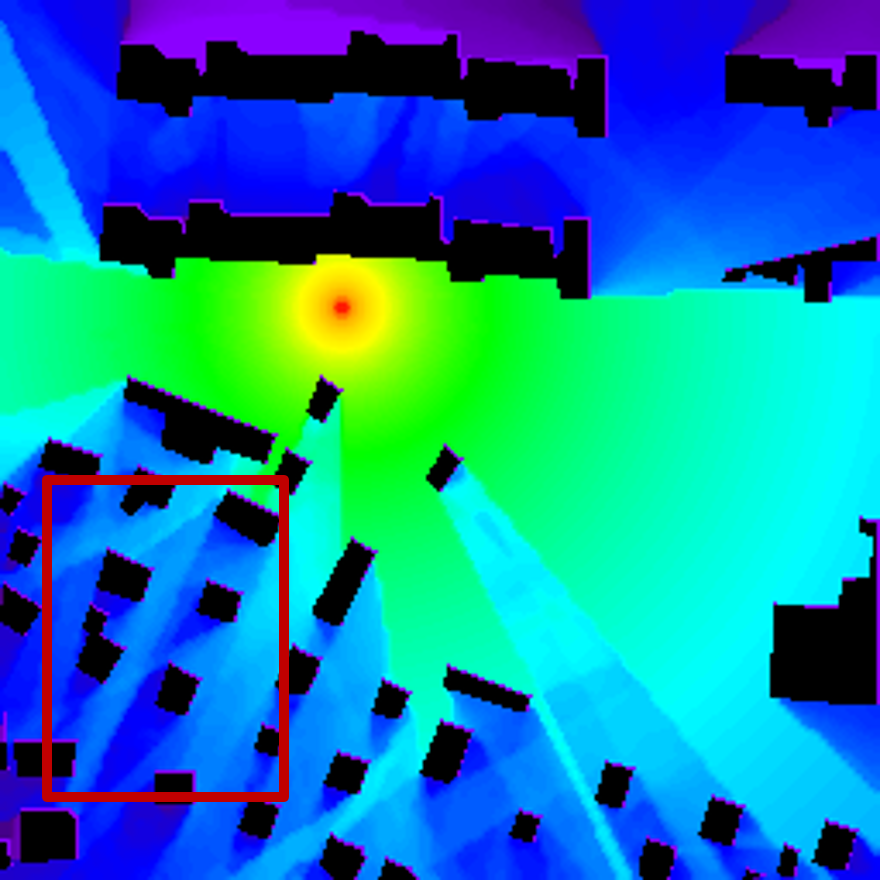}} \\
  \subcaptionbox{}{\includegraphics[width=0.13\linewidth]{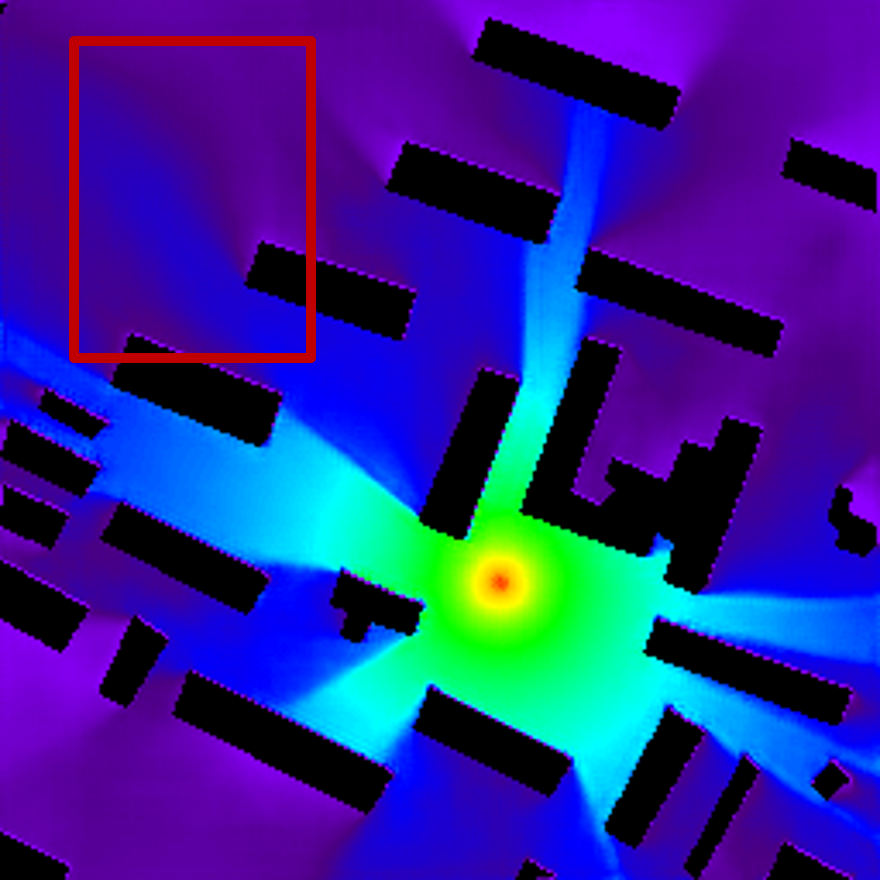}} &
  \subcaptionbox{}{\includegraphics[width=0.13\linewidth]{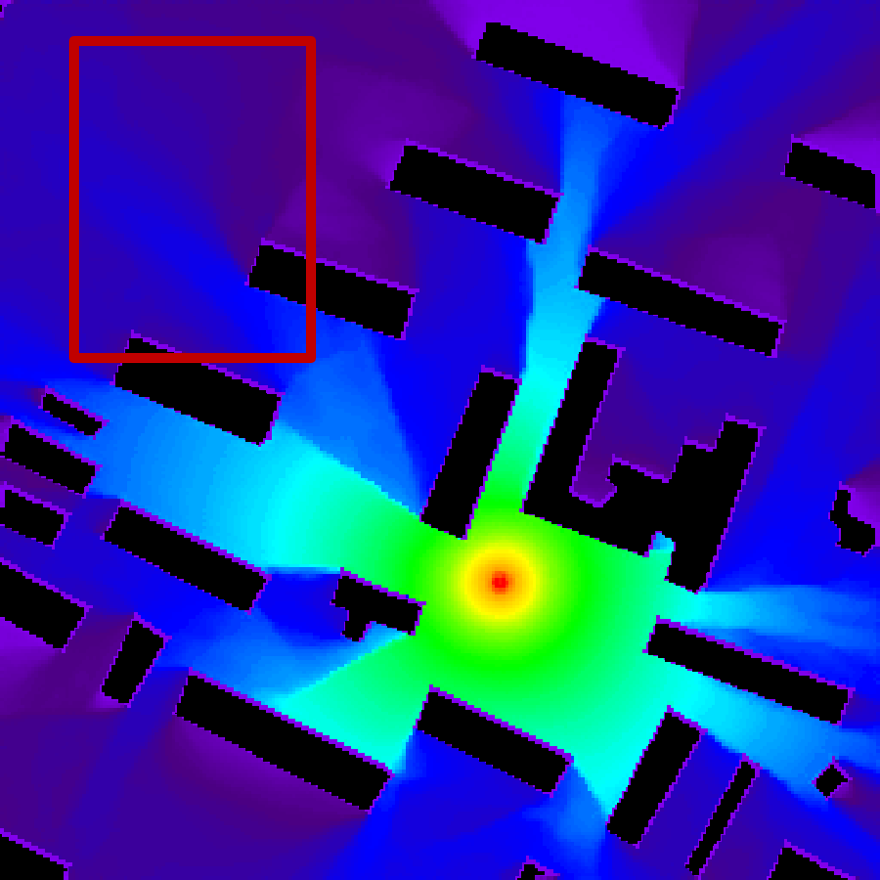}} &
  \subcaptionbox{}{\includegraphics[width=0.13\linewidth]{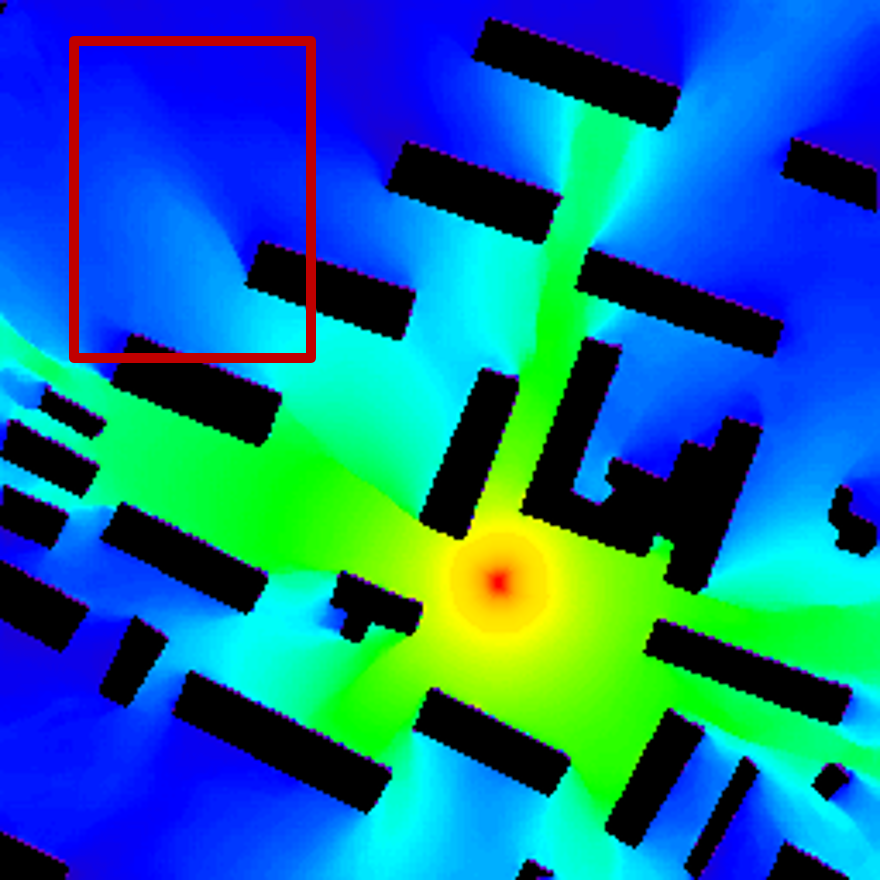}} &
  \subcaptionbox{}{\includegraphics[width=0.13\linewidth]{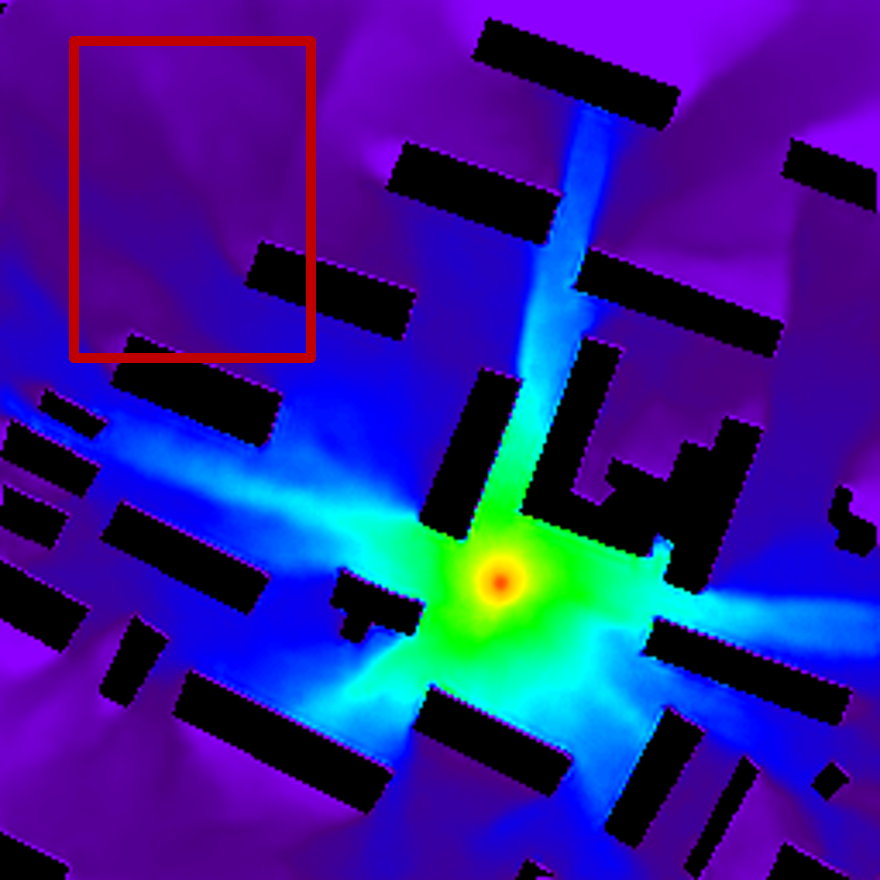}} &
  \subcaptionbox{}{\includegraphics[width=0.13\linewidth]{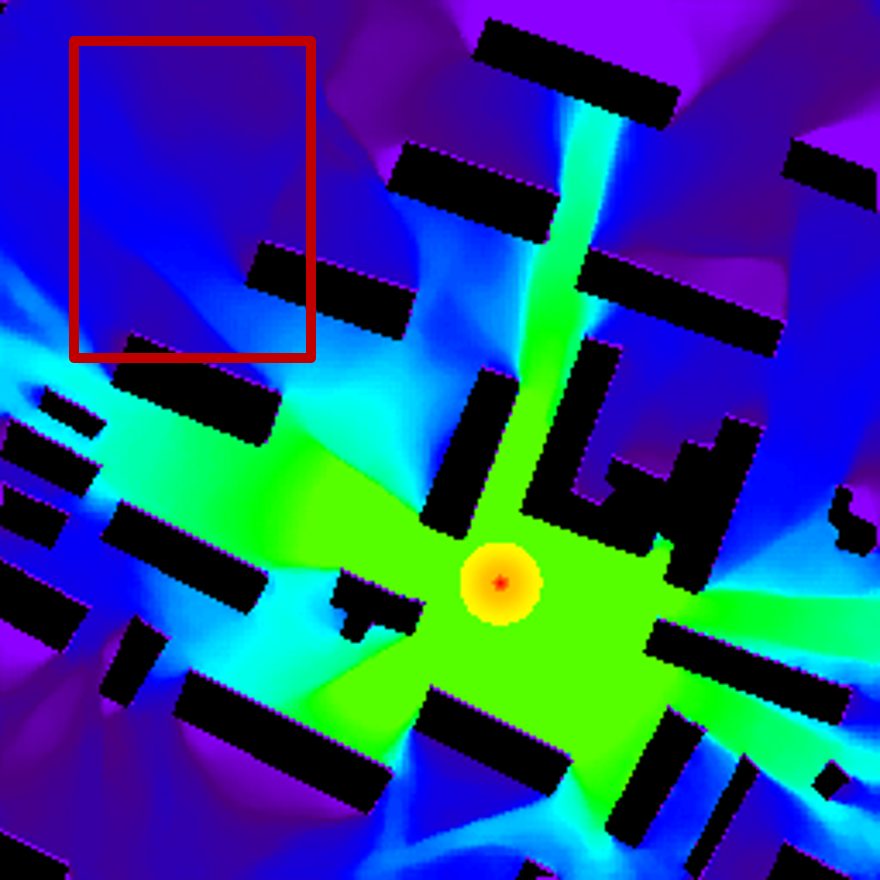}} &
  \subcaptionbox{}{\includegraphics[width=0.13\linewidth]{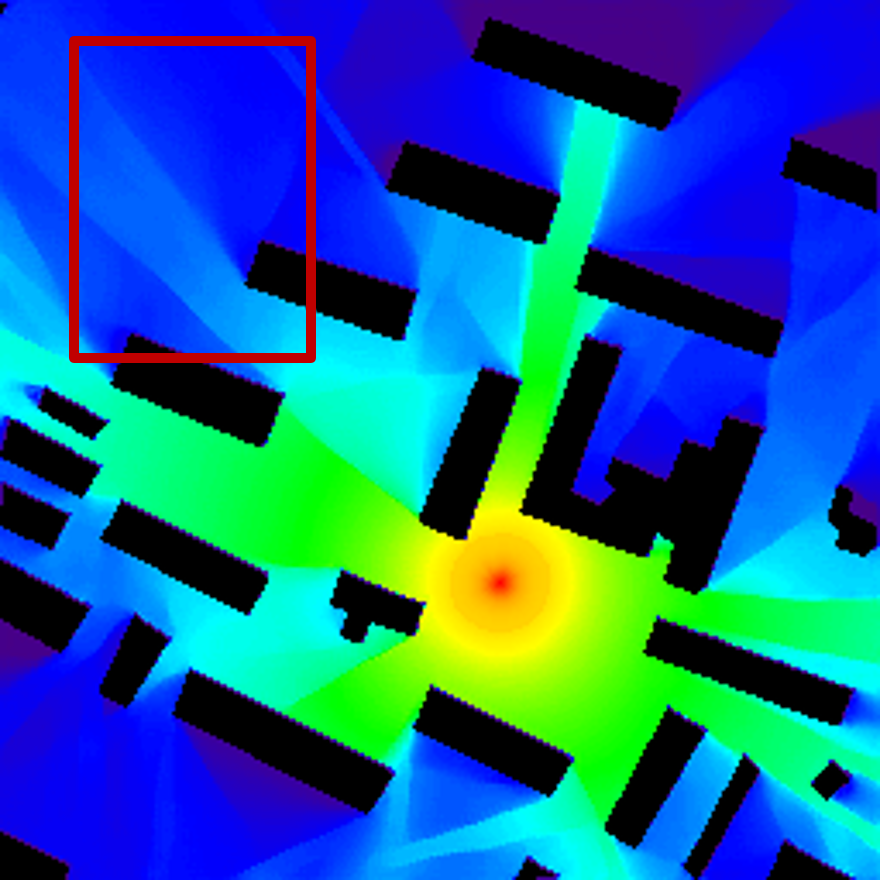}} &
  \subcaptionbox{}{\includegraphics[width=0.13\linewidth]{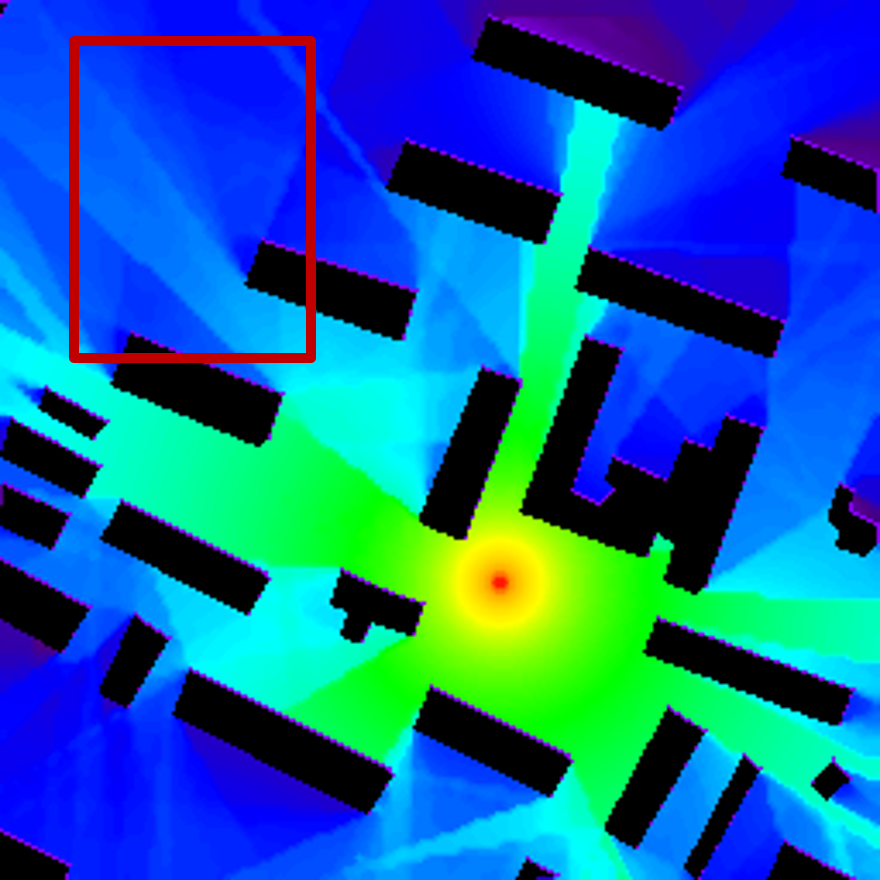}} \\
  \subcaptionbox{}{\includegraphics[width=0.13\linewidth]{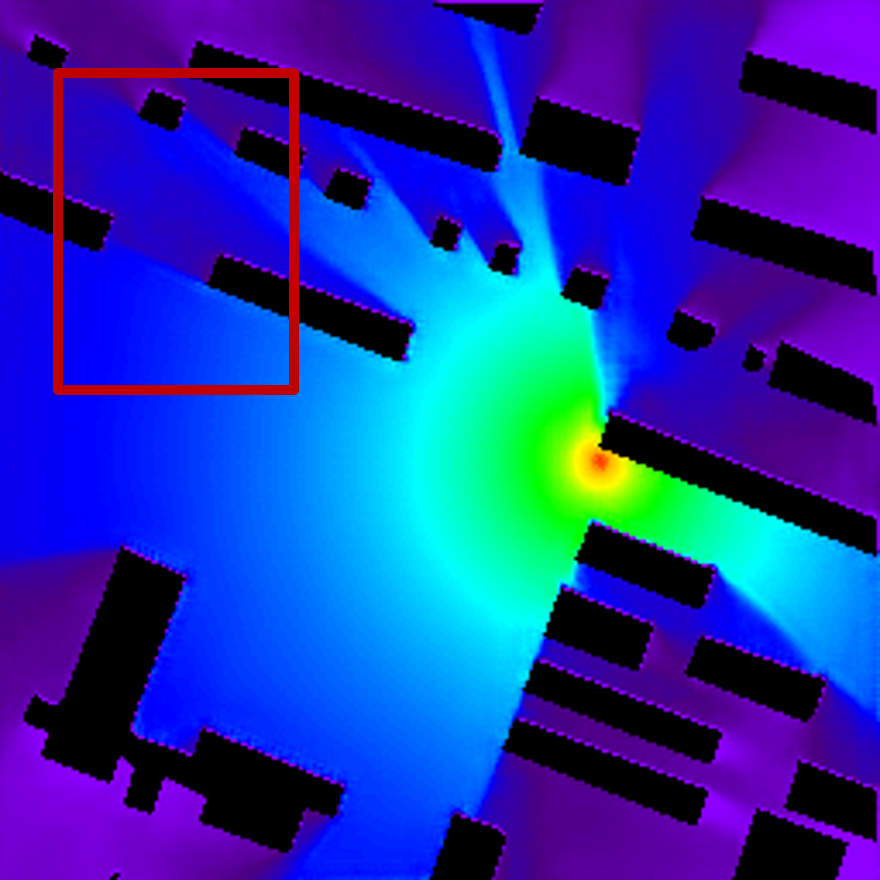}} &
  \subcaptionbox{}{\includegraphics[width=0.13\linewidth]{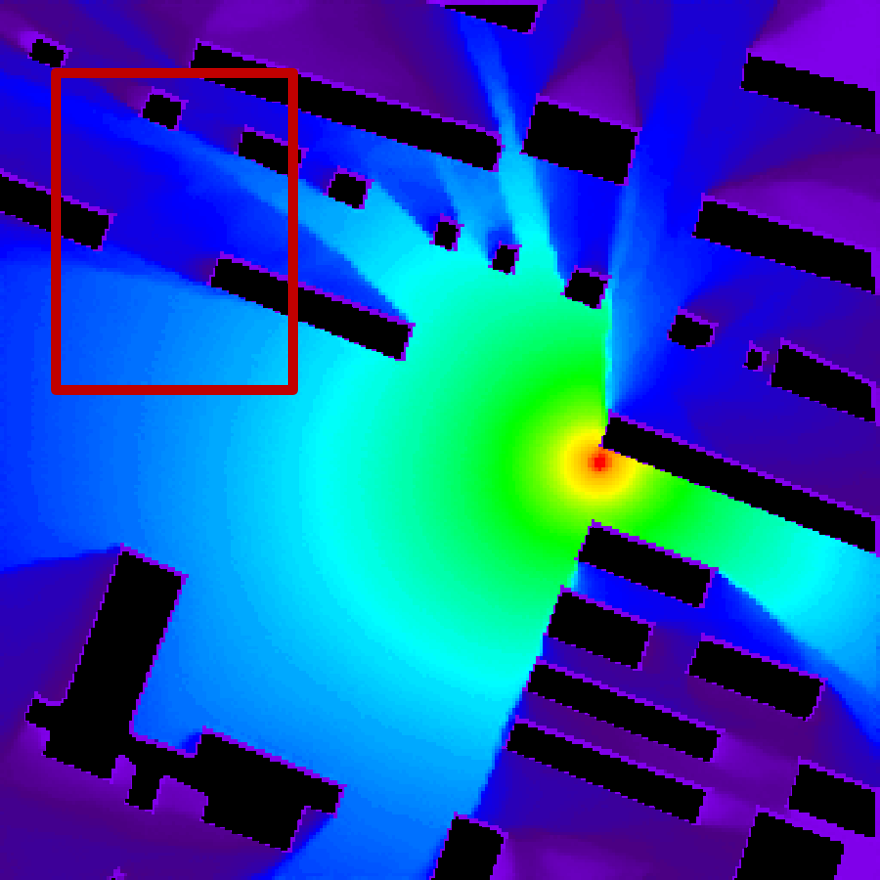}} &
  \subcaptionbox{}{\includegraphics[width=0.13\linewidth]{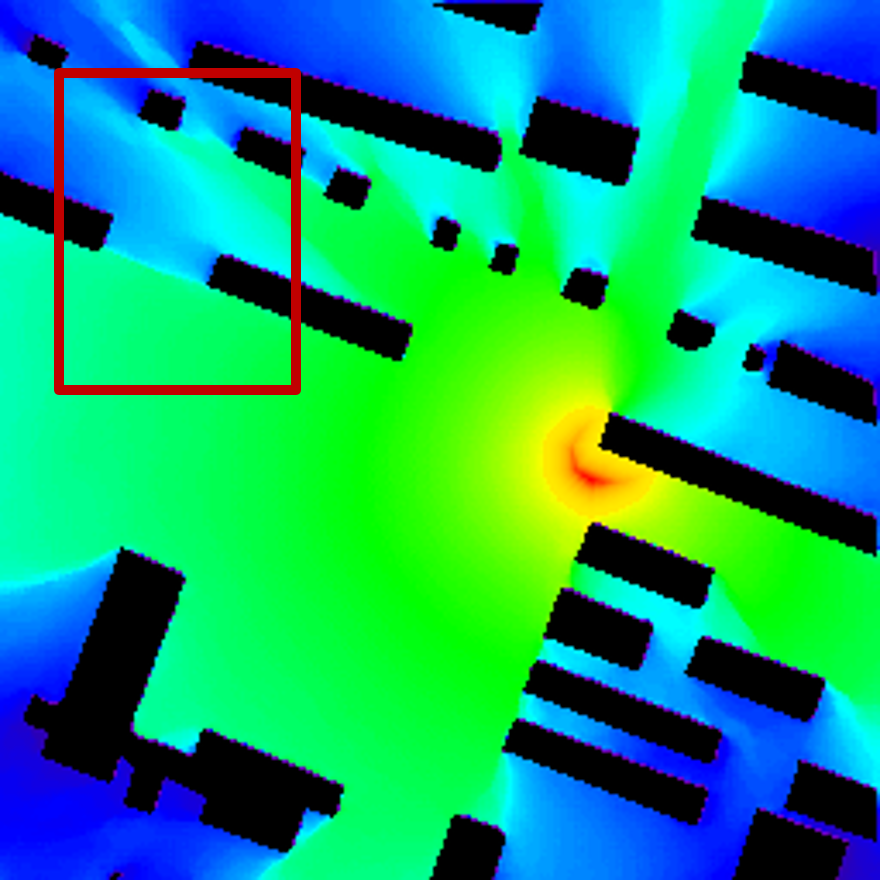}} &
  \subcaptionbox{}{\includegraphics[width=0.13\linewidth]{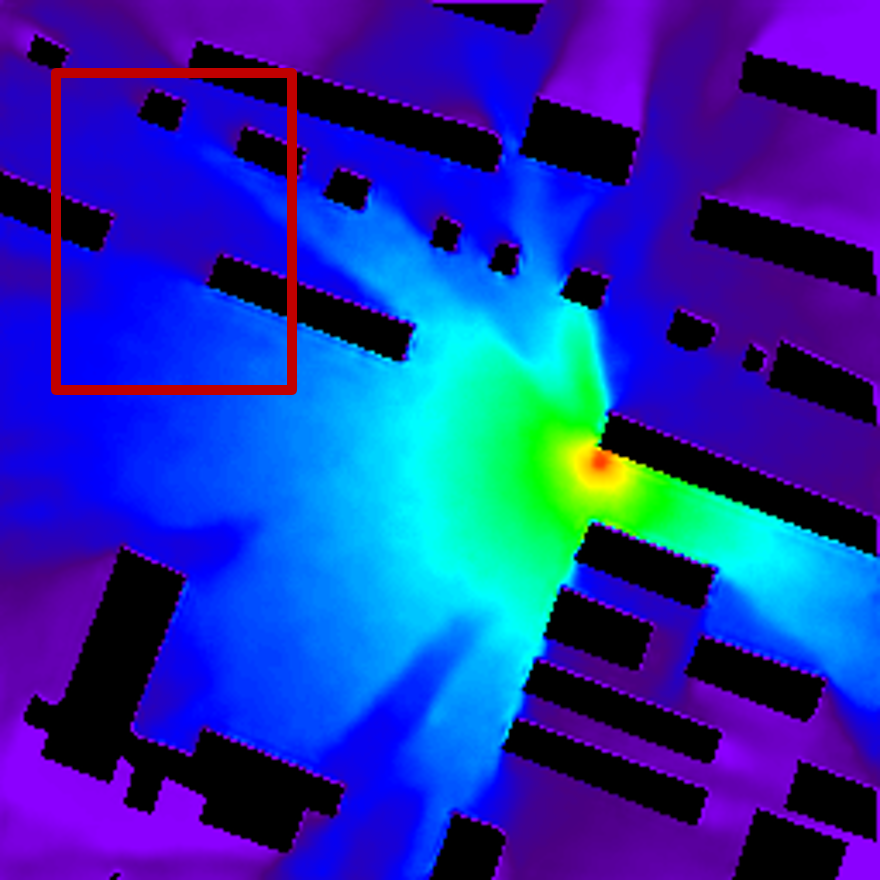}} &
  \subcaptionbox{}{\includegraphics[width=0.13\linewidth]{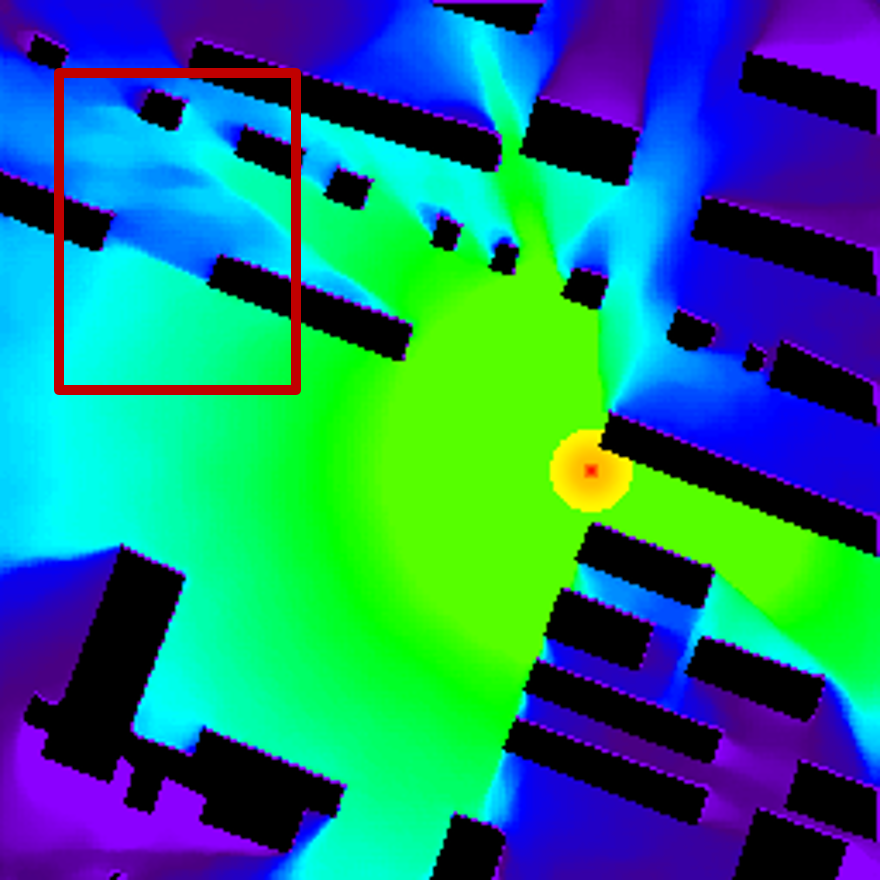}} &
  \subcaptionbox{}{\includegraphics[width=0.13\linewidth]{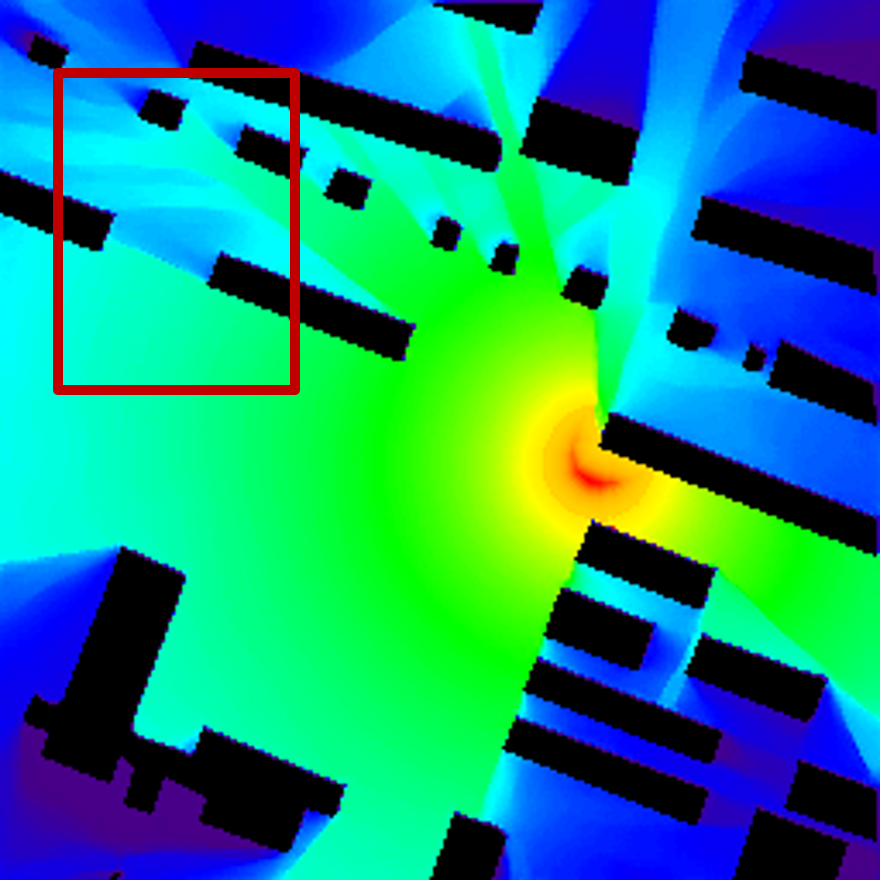}} &
  \subcaptionbox{}{\includegraphics[width=0.13\linewidth]{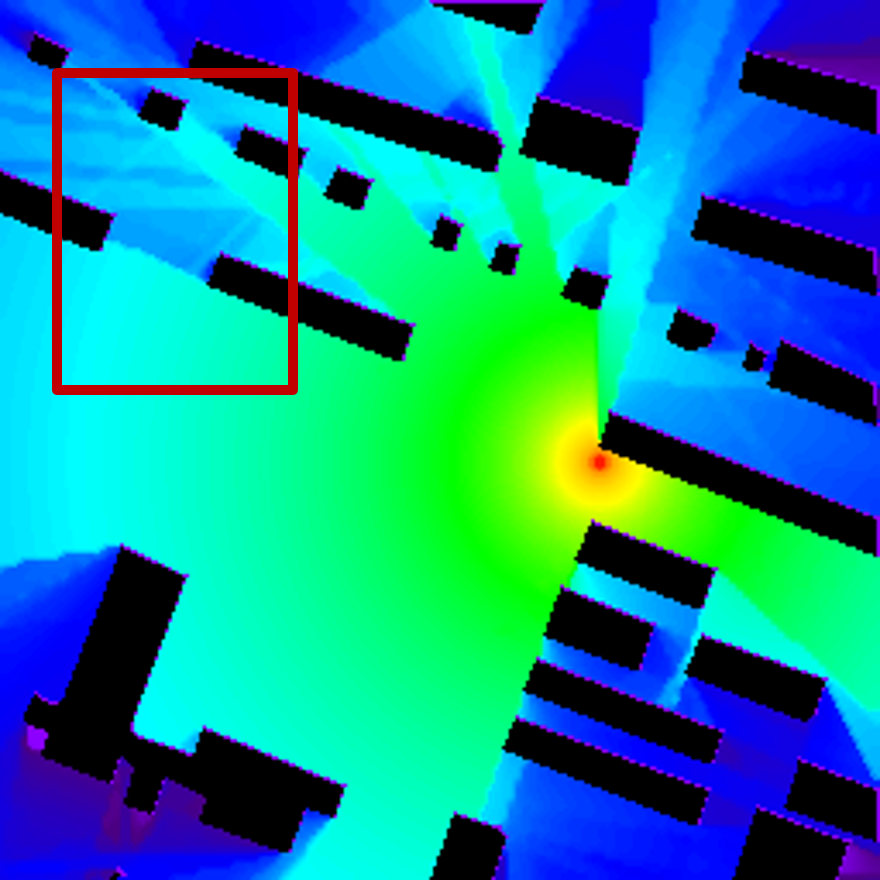}} \\
  \small\bfseries RadioUNet &
  \small\bfseries PhyRMDM &
  \small\bfseries RadioDiff &
  \small\bfseries RME-GAN &
  \small\bfseries \begin{tabular}{@{}c@{}}RadioDiff-FS \\ (one-shot)\end{tabular} &
  \small\bfseries RadioDiff-FS &
  \small\bfseries Ground Truth
\end{tabular}
\caption{Visual comparison of constructed SRM across different methods.}
\label{fig:IRT4_CP}
\end{figure*}
 
We evaluate on RadioMapSeer~\cite{yapar2023first}, a large-scale urban benchmark for RM prediction. Each scene is represented by a $256\times256$ building occupancy grid with $1$~m spatial resolution and a unified building height of $25$~m. We consider the device-to-device (D2D) setting with transmitter and receiver heights of $1.5$~m, a carrier frequency of $5.9$~GHz, and a transmitter power of $23$~dBm~\cite{yapar2023first}. RadioMapSeer provides subsets with different propagation complexities. We use the multipath-rich subset with up to four interactions, denoted as IRT4 in the dataset release, as the target MU-RM domain. A key characteristic of IRT4 is its limited supervision density: each scene contains only two transmitter placements and thus only two MU-RM samples per scene. This results in $1402$ MU-RM samples over $701$ scenes, which naturally forms a few-shot adaptation setting when fine-tuning a pretrained generator. We evaluate both static RMs (SRM) that account for building-induced attenuation and dynamic RMs (DRM) that additionally include vehicle-induced blockage~\cite{yapar2023first}. We adopt a scene-disjoint split with $500$ scenes for training and validation and $200$ unseen scenes for testing, ensuring generalization to new urban layouts. Unless otherwise stated, few-shot fine-tuning uses only a small subset of the IRT4 training samples, while the remaining samples are used for validation and ablation.

\subsection{Metrics and Baselines}
 
We evaluate RM reconstruction using pixel-level fidelity and structural consistency metrics, following common practice in RM prediction~\cite{levie2021radiounet, wang2024radiodiff}. Let $Y,\widehat Y\in\mathbb{R}^{H\times W}$ denote the ground-truth and predicted maps. We report normalized mean squared error (NMSE) $\mathrm{NMSE}(Y,\widehat Y)={\sum_{i=1}^{H}\sum_{j=1}^{W}(Y_{i,j}-\widehat Y_{i,j})^2}/{\sum_{i=1}^{H}\sum_{j=1}^{W}Y_{i,j}^2}$ and root mean squared error (RMSE) $\mathrm{RMSE}(Y,\widehat Y)=\sqrt{{1}/{(HW)}\sum_{i=1}^{H}\sum_{j=1}^{W}(Y_{i,j}-\widehat Y_{i,j})^2}$, where smaller values indicate higher accuracy. To quantify structure preservation, we compute structural similarity index measure (SSIM)~\cite{wang2004image}, $\mathrm{SSIM}(Y,\widehat Y)={(2\mu_Y\mu_{\widehat Y}+C_1)(2\sigma_{Y\widehat Y}+C_2)}/{((\mu_Y^2+\mu_{\widehat Y}^2+C_1)(\sigma_Y^2+\sigma_{\widehat Y}^2+C_2))}$, and peak signal-to-noise ratio (PSNR), $\mathrm{PSNR}(Y,\widehat Y)=10\log_{10}\!\left({\mathrm{MAX}_Y^2}/{\mathrm{MSE}(Y,\widehat Y)}\right)$ with $\mathrm{MSE}(Y,\widehat Y)={1}/{(HW)}\sum_{i,j}(Y_{i,j}-\widehat Y_{i,j})^2$. Higher SSIM and PSNR indicate better structural consistency and visual fidelity. All metrics are averaged over the test set.

\begin{figure*}[!ht]
\centering
\captionsetup{font={small}, skip=16pt}
\renewcommand{\thesubfigure}{\arabic{subfigure}}
\renewcommand{\arraystretch}{0.8}
\setlength{\tabcolsep}{1.5pt}
\begin{tabular}{c c c c c c c}
  \subcaptionbox{}{\includegraphics[width=0.13\linewidth]{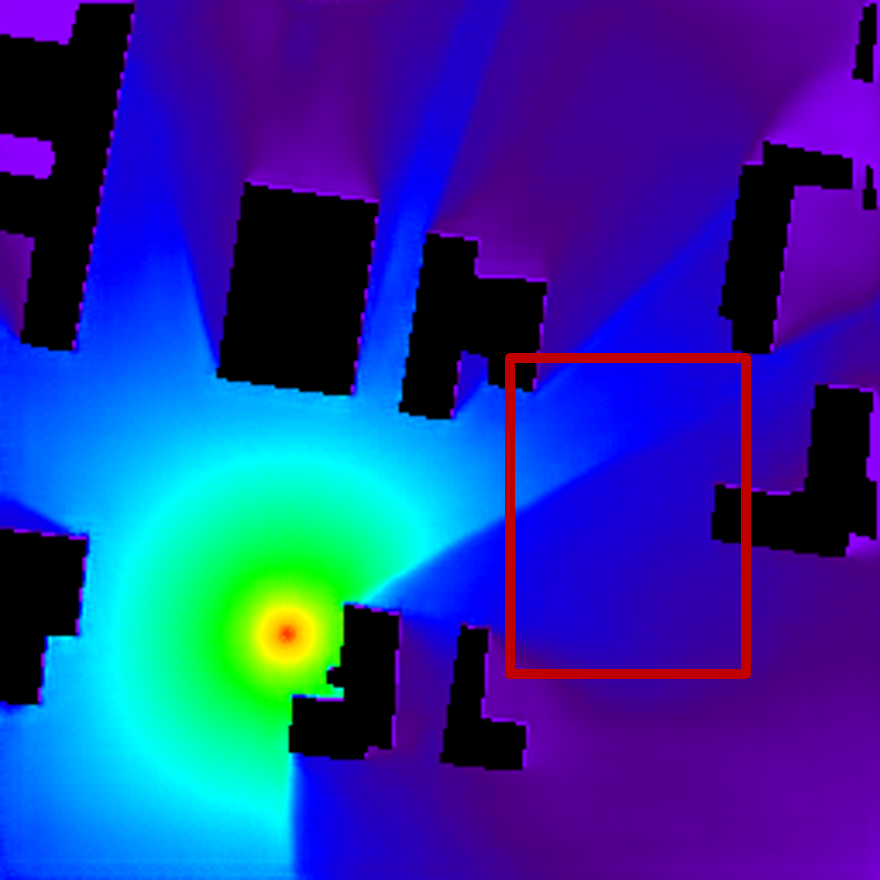}} &  
  \subcaptionbox{}{\includegraphics[width=0.13\linewidth]{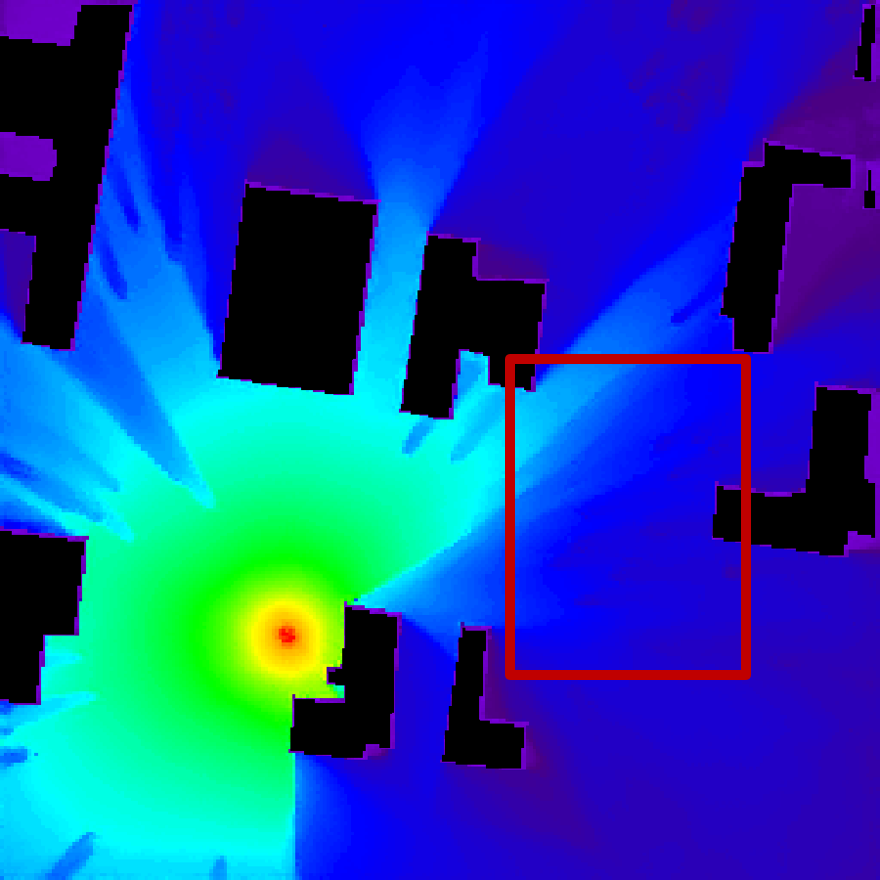}} &  
  \subcaptionbox{}{\includegraphics[width=0.13\linewidth]{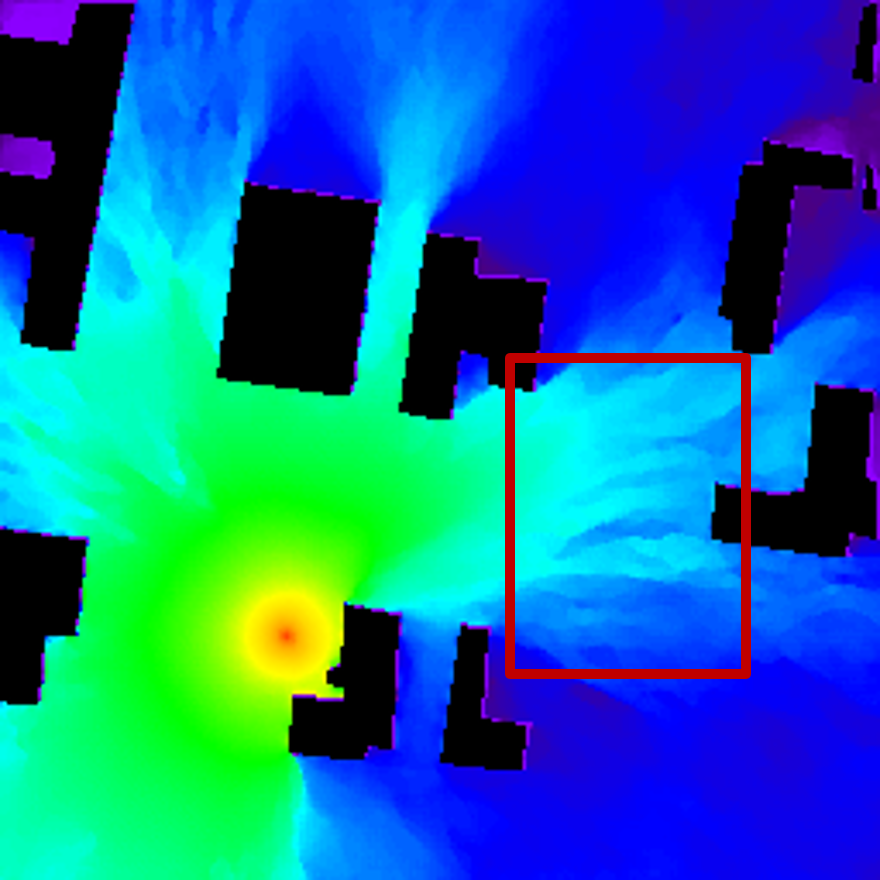}} &  
  \subcaptionbox{}{\includegraphics[width=0.13\linewidth]{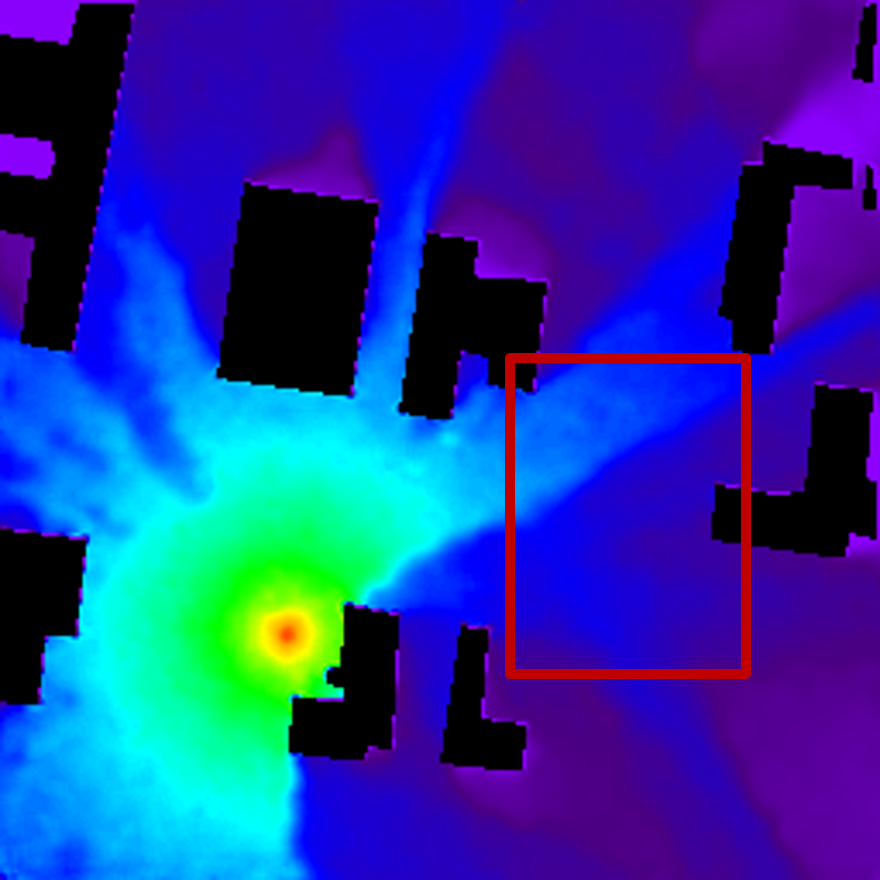}} & 
  \subcaptionbox{}{\includegraphics[width=0.13\linewidth]{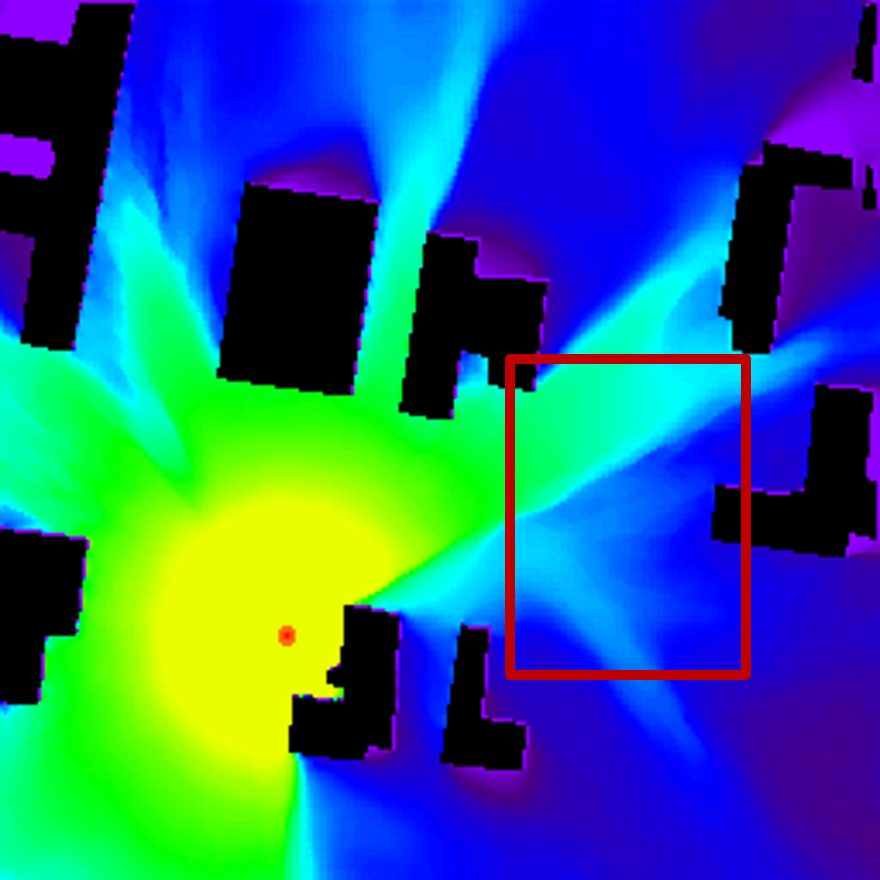}} &
  \subcaptionbox{}{\includegraphics[width=0.13\linewidth]{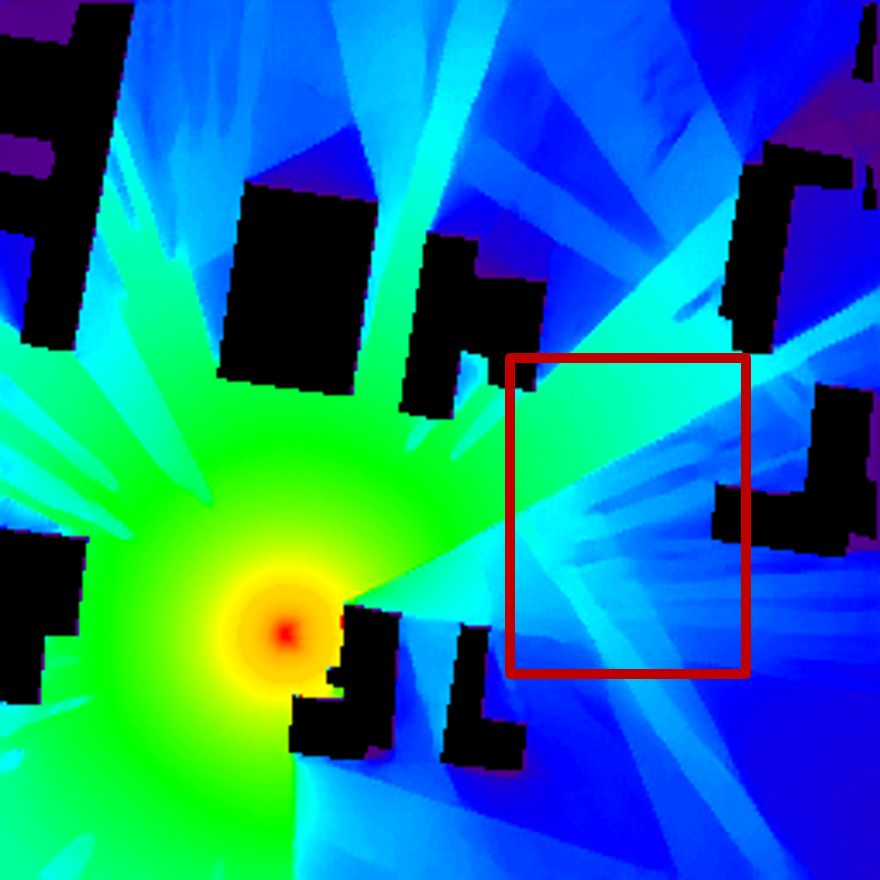}} &  
  \subcaptionbox{}{\includegraphics[width=0.13\linewidth]{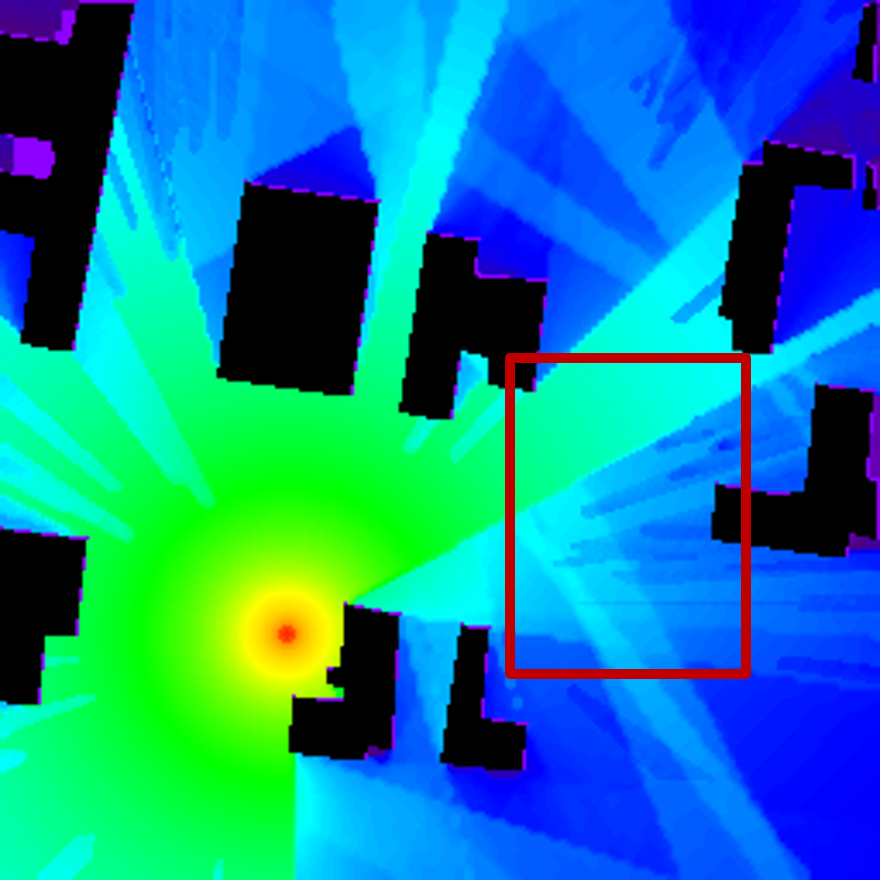}} \\  
  \subcaptionbox{}{\includegraphics[width=0.13\linewidth]{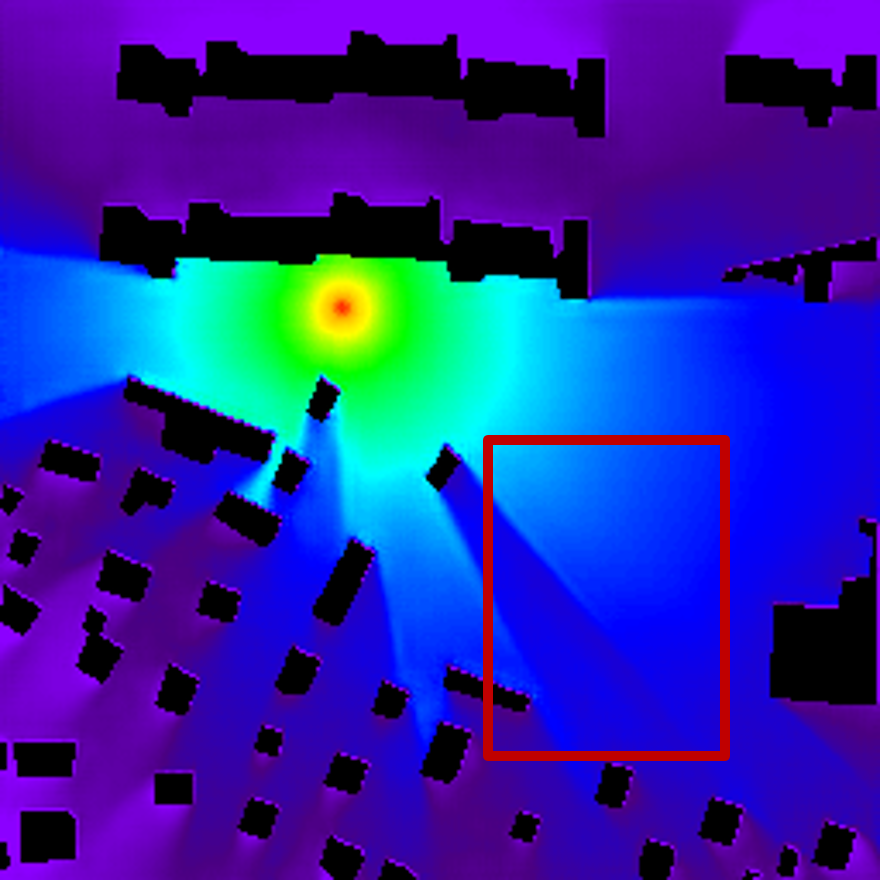}} &
  \subcaptionbox{}{\includegraphics[width=0.13\linewidth]{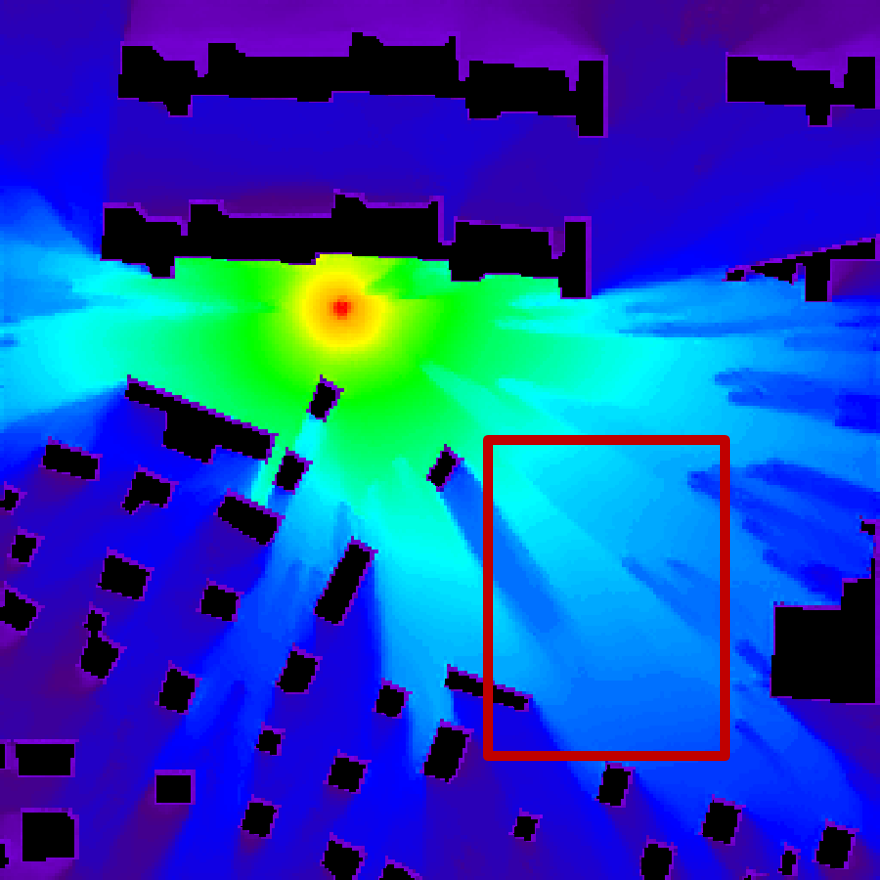}} &
  \subcaptionbox{}{\includegraphics[width=0.13\linewidth]{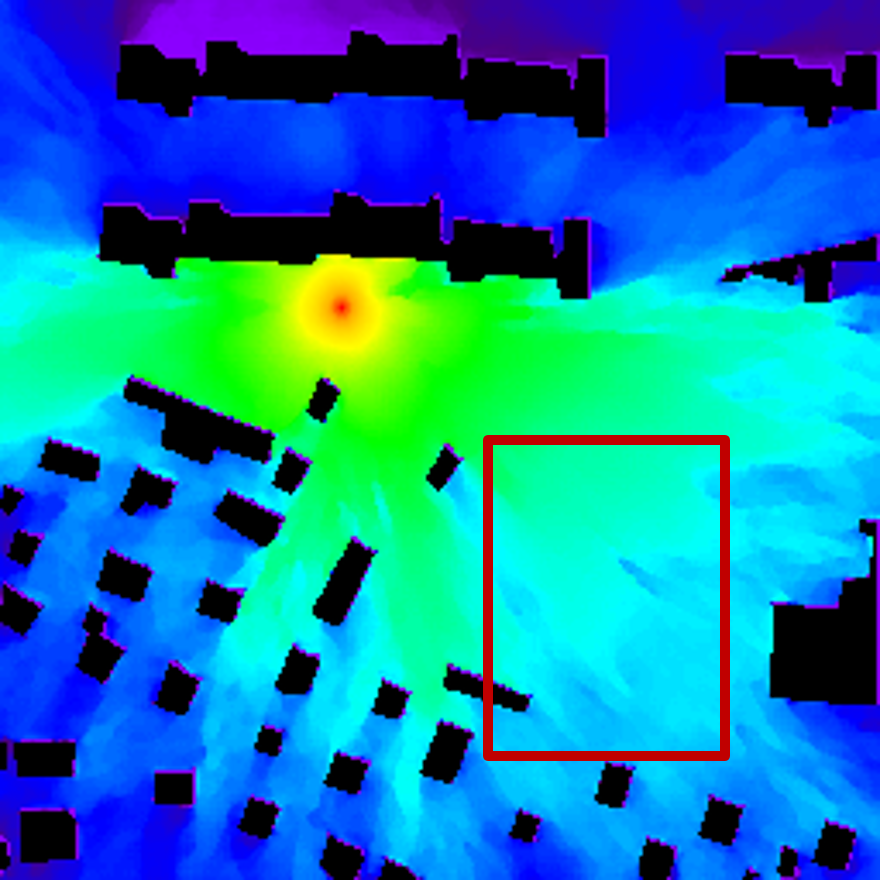}} &
  \subcaptionbox{}{\includegraphics[width=0.13\linewidth]{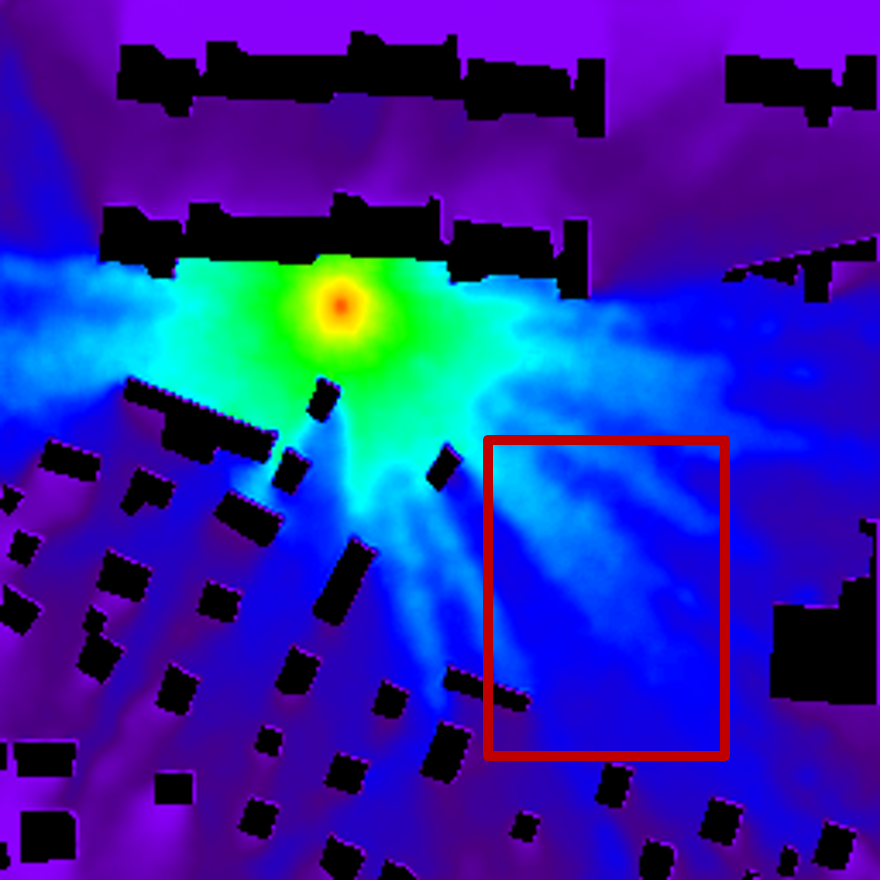}} &
  \subcaptionbox{}{\includegraphics[width=0.13\linewidth]{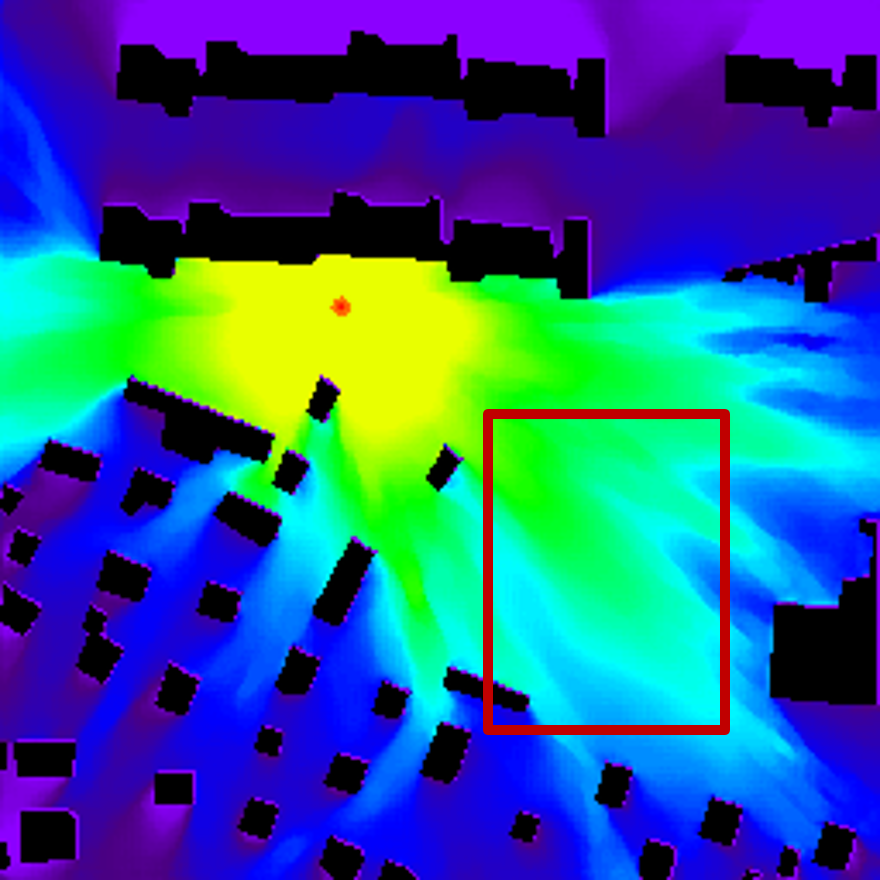}} &
  \subcaptionbox{}{\includegraphics[width=0.13\linewidth]{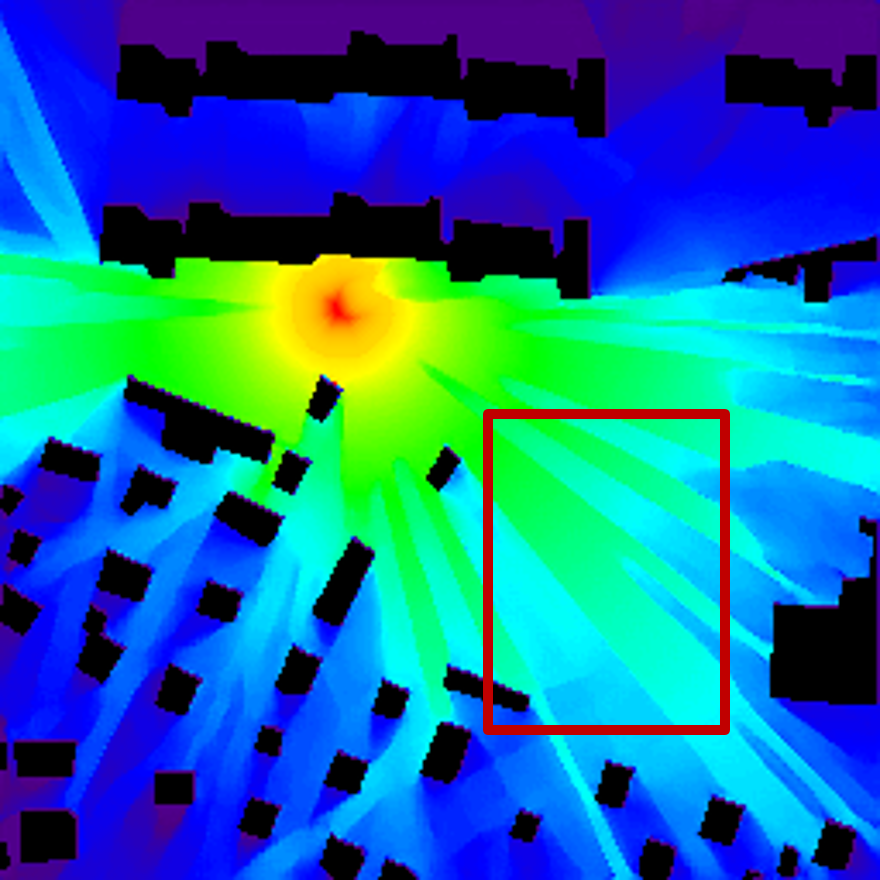}} &
  \subcaptionbox{}{\includegraphics[width=0.13\linewidth]{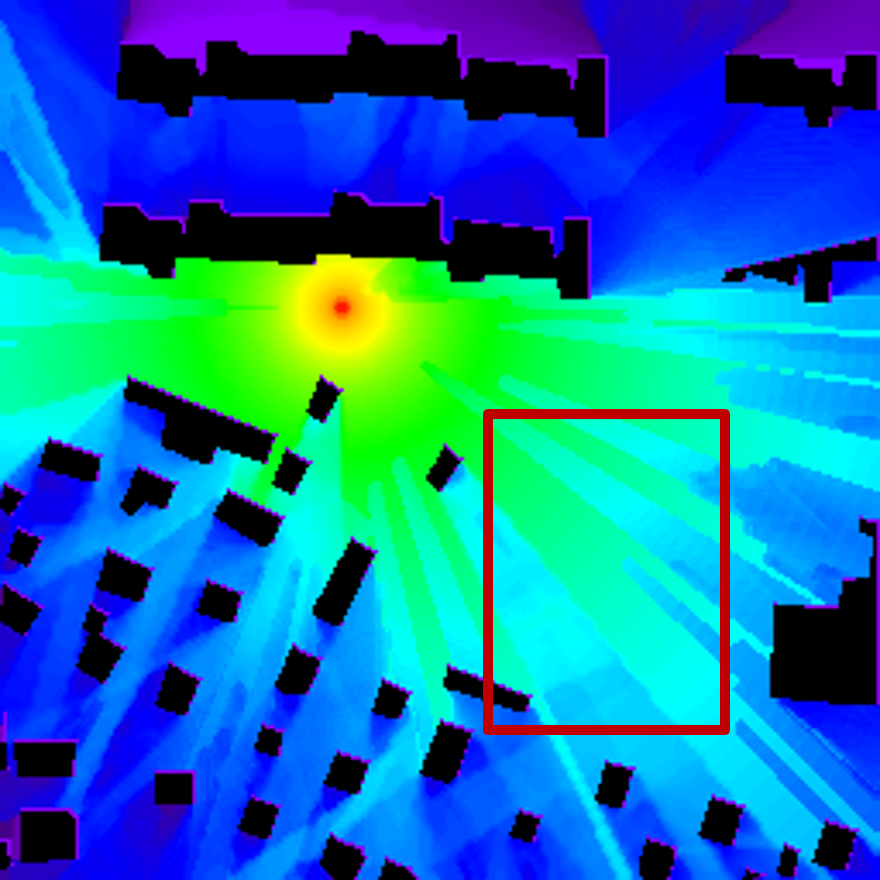}} \\  
  \subcaptionbox{}{\includegraphics[width=0.13\linewidth]{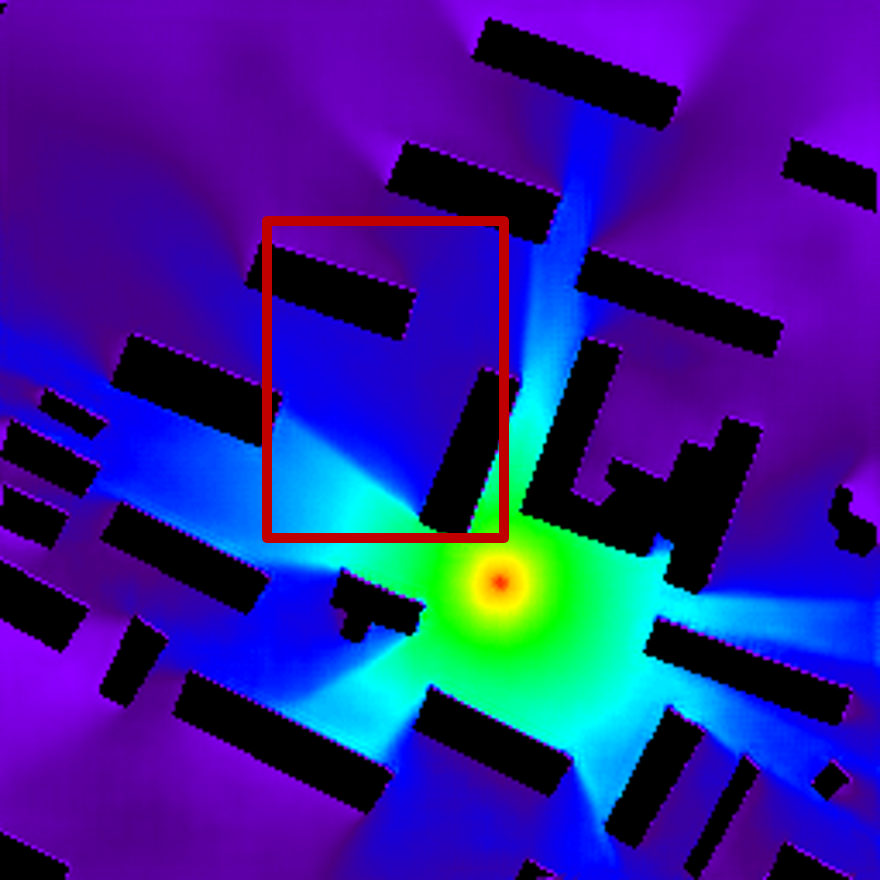}} &
  \subcaptionbox{}{\includegraphics[width=0.13\linewidth]{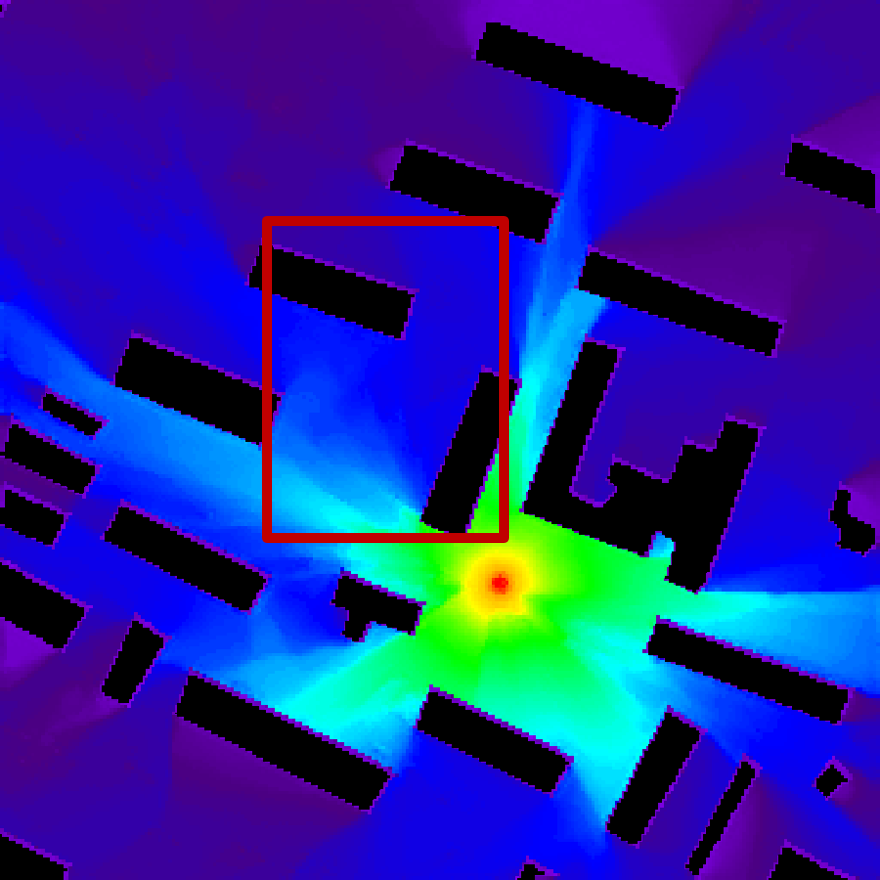}} &
  \subcaptionbox{}{\includegraphics[width=0.13\linewidth]{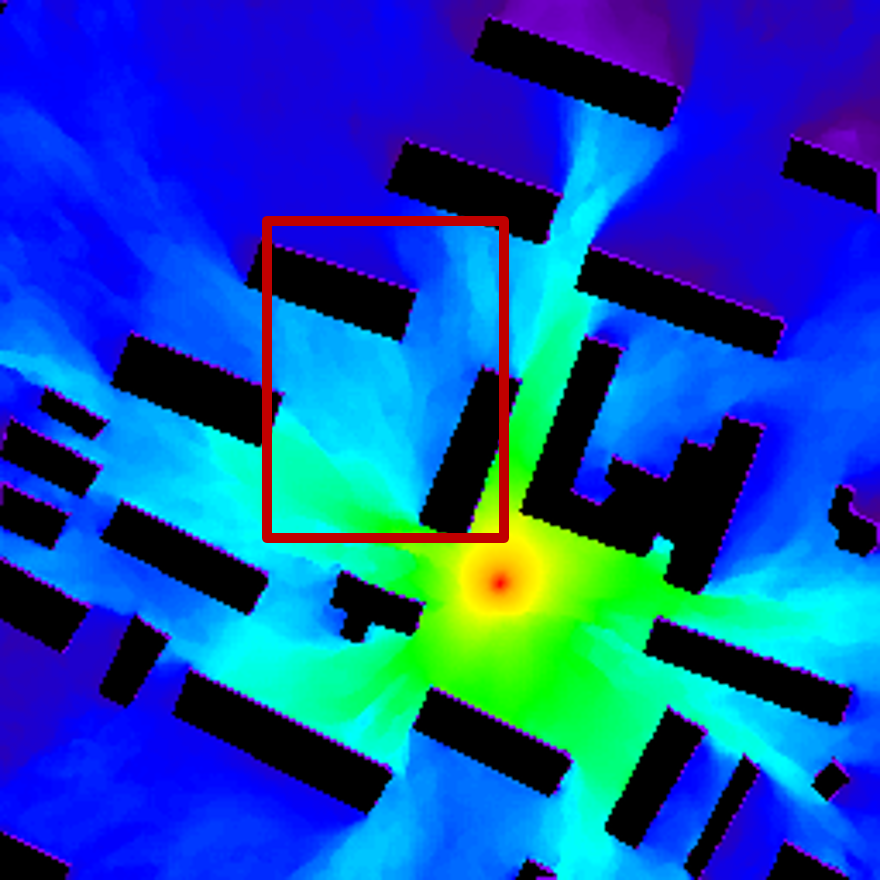}} &
  \subcaptionbox{}{\includegraphics[width=0.13\linewidth]{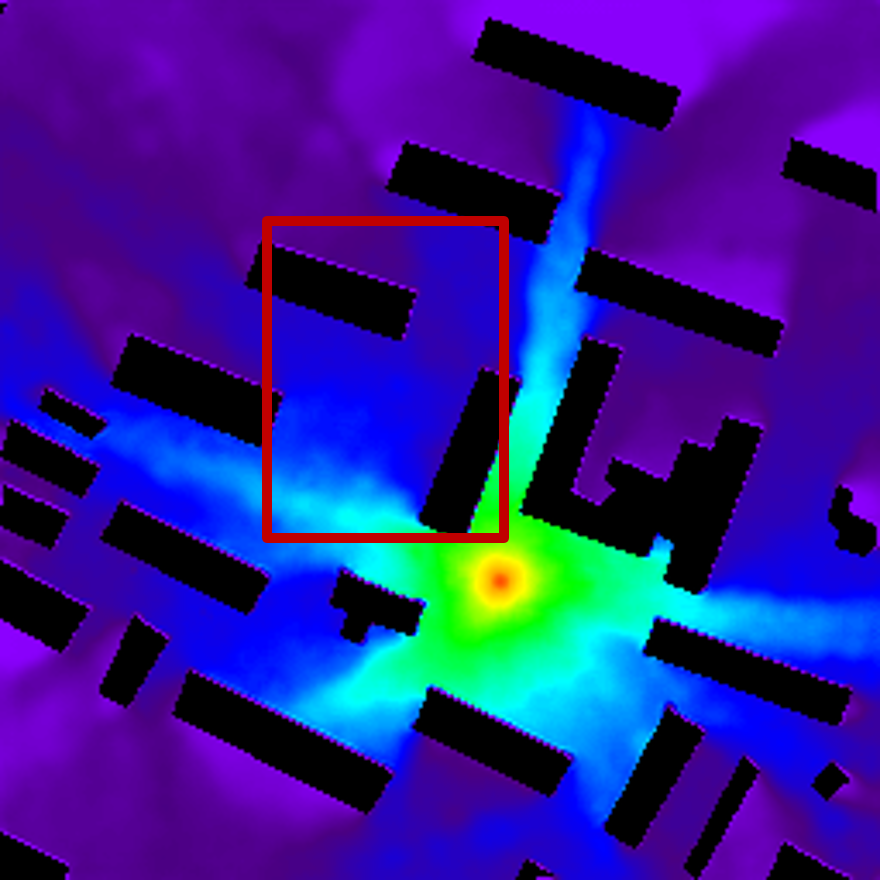}} &
  \subcaptionbox{}{\includegraphics[width=0.13\linewidth]{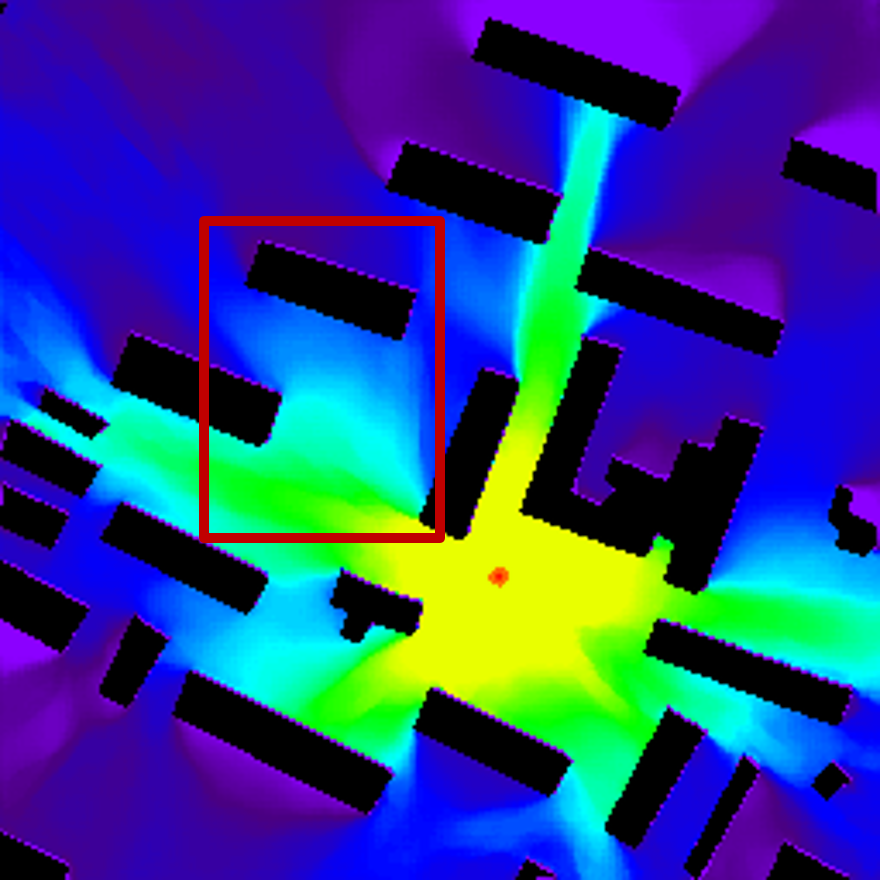}} &
  \subcaptionbox{}{\includegraphics[width=0.13\linewidth]{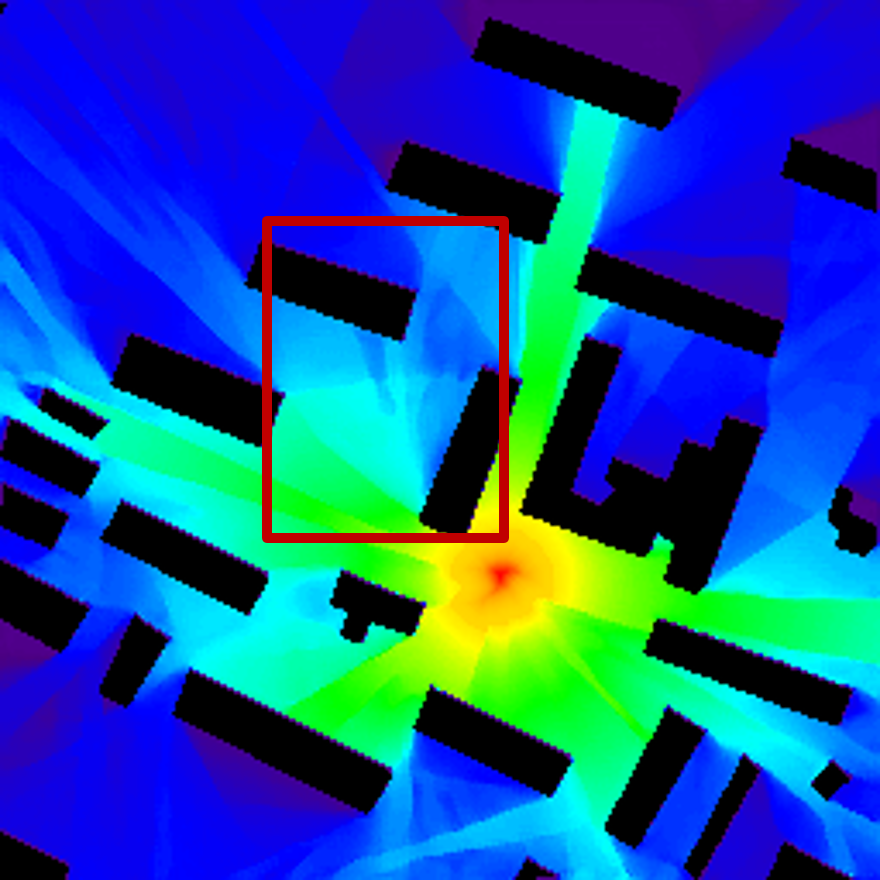}} &
  \subcaptionbox{}{\includegraphics[width=0.13\linewidth]{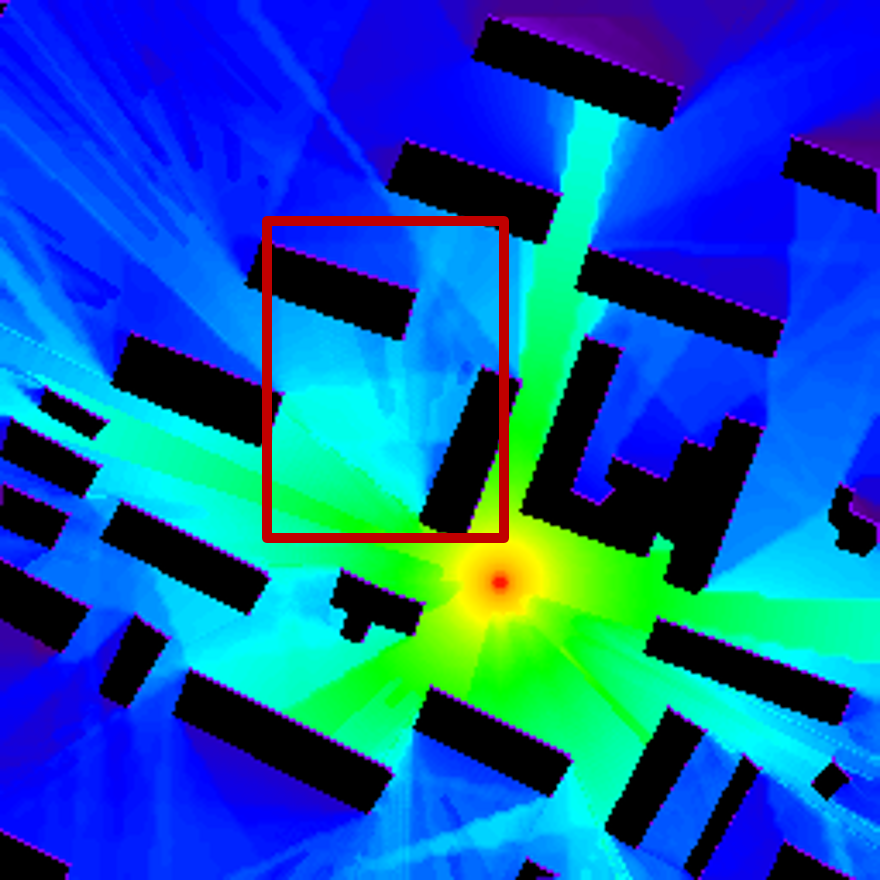}} \\
  \subcaptionbox{}{\includegraphics[width=0.13\linewidth]{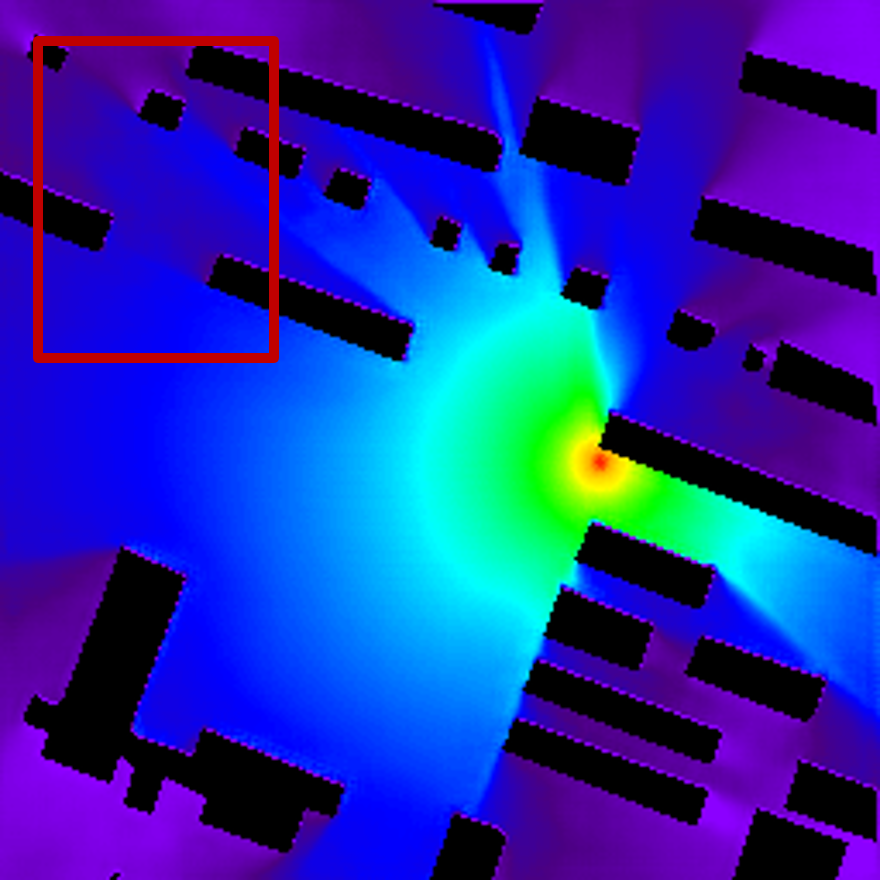}} &
  \subcaptionbox{}{\includegraphics[width=0.13\linewidth]{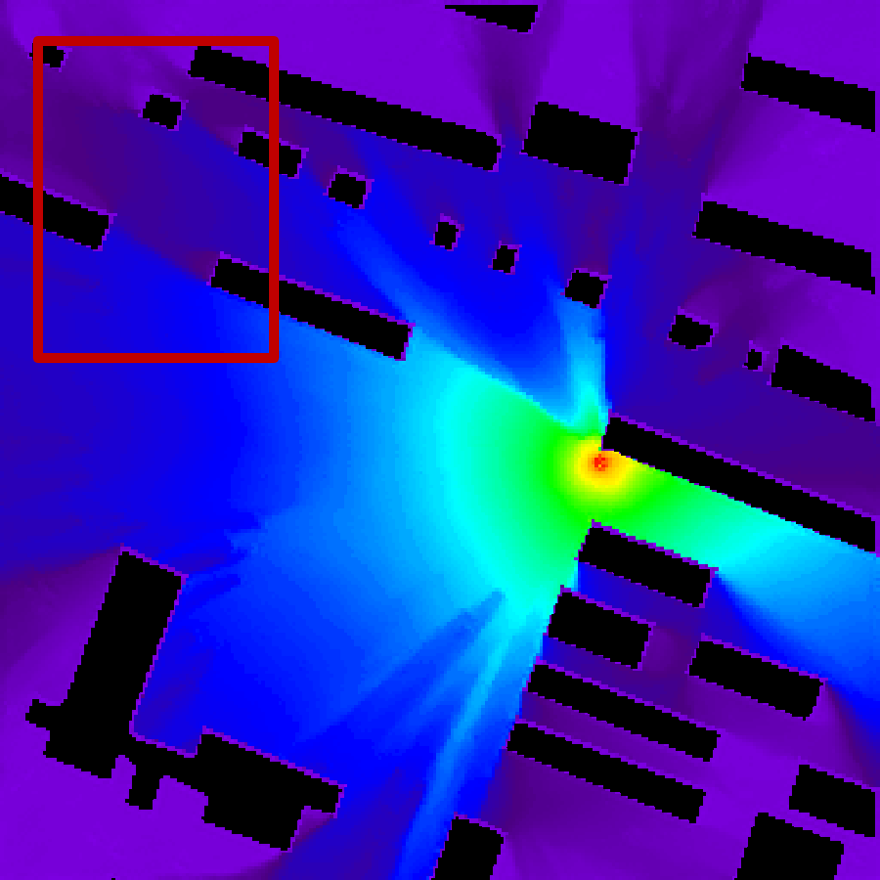}} &
  \subcaptionbox{}{\includegraphics[width=0.13\linewidth]{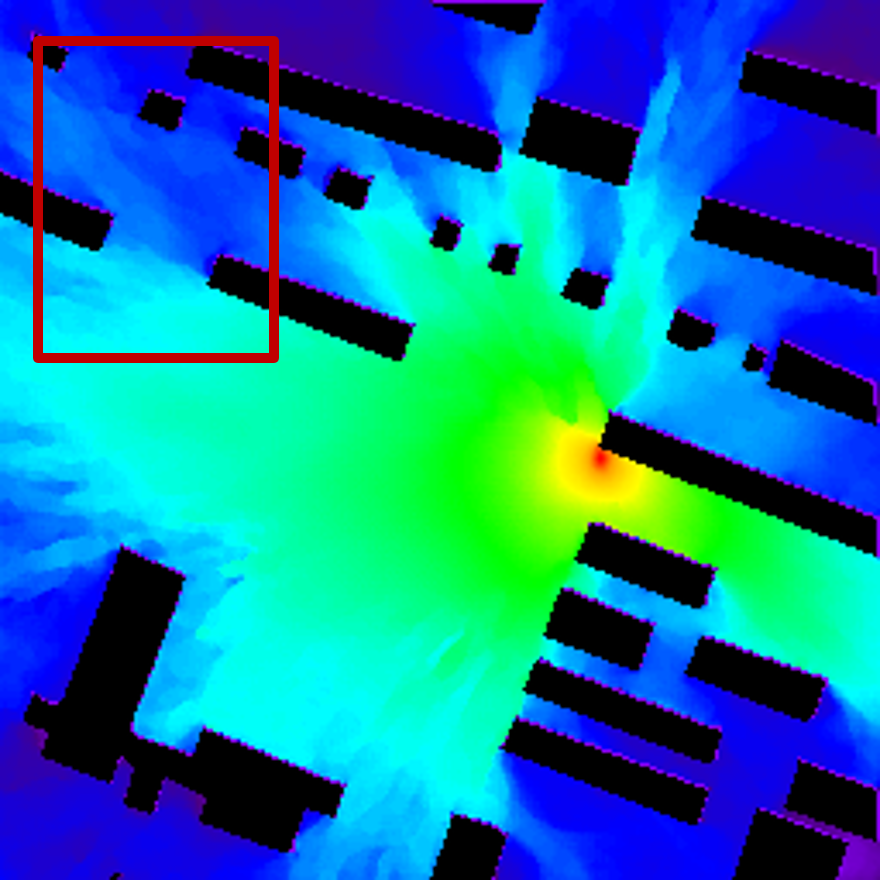}} &
  \subcaptionbox{}{\includegraphics[width=0.13\linewidth]{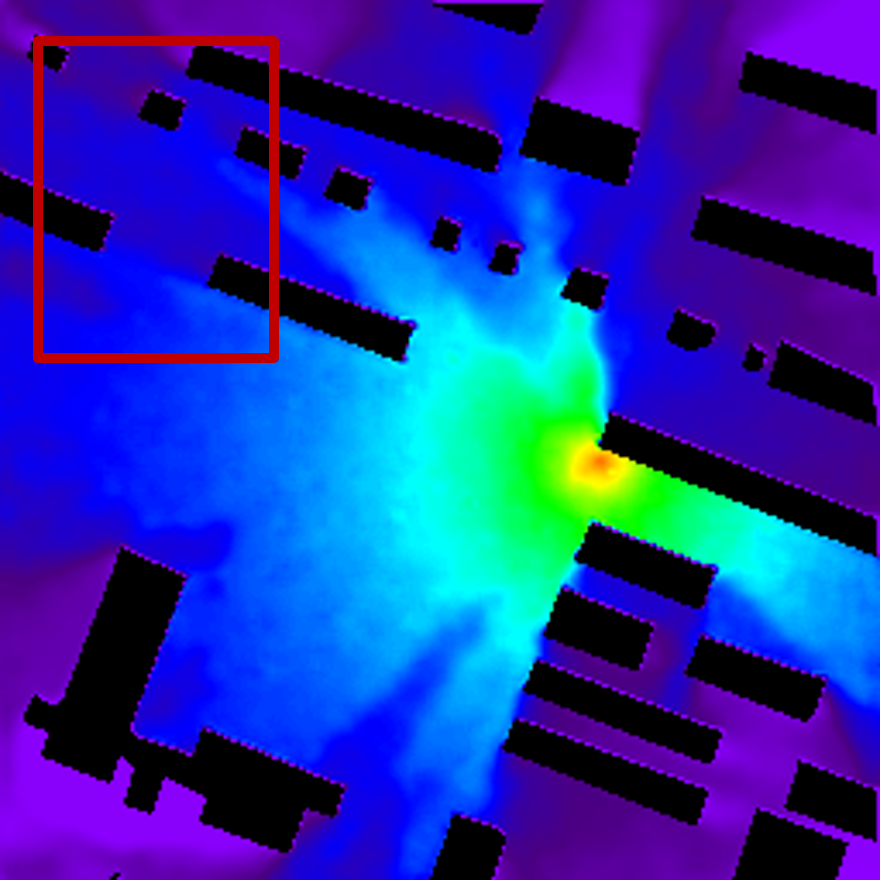}} &
  \subcaptionbox{}{\includegraphics[width=0.13\linewidth]{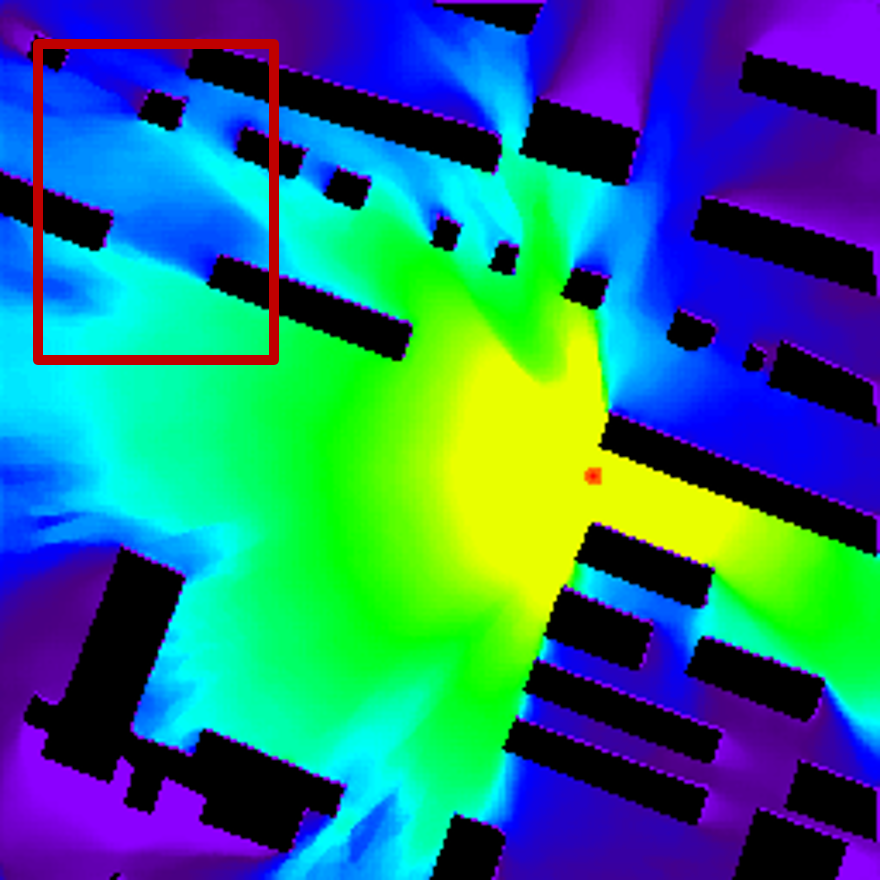}} &
  \subcaptionbox{}{\includegraphics[width=0.13\linewidth]{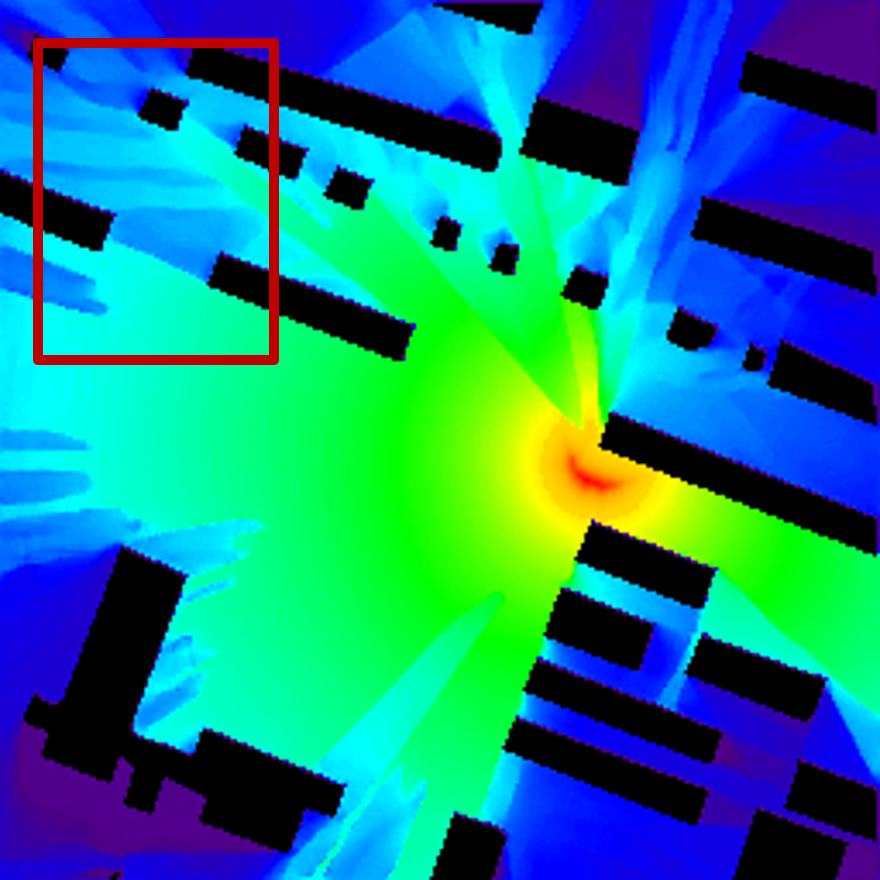}} &
  \subcaptionbox{}{\includegraphics[width=0.13\linewidth]{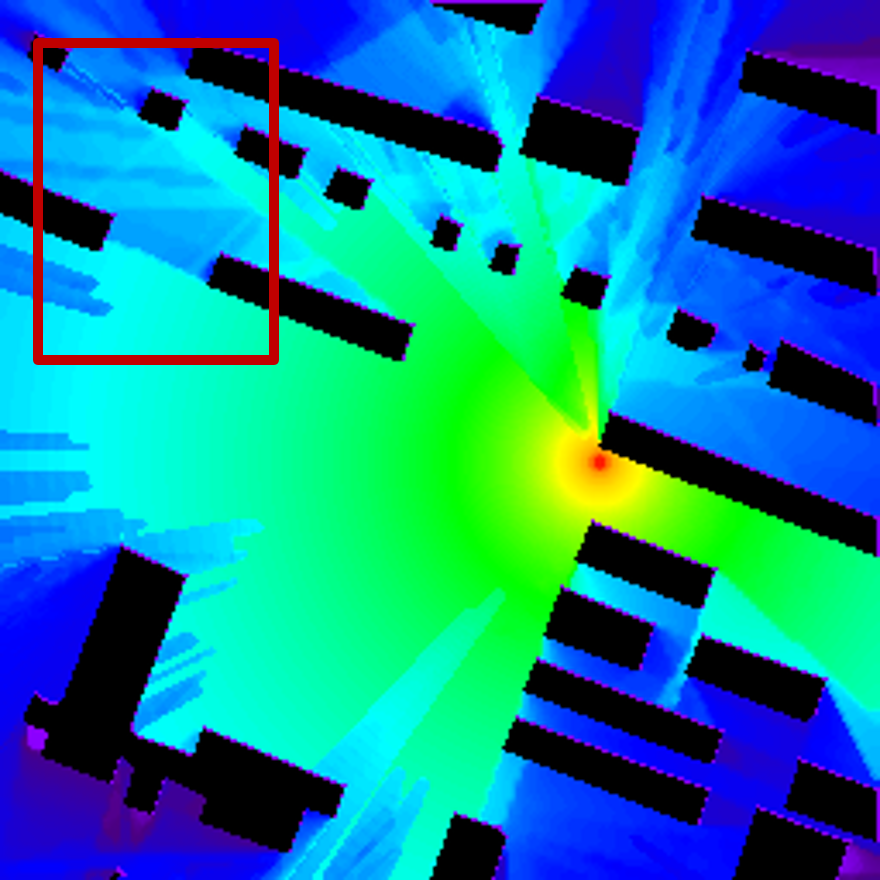}} \\
  \small\bfseries RadioUNet &  
  \small\bfseries PhyRMDM &          
  \small\bfseries RadioDiff &      
  \small\bfseries RME-GAN &       
  \small\bfseries \begin{tabular}{@{}c@{}}RadioDiff-FS \\ (one-shot)\end{tabular} & 
  \small\bfseries RadioDiff-FS &     
  \small\bfseries Ground Truth    
\end{tabular}
\caption{Visual comparison of constructed DRM across different methods.}
\label{fig:IRT4_Car_CP} 
\end{figure*}

We compare RadioDiff-FS with representative baselines covering supervised regression, adversarial generation, and diffusion-based generation. To isolate the effect of the proposed few-shot adaptation, all methods are trained from scratch on the same few-shot supervision from the IRT4 subset. Specifically, all methods use $500$ training scenes with $2$ transmitter placements per scene, and are evaluated on the identical scene-disjoint test split. We additionally report RadioDiff-FS in a one-shot setting, where only a single MU-RM sample per scene is used for fine-tuning, to examine performance under extreme data scarcity. The baselines are summarized as follows.
\begin{enumerate}
    \item \textbf{RadioUNet}~\cite{levie2021radiounet}: A sampling-free convolutional regressor based on a U-Net backbone that predicts RMs directly from environment and transmitter inputs.
    \item \textbf{RME-GAN}~\cite{zhang2023rme}: An adversarial generator adapted to the sampling-free protocol by using only environmental features and transmitter metadata for direct RM synthesis.
    \item \textbf{PhyRMDM}~\cite{jia2025rmdm}: A conditional diffusion framework that generates high-resolution RMs in the image domain, augmented with a physics-alignment loss based on a Helmholtz-equation residual.
    \item \textbf{RadioDiff}~\cite{wang2024radiodiff}: A diffusion-based sampling-free RM generator that serves as a strong baseline under the same conditioning inputs but without the proposed directional regularization.
\end{enumerate}
RadioDiff is particularly relevant because it belongs to the same backbone family as our approach. It directly maps the environment and transmitter inputs to the RM without the proposed main-path to multipath guidance mechanism. It therefore serves as a controlled ablation baseline that isolates the contribution of the directional consistency constraint while keeping the generative capacity unchanged.

\subsection{Performance on SRM}
 
Quantitative results on the SRM are reported in Table~\ref{tab:comp_irt4}. RadioDiff-FS achieves the best performance across all metrics, attaining an NMSE of $0.0049$ and a PSNR of $36.37$~dB. Compared with the vanilla RadioDiff baseline at NMSE $0.0121$ and PhyRMDM at NMSE $0.0100$, RadioDiff-FS reduces the reconstruction error by $59.50\%$. This result indicates that few-shot training alone is insufficient for high-fidelity multipath reconstruction without additional adaptation guidance. RadioDiff-FS also yields the highest SSIM of $0.9752$, which substantially exceeds all other methods and suggests improved preservation of spatial structures such as diffraction edges and shadow boundaries. The one-shot variant of RadioDiff-FS also demonstrates competitive performance. With only a single MU-RM sample per scene for fine-tuning, RadioDiff-FS achieves the second-best SSIM of $0.9285$ and the second-best PSNR of $31.77$~dB, outperforming all baselines except the full RadioDiff-FS. This result validates that the directional consistency constraint provides a strong inductive bias even under extreme data scarcity, as it leverages the pretrained MP-RM prior to guide adaptation with minimal supervision.
 
These quantitative trends are consistent with the qualitative results in Fig.~\ref{fig:IRT4_CP}. RadioUNet produces smooth attenuation fields that capture coarse pathloss gradients but blur sharp transitions near building silhouettes. These transitions are critical in multipath-dominant settings. RME-GAN generates sharper textures, yet it exhibits intermittent artifacts in low-signal regions where adversarial objectives may amplify visually plausible but physically inconsistent patterns. Diffusion-based baselines improve perceptual realism. However, without an explicit cross-domain constraint, they still under-represent boundary sharpness and local contrast in deep-shadow areas. In contrast, RadioDiff-FS better matches the ground truth both globally and locally. It preserves the large-scale attenuation skeleton while recovering fine-grained transition bands around occluded corridors, aligning with the intended main-path to multipath generalization behavior.

\subsection{Performance on DRM}
 
The DRM further introduces time-varying occlusions from vehicles, which produce localized attenuation spots and small-scale shadowing patterns along roads. As shown in Table~\ref{tab:comp_irt4_car}, most baselines degrade noticeably when moving from SRM to DRM. PhyRMDM suffers a sharp deterioration with NMSE increasing to $0.1113$, indicating that physics-alignment alone does not guarantee robustness when the target domain contains sparse and configuration-dependent perturbations under limited supervision. The vanilla RadioDiff baseline also degrades significantly, reflecting that the pretrained generative prior may not correctly allocate capacity to vehicle-induced details when trained with few DRM samples. RadioDiff-FS remains substantially more stable in this regime, achieving an NMSE of $0.0121$ and an SSIM of $0.9510$, which are markedly better than all baselines. The one-shot variant again delivers strong results, attaining the second-best RMSE of $0.0296$ and the second-best SSIM of $0.8969$. This confirms that the directional consistency constraint is effective even with a single fine-tuning sample per scene.
 
The qualitative comparison in Fig.~\ref{fig:IRT4_Car_CP} explains this stability. Vehicle occlusions manifest as compact attenuation patches whose locations are coupled to road geometry and line-of-sight corridors. Regression and diffusion baselines tend to smooth these patches into the background, leading to systematic underestimation of localized blockage. RadioDiff-FS more consistently recovers these attenuation signatures and their spatial extent, producing DRM reconstructions that are closer to the ground truth in both contrast and location. This behavior is particularly important for downstream tasks that depend on local reliability margins, where missing a small occlusion-induced dip can lead to incorrect link-quality assessment.

\subsection{Ablation Study}
 
We ablate the proposed direction-consistency loss (DCL) to quantify its contribution under both full fine-tuning and low-rank adaptation (LoRA). The DCL constrains the feature displacement of generated maps to follow the dominant main-path to multipath shift while suppressing orthogonal distortions. This empirically translates to sharper boundaries and fewer spurious artifacts in low-signal regions. Table~\ref{tab:ablation_irt4_car} presents results on the more challenging DRM setting. Removing DCL degrades performance consistently. Under full fine-tuning, adding DCL reduces NMSE from $0.0163$ to $0.0121$ and improves SSIM from $0.9105$ to $0.9510$. The SSIM gain indicates that DCL primarily improves structural consistency rather than merely rescaling pixel intensities, which matches the goal of preserving interaction-induced patterns. The improvement remains visible under LoRA, where DCL reduces NMSE from $0.0259$ to $0.0195$ and improves SSIM from $0.8750$ to $0.8982$. This suggests that DCL is complementary to parameter-efficient adaptation and helps concentrate limited update capacity on multipath-relevant corrections.
 
Similar trends are observed on SRM in Table~\ref{tab:ablation_irt4}. Although SRM is less challenging than DRM, DCL still yields consistent gains for both adaptation strategies. Under full fine-tuning, DCL improves PSNR from $35.79$~dB to $36.37$~dB and SSIM from $0.9715$ to $0.9752$. Under LoRA, the gains follow the same direction with PSNR improving from $31.88$~dB to $32.65$~dB. Taken together, these ablations confirm that the advantage of RadioDiff-FS does not stem solely from additional training steps. The proposed directional constraint provides a principled and effective inductive bias for few-shot generalization from main-path structure to multipath detail.

\begin{table}[H]
\centering
\caption{\textbf{Ablation Study on IRT4.} Results highlight the impact of DCL across different fine-tuning strategies.}
\vspace{-4pt}
\captionsetup{font={small}, skip=8pt}
\resizebox{0.95\linewidth}{!}{
\begin{tabular}{@{}ll|cccc@{}}
\toprule
\multicolumn{2}{c|}{\textbf{Configuration}} & \textbf{NMSE} & \textbf{RMSE} & \textbf{PSNR (dB)} & \textbf{SSIM} \\ \midrule
\multirow{2}{*}{\textbf{Full FT}} 
  & w/o DCL & 0.0053 & 0.0204 & 35.79 & 0.9715 \\
  & w/ DCL  & {\color[HTML]{9A0000} \textbf{0.0049}} & {\color[HTML]{9A0000} \textbf{0.0196}} & {\color[HTML]{9A0000} \textbf{36.37}} & {\color[HTML]{9A0000} \textbf{0.9752}} \\ \midrule
\multirow{2}{*}{\textbf{LoRA}} 
  & w/o DCL & 0.0069 & 0.0233 & 31.88 & 0.9550 \\
  & w/ DCL  & 0.0064 & 0.0225 & 32.65 & 0.9582 \\ \bottomrule
\end{tabular}
}
\vspace{-12pt}
\label{tab:ablation_irt4}
\end{table}

\begin{table}[H]
\centering
\caption{\textbf{Ablation Study on IRT4 With Car.} Comparison of full fine-tuning and LoRA strategies with and without direction-consistency loss (DCL).}
\vspace{-4pt}
\captionsetup{font={small}, skip=8pt}
\resizebox{0.95\linewidth}{!}{
\begin{tabular}{@{}ll|cccc@{}}
\toprule
\multicolumn{2}{c|}{\textbf{Configuration}} & \textbf{NMSE} & \textbf{RMSE} & \textbf{PSNR (dB)} & \textbf{SSIM} \\ \midrule
\multirow{2}{*}{\textbf{Full FT}} 
  & w/o DCL & 0.0163 & 0.0264 & 31.79 & 0.9105 \\
  & w/ DCL  & {\color[HTML]{9A0000} \textbf{0.0121}} & {\color[HTML]{9A0000} \textbf{0.0214}} & {\color[HTML]{9A0000} \textbf{32.68}} & {\color[HTML]{9A0000} \textbf{0.9510}} \\ \midrule
\multirow{2}{*}{\textbf{LoRA}} 
  & w/o DCL & 0.0259 & 0.0302 & 28.88 & 0.8750 \\
  & w/ DCL  & 0.0195 & 0.0270 & 30.65 & 0.8982 \\ \bottomrule
\end{tabular}
}
\vspace{-12pt}
\label{tab:ablation_irt4_car}
\end{table}

\section{Conclusion}
We have presented RadioDiff-FS, a few-shot adaptation framework for multipath RM construction that transfers a pretrained main-path prior to multipath-rich environments via a direction-consistency constraint in a frozen feature space. Experiments have demonstrated that the proposed adaptation improves both pixel-level accuracy and structural fidelity on static and dynamic RMs, which can translate into more reliable coverage assessment and blockage-aware resource planning in practical communication systems under scarce high-fidelity supervision. Future work will extend the framework to multi-frequency and multi-antenna RM targets and validate the adaptation with real-world measurement-driven datasets.

\bibliography{ref}

@inproceedings{radiogat,
  title={RadioGAT: A Model-Based Learning Framework for Radio Map Reconstruction via Graph Attention Networks},
  author={Li, Xiaojie and Li, Hang and Li, Xiaoyang and Zhu, Guangxu and Qi, Nan and Xiao, Ming},
  booktitle={2025 IEEE Wireless Communications and Networking Conference (WCNC)},
  pages={1--6},
  year={2025},
  organization={IEEE}
}

@article{wang2025radiodiff,
  title={RadioDiff-Inverse: Diffusion Enhanced Bayesian Inverse Estimation for ISAC Radio Map Construction},
  author={Wang, Xiucheng and Fang, Zhongsheng and Cheng, Nan and Sun, Ruijin and Li, Zan and others},
  journal={arXiv preprint arXiv:2504.14298},
  year={2025}
}

@article{deeprem,
  title={DeepREM: Deep-learning-based radio environment map estimation from sparse measurements},
  author={Chaves-Villota, Andrea and Viteri-Mera, Carlos A},
  journal={IEEE Access},
  volume={11},
  pages={48697--48714},
  year={2023},
  publisher={IEEE}
}

@article{bhalla1996three,
  title={Three-dimensional scattering center extraction using the shooting and bouncing ray technique},
  author={Bhalla, Rajan and Ling, Hao},
  journal={IEEE Transactions on Antennas and Propagation},
  volume={44},
  number={11},
  pages={1445--1453},
  year={1996},
  publisher={IEEE}
}

@inproceedings{wolfle1997intelligent,
  title={Intelligent ray tracing-a new approach for field strength prediction in microcells},
  author={Wolfle, G and Gschwendtner, BE and Landstorfer, FM},
  booktitle={1997 IEEE 47th Vehicular Technology Conference. Technology in Motion},
  volume={2},
  pages={790--794},
  year={1997},
  organization={IEEE}
}

@inproceedings{zhang2024metadiff,
  title={Metadiff: Meta-learning with conditional diffusion for few-shot learning},
  author={Zhang, Baoquan and Luo, Chuyao and Yu, Demin and Li, Xutao and Lin, Huiwei and Ye, Yunming and Zhang, Bowen},
  booktitle={Proceedings of the AAAI conference on artificial intelligence},
  volume={38},
  number={15},
  pages={16687--16695},
  year={2024}
}

@book{goodfellow2016deep,
  title={Deep learning},
  author={Goodfellow, Ian and Bengio, Yoshua and Courville, Aaron},
  year={2016},
  publisher={MIT press}
}

@book{vapnik1998statistical,
  title={Statistical learning theory},
  author={Vapnik, Vladimir Naumovich and Vapnik, Vlamimir and others},
  year={1998},
  publisher={wiley New York}
}

@article{ho2020denoising,
  title={Denoising diffusion probabilistic models},
  author={Ho, Jonathan and Jain, Ajay and Abbeel, Pieter},
  journal={Advances in neural information processing systems},
  volume={33},
  pages={6840--6851},
  year={2020}
}

@inproceedings{yapar2023first,
   title={The First Pathloss Radio Map Prediction Challenge},
  author={Yapar, {\c{C}}a{\u{g}}kan and Jaensch, Fabian and Levie, Ron and Kutyniok, Gitta and Caire, Giuseppe},
  booktitle={Proceedings of the 2023 IEEE International Conference on Acoustics, Speech and Signal Processing (ICASSP)},
  pages={1--2},
  year={2023},
}

@article{shen2023toward,
  title={Toward immersive communications in {6G}},
  author={Shen, Xuemin and Gao, Jie and Li, Mushu and Zhou, Conghao and Hu, Shisheng and He, Mingcheng and Zhuang, Weihua},
  journal={Frontiers in Computer Science},
  volume={4},
  pages={1068478},
  year={2023},
  publisher={Frontiers Media SA}
}

@article{zhang2023rme,
  title={{RME}-{GAN}: A learning framework for radio map estimation based on conditional generative adversarial network},
  author={Zhang, Songyang and Wijesinghe, Achintha and Ding, Zhi},
  journal={IEEE Internet Things J.},
  year={2023},
  volume={10},
  number={20},
  pages={18016-18027},
  publisher={IEEE}
}

@article{chen2023graph,
  title={A graph neural network based radio map construction method for urban environment},
  author={Chen, Guokai and Liu, Yongxiang and Zhang, Tao and Zhang, Jianzhao and Guo, Xiye and Yang, Jun},
  journal={IEEE Commun. Lett.},
  year={2023},
  publisher={IEEE}
}

@article{wang2022joint,
  title={Joint flying relay location and routing optimization for {6G} {UAV}--{IoT} networks: A graph neural network-based approach},
  author={Wang, Xiucheng and Fu, Lianhao and Cheng, Nan and Sun, Ruijin and Luan, Tom and Quan, Wei and Aldubaikhy, Khalid},
  journal={Remote Sens.},
  volume={14},
  number={17},
  pages={4377},
  year={2022},
  publisher={MDPI}
}

@article{wang2004image,
  title={Image quality assessment: from error visibility to structural similarity},
  author={Wang, Zhou and Bovik, Alan C and Sheikh, Hamid R and Simoncelli, Eero P},
  journal={IEEE Trans. Image Processing},
  volume={13},
  number={4},
  pages={600--612},
  year={2004},
  publisher={IEEE}
}

@article{song2020score,
  title={Score-based generative modeling through stochastic differential equations},
  author={Song, Yang and Sohl-Dickstein, Jascha and Kingma, Diederik P and Kumar, Abhishek and Ermon, Stefano and Poole, Ben},
  journal={arXiv preprint arXiv:2011.13456},
  year={2020}
}

@article{deschamps1972ray,
  title={Ray techniques in electromagnetics},
  author={Deschamps, Georges A},
  journal={Proc. {IEEE}},
  volume={60},
  number={9},
  pages={1022--1035},
  year={1972},
  publisher={IEEE}
}

@article{wang2024tutorial,
  title={A tutorial on extremely large-scale {MIMO} for {6G}: Fundamentals, signal processing, and applications},
  author={Wang, Zhe and Zhang, Jiayi and Du, Hongyang and Niyato, Dusit and Cui, Shuguang and Ai, Bo and Debbah, M{\'e}rouane and Letaief, Khaled B and Poor, H Vincent},
  journal={IEEE Commun. Surveys Tuts.},
  volume={26},
  number={3},
  pages={1560--1605},
  year={2024},
  publisher={IEEE}
}

@inproceedings{irt,
  title={Verifying path loss and delay spread predictions of a {3D} ray tracing propagation model in urban environment},
  author={Rautiainen, Terhi and Wolfle, G and Hoppe, Reiner},
  booktitle={Proceedings IEEE 56th Vehicular Technology Conference},
  volume={4},
  pages={2470--2474},
  year={2002},
  organization={IEEE}
}

@inproceedings{dpm,
  title={Dominant path prediction model for urban scenarios},
  author={Wahl, Ren{\'e} and W{\"o}lfle, Gerd and Wertz, Philipp and Wildbolz, Pascal and Landstorfer, Friedrich},
  booktitle={Proceedings of 14th IST mobile and wireless communications summit},
  pages={1--5},
  year={2005}
}

@inproceedings{jia2025rmdm,
  title={Physics-informed representation alignment for sparse radio-map reconstruction},
  author={Jia, Haozhe and Chen, Wenshuo and Huang, Zhihui and Wang, Lei and Xiao, Hongru and Jia, Nanqian and Wu, Keming and Lai, Songning and Tian, Bowen and Yue, Yutao},
  booktitle={Proceedings of the 33rd ACM International Conference on Multimedia},
  pages={12352--12360},
  year={2025}
}

@book{jones2013theory,
  title={The theory of electromagnetism},
  author={Jones, Douglas Samuel},
  year={2013},
  publisher={Elsevier}
}

@article{levie2021radiounet,
  title={{RadioUNet}: Fast radio map estimation with convolutional neural networks},
  author={Levie, Ron and Yapar, {\c{C}}a{\u{g}}kan and Kutyniok, Gitta and Caire, Giuseppe},
  journal={ IEEE Trans. Wireless Commun.},
  volume={20},
  number={6},
  pages={4001--4015},
  year={2021},
  publisher={IEEE}
}

@ARTICLE{11083758,
  author={Wang, Xiucheng and Zhang, Qiming and Cheng, Nan and Chen, Junting and Zhang, Zezhong and Li, Zan and Cui, Shuguang and Shen, Xuemin},
  journal={IEEE Transactions on Network Science and Engineering}, 
  title={RadioDiff-3D: A 3D× 3D Radio Map Dataset and Generative Diffusion Based Benchmark for 6G Environment-Aware Communication}, 
  year={2026},
  volume={13},
  number={},
  pages={3773-3789},
  doi={10.1109/TNSE.2025.3590545}}

@ARTICLE{11278649,
  author={Wang, Xiucheng and Zhang, Qiming and Cheng, Nan and Sun, Ruijin and Li, Zan and Cui, Shuguang and Shen, Xuemin},
  journal={IEEE Journal on Selected Areas in Communications}, 
  title={RadioDiff-k2: Helmholtz Equation Informed Generative Diffusion Model for Multi-Path Aware Radio Map Construction}, 
  year={2025},
  volume={},
  number={},
  pages={1-1},
  doi={10.1109/JSAC.2025.3641105}}

@ARTICLE{11314850,
  author={Cheng, Nan and Yang, Shuangyu and Sun, Ruijin and Yin, Zhisheng and Shao, Xiaodan and Zhuang, Weihua and Shen, Xuemin},
  journal={IEEE Transactions on Wireless Communications}, 
  title={Channel Knowledge Map-Enabled 6D Movable Antenna Systems With Kinematic Constraints: A Manifold Optimization Approach}, 
  year={2026},
  volume={25},
  number={},
  pages={8968-8981},
  doi={10.1109/TWC.2025.3645480}}

@INPROCEEDINGS{11152929,
  author={Wang, Xiucheng and Zheng, Peilin and Cheng, Nan and Sun, Ruijin and Chen, Junting and Tao, Keda and Yin, Zhisheng and Liu, Zhiquan and Zeng, Yong},
  booktitle={IEEE INFOCOM 2025 - IEEE Conference on Computer Communications Workshops (INFOCOM WKSHPS)}, 
  title={RadioDiff-Turbo: Lightweight Generative Large Electromagnetic Model for Wireless Digital Twin Construction}, 
  year={2025},
  volume={},
  number={},
  pages={1-6},
  doi={10.1109/INFOCOMWKSHPS65812.2025.11152929}}

@ARTICLE{8373698,
  author={Wang, Xiong and Kong, Linghe and Kong, Fanxin and Qiu, Fudong and Xia, Mingyu and Arnon, Shlomi and Chen, Guihai},
  journal={IEEE Communications Surveys \& Tutorials}, 
  title={Millimeter Wave Communication: A Comprehensive Survey}, 
  year={2018},
  volume={20},
  number={3},
  pages={1616-1653},
  doi={10.1109/COMST.2018.2844322}}

@article{liu2025foundation,
  title={Foundation Model for Intelligent Wireless Communications},
  author={Liu, Boxun and Liu, Xuanyu and Gao, Shijian and Cheng, Xiang and Yang, Liuqing},
  journal={arXiv preprint arXiv:2511.22222},
  year={2025}
}

@INPROCEEDINGS{9753644,
  author={Tian, Yu and Yuan, Shuai and Chen, Weisheng and Liu, Naijin},
  booktitle={2021 13th International Symposium on Antennas, Propagation and EM Theory (ISAPE)}, 
  title={Transformer based Radio Map Prediction Model for Dense Urban Environments}, 
  year={2021},
  volume={Volume1},
  number={},
  pages={1-3},
  doi={10.1109/ISAPE54070.2021.9753644}}

@ARTICLE{11036155,
  author={Yang, Tingting and Zhang, Ping and Zheng, Mengfan and Shi, Yuxuan and Jing, Liwen and Huang, Jianbo and Li, Nan},
  journal={IEEE Network}, 
  title={WirelessGPT: A Generative Pre-Trained Multi-Task Learning Framework for Wireless Communication}, 
  year={2025},
  volume={39},
  number={5},
  pages={58-65},
  doi={10.1109/MNET.2025.3579496}}

@ARTICLE{11282987,
  author={Wang, Xiucheng and Zheng, Peilin and Jia, Honggang and Cheng, Nan and Sun, Ruijin and Zhou, Conghao and Shen, Xuemin},
  journal={IEEE Transactions on Cognitive Communications and Networking}, 
  title={RadioDiff-Flux: Efficient Radio Map Construction via Generative Denoise Diffusion Model Trajectory Midpoint Reuse}, 
  year={2026},
  volume={12},
  number={},
  pages={4882-4895},
  doi={10.1109/TCCN.2025.3641513}}

@article{wang2024radiodiff,
  title={{RadioDiff}: An Effective Generative Diffusion Model for Sampling-Free Dynamic Radio Map Construction},
  author={Wang, Xiucheng and Tao, Keda and Cheng, Nan and Yin, Zhisheng and Li, Zan and Zhang, Yuan and Shen, Xuemin},
  journal={IEEE Trans. Cognit. Commun. Networking, Early access},
  year={2024},
  pages={1-13},
  publisher={IEEE}
}

@article{wang2026tutorial,
  title={A Tutorial on Learning-Based Radio Map Construction: Data, Paradigms, and Physics-Awarenes},
  author={Wang, Xiucheng and Pan, Yuhao and Cheng, Nan},
  journal={arXiv preprint arXiv:2603.17499},
  year={2026}
}

@article{Yang2023a,
    title = {{6G Network AI Architecture for Everyone-Centric Customized Services}},
    year = {2023},
    journal = {IEEE Network},
    author = {Yang, Yang and Ma, Mulei and Wu, Hequan and Yu, Quan and You, Xiaohu and Wu, Jianjun and Peng, Chenghui and Yum, Tak Shing Peter and Aghvami, A. Hamid and Li, Geoffrey Y. and Wang, Jiangzhou and Liu, Guangyi and Gao, Peng and Tang, Xiongyan and Cao, Chang and Thompson, John and Wong, Kat Kit and Chen, Shanzhi and Wang, Zhiqin and Debbah, Merouane and Dustdar, Schahram and Eliassen, Frank and Chen, Tao and Duan, Xiangyang and Sun, Shaohui and Tao, Xiaofeng and Zhang, Qinyu and Huang, Jianwei and Zhang, Wenjun and Li, Jie and Gao, Yue and Zhang, Honggang and Chen, Xu and Ge, Xiaohu and Xiao, Yong and Wang, Cheng Xiang and Zhang, Zaichen and Ci, Song and Mao, Guoqiang and Li, Changle and Shao, Ziyu and Zhou, Yong and Liang, Junrui and Li, Kai and Wu, Liantao and Sun, Fanglei and Wang, Kunlun and Liu, Zening and Yang, Kun and Wang, Jun and Gao, Teng and Shu, Hongfeng},
    number = {5},
    pages = {71--80},
    volume = {37},
    publisher = {IEEE},
    doi = {10.1109/MNET.124.2200241},
    issn = {1558156X}
}
\bibliographystyle{IEEEtran}
\ifCLASSOPTIONcaptionsoff
  \newpage·
\fi
\end{document}